\documentclass[letterpaper,11pt,tutorial]{article}   

\usepackage{aop}

\usepackage[pdftex,colorlinks=true,bookmarks=false,linkcolor=AOP,citecolor=blue,filecolor=blue,urlcolor=blue,pdfstartview=FitH]{hyperref}

\usepackage{amsmath}
\usepackage{amsfonts}
\usepackage{wrapfig}
\def\omex{ Opt.\ Mat.\ Express }
\def\pr{ Phys.\ Rev. }
\def\jlt{ J.\ Lightwave\ Technol.\ }

\def\claop{\color{AOP}}

\begin{document}
\title{Transverse Anderson localization of light: a tutorial review}

\author{Arash Mafi}
\address{Department of Physics and Astronomy\\ Center for High Technology Materials\\ University of New Mexico \\Albuquerque, NM 87131, USA}
\address{mafi@unm.edu}

\begin{abstract}{
This tutorial review gives an overview of the transverse Anderson localization of light in one and two transverse dimensions. A pedagogical
approach is followed throughout the presentation, where many aspects of localization are illustrated
by means of a few simple models. The tutorial starts with some basic aspects of random matrix theory, and light propagation through and
reflection from a random stack of dielectric slabs. Transverse Anderson localization of light in one- and two-dimensional coupled waveguide 
arrays is subsequently established and discussed. Recent experimental observations of localization and image transport in disordered optical
fibers are discussed. More advanced topics, such as hyper-transport in longitudinally varying disordered waveguides, the impact of nonlinearity, 
and propagation of partially coherent and quantum light, are also examined.}
\end{abstract}
\newpage
\tableofcontents
\setcounter{tocdepth}{2}

\newpage
\section{Introduction}
\label{sec:overview}
Anderson localization is the absence of diffusive wave transport in highly disordered 
scattering media~\cite{Anderson1,EAbrahams,John1,Sheng}. It was first introduced in a pioneering
theoretical study in 1958 by Philip Warren Anderson~\cite{Anderson1}, who investigated the behavior 
of spin diffusion and electronic conduction in random lattices. 
It took more than ten years for the scientific community to recognize the importance of Anderson's work.
However, it has remained at the forefront of physics research since 1968. There are still many
uncertainties and unanswered questions in the linear and nonlinear behavior of disordered systems in 
various dimensions.

The model that Anderson studied involved an electron on a potential lattice with a random spread in the energies 
of the sites caused by a source of disorder. The electron was allowed to hop between sites via nearest 
neighbor potential coupling terms. Anderson showed that the wavefunction of electron localizes to only few 
sites at all times, provided that the amount of randomness is sufficiently large.

It did not take long for Anderson and others to realize 
that the novel localization phenomenon was due to the wave nature of the quantum mechanical electrons
scattering in a disordered potential, and that similar behavior should also be observed in other coherent wave systems, 
including classical ones~\cite{John1,John2,Anderson2,John3,Lagendijk1}. 
The disorder-induced localization 
in electronic systems was shown to be inhibited by thermal fluctuations and nonlinear effects; therefore,
it was encouraging to find other avenues in which the disorder-induced Anderson localization could be observed.
Subsequently, localization was studied in various classical wave systems including acoustics, elastics,
electromagnetics, optics~\cite{John1,John2,Anderson2,Lagendijk1,Graham,Hu,Chabanov,John3}, and various quantum optical systems, such as atomic 
lattices~\cite{Billy} and propagating photons~\cite{Lahini2,Lahini3,Abouraddy,Thompson}.

Optical systems have played a unique role in the fundamental understanding and experimental observation of 
Anderson localization. Optical phenomena are easy to ``visualize,'' and there are many advanced tools and 
techniques in optics that can be used to study the physics of localization. Optical studies of Anderson 
localization can often be done with tools that are widely accessible and can be performed in a single laboratory. 
In addition, Anderson localization already has device-level applications in the optics~\cite{SalmanOL,SalmanMB,SalmanNature},
and optics can ``illuminate'' the path to localization-based devices in other disordered classical and quantum wave systems.

It has been shown that coherent waves in one-dimensional (1D) and and two-dimensional (2D) unbounded disordered systems
are always localized~\cite{Abrahams}. For bounded 1D and 2D systems, if the sample size is considerably larger than the
localization radius, the boundary effects are minimal and can often be ignored~\cite{Szameit2010,Jovic}. However, in three-dimensional (3D)
coherent wave systems, the scattering strength needs to be larger than a threshold value for the localization to happen~\cite{Sperling}.
The scattering strength is characterized by the wave scattering transport length $l^\ast$ (shorter $l^\ast$ means 
stronger scattering), and the Ioffe-Regel condition~\cite{Ioffe} states that in order to observe Anderson 
localization, the disorder must be strong enough that the wave scattering transport length becomes on the order 
of the wavelength. The Ioffe-Regel condition is often cast in the form of $kl^\ast\sim 1$, where $k$ is the effective 
wavevector in the medium.

It is notoriously difficult to satisfy in 3D disordered-media. For example, for the optical field
to localize in 3D, very large refractive index contrasts are required that are not generally
available in low-loss optical materials~\cite{John3}. ``{\em The fact that Anderson localization is hard to achieve 
in 3D optical systems may be a blessing in disguise; otherwise, no sunlight would reach the earth 
on highly cloudy~days''}~\cite{MafiMDPI}.
In order to observe Anderson localization of light, strongly scattering materials at optical
and near infrared frequencies such as TiO$_2$, GaAs, GaP, Si, and Ge nanoparticles can be used. Careful
measurements are required because bulk absorption can easily lead to experimental signatures similar to Anderson localization
\cite{Wiersma,Scheffold,vanderBeek}.

Unlike 3D lightwave systems, in which observation of localization is prohibitively difficult, 
observation of Anderson localization in quasi-2D and -1D optical systems (transverse Anderson localization) 
is readily possible, as was first shown by Abdullaev \textit {et al}.~\cite{Abdullaev} and De~Raedt \textit {et al}.~\cite{DeRaedt}.
There have since been many reports on the observation of transverse Anderson localization of light in 1D and 2D, 
which is the main focus of this tutorial review, as well. Transverse Anderson localization is attractive because
of its relative simplicity, ease of experimentation, and the rewarding physical insights it brings about on many fronts.  
Moreover, the longitudinal coordinate along the direction of propagation plays the role of time in a 2D 
disordered system; therefore, controlled temporal variations can also be studied in these systems. 

There are many excellent reviews that cover various aspects of wave propagation in disordered 
systems as well as Anderson localization~\cite{John3,Wiersma,Sheng,Ziman,Lifshits,PALee,BKramer,Beenakker,LagendijkBook,SegevNaturePhotonicsReview}.
This tutorial review is neither intended to be comprehensive, nor is it intended to explore each area it covers in great depth. Rather,
the intent is to provide a pedestrian and intuitive approach to Anderson localization, mainly focused on the transverse localization of 
light. The coverage of topics is inevitably slanted toward those of particular interest to the author. The pedagogical approach 
is intended to benefit both newcomers to this rewarding research field, as well as outsiders who are interested to learn
about Anderson localization. The author seeks forgiveness from those whose work is not mentioned here, as well as for any 
technical errors or omissions. 
\section{A random matrix example}
\label{sec:randommatrix}
A good way to build an intuition about the relationship between randomness and localization is to use
random matrices~\cite{Beenakker,RandomMatrix}. The following example shows that the extended eigenvectors of an ordered matrix 
become very localized when some randomness is added to the elements of a matrix.

Consider a symmetric tridiagonal $N\times N$ matrix ${\mathbb M}$ defined as
\begin{align}
{\mathbb M}_{i,i}=1,\quad {\mathbb M}_{i,i+1}={\mathbb M}_{i+1,i}=0.1,
\label{eq:defMorederd}
\end{align} 
for all possible values of $i$. For definiteness in this numerical example, we consider $N=200$.
Matrix ${\mathbb M}$ has $N$ real eigenvectors, and each eigenvector is an $N$-element vector.
We identify the $i$th eigenvector as ${\mathbb V}^{(i)}$, and ${\mathbb V}^{(i)}_j$ represents
its $j$th element.

\begin{figure}[b]
\centering\includegraphics[width=5in]{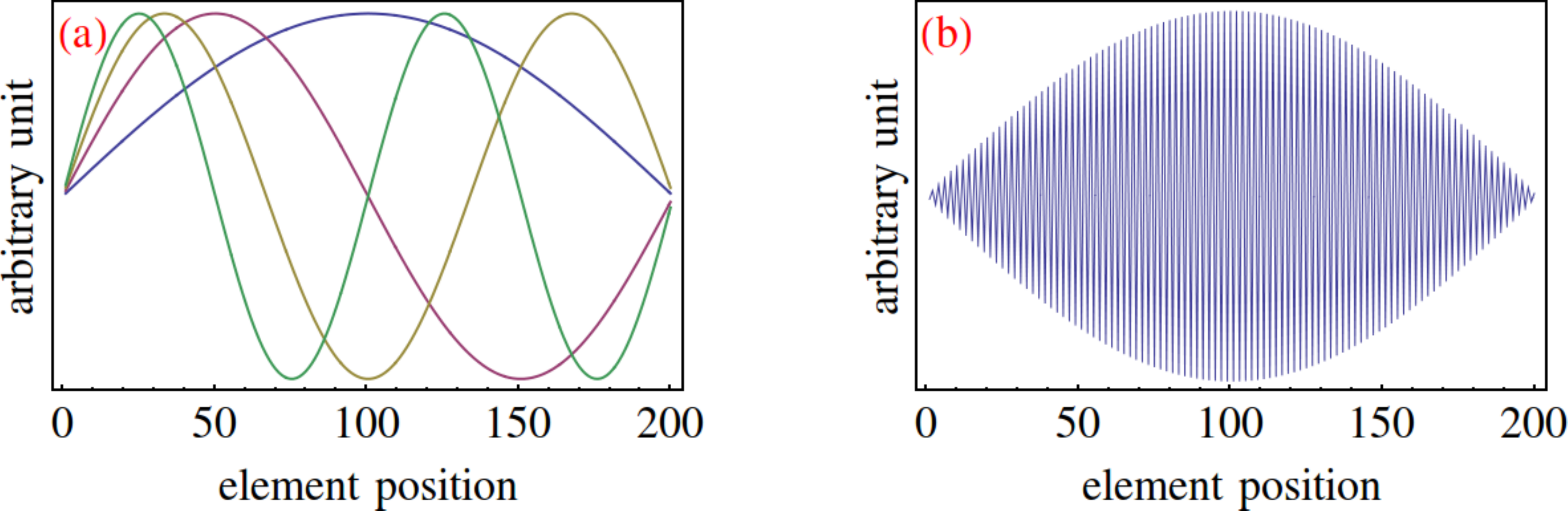}
\caption{
This figure shows that the eigenvectors of the ordered matrix ${\mathbb M}$ defined in 
Eq.~\ref{eq:defMorederd} are extended over the entire element-position domain. 
Eigenvectors ${\mathbb V}^{(1)}_j, {\mathbb V}^{(2)}_j, {\mathbb V}^{(3)}_j$, 
and ${\mathbb V}^{(4)}_j$ are plotted in {\claop (a)}; and  ${\mathbb V}^{(200)}_j$ is plotted in {\claop (b)} as a function of the element position $j$.
}
\label{fig:orderedMatrix1}
\end{figure}
In Figure~\ref{fig:orderedMatrix1}, a few eigenvectors ${\mathbb V}^{(i)}_j$ are plotted as a function 
of their element number $j$. In Figure~\ref{fig:orderedMatrix1}{\claop (a)}, ${\mathbb V}^{(1)}_j, {\mathbb V}^{(2)}_j, {\mathbb V}^{(3)}_j$, 
and ${\mathbb V}^{(4)}_j$ are plotted and are all oscillatory functions of $j$, where the $i$th eigenvector 
is identified by $i+1$ oscillation nodes. ${\mathbb V}^{(200)}_j$ is also plotted in 
Figure~\ref{fig:orderedMatrix1}{\claop (b)}. The relevant observation here is that all of these eigenvectors are extended
over the entire element-position domain.

\begin{figure}[t]
\centering\includegraphics[width=5in]{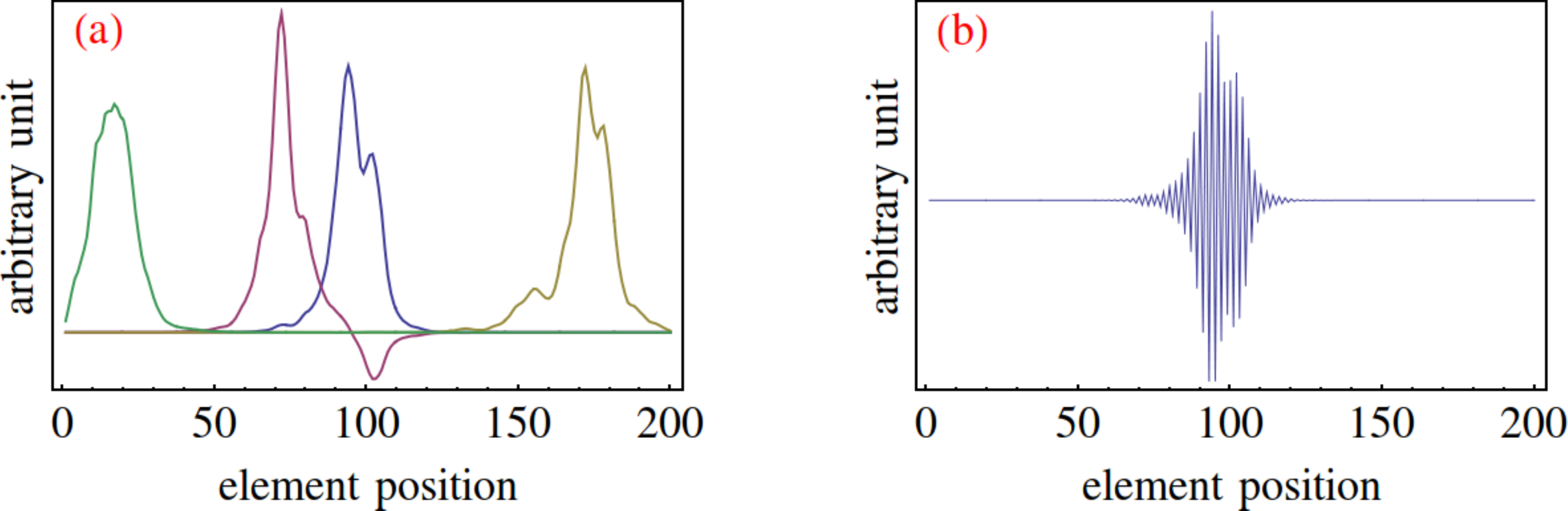}
\caption{Similar to Figure~\ref{fig:orderedMatrix1}, except the matrix ${\mathbb M}$ is {\bf slightly} randomized 
according to Eq.~\ref{eq:defMrandom1} and the eigenvectors are localized.}
\label{fig:randomMatrix1}
\end{figure}

We now would like to show that randomness can localize the eigenvectors of ${\mathbb M}$ over the element-position 
domain. For this part, lets preserve the tridiagonal character of ${\mathbb M}$ as presented 
in Eq.~\ref{eq:defMorederd}, but add a small random number to each off-diagonal element. The new, slightly 
randomized ${\mathbb M}$ is formally defined as
\begin{align}
{\mathbb M}_{i,i}=1,\quad {\mathbb M}_{i,i+1}={\mathbb M}_{i+1,i}=0.1+r_i, \quad r_i\in{\rm unif}[-0.01,0.01],
\label{eq:defMrandom1}
\end{align} 
where $r_i$ is a number randomly selected from a real uniform distribution in the range $[-0.01,0.01]$. 
Similar to the case of the ordered ${\mathbb M}$ in Figure~\ref{fig:orderedMatrix1}, ${\mathbb V}^{(1)}_j, {\mathbb V}^{(2)}_j, {\mathbb V}^{(3)}_j$, 
and ${\mathbb V}^{(4)}_j$ are plotted in Figure~\ref{fig:randomMatrix1}{\claop(a)} and ${\mathbb V}^{(200)}_j$ is plotted 
in Figure~\ref{fig:randomMatrix1}{\claop (b)}.
This time, all of these eigenvectors appear to be localized over the element-position domain. 

\begin{figure}[b]
\centering\includegraphics[width=5in]{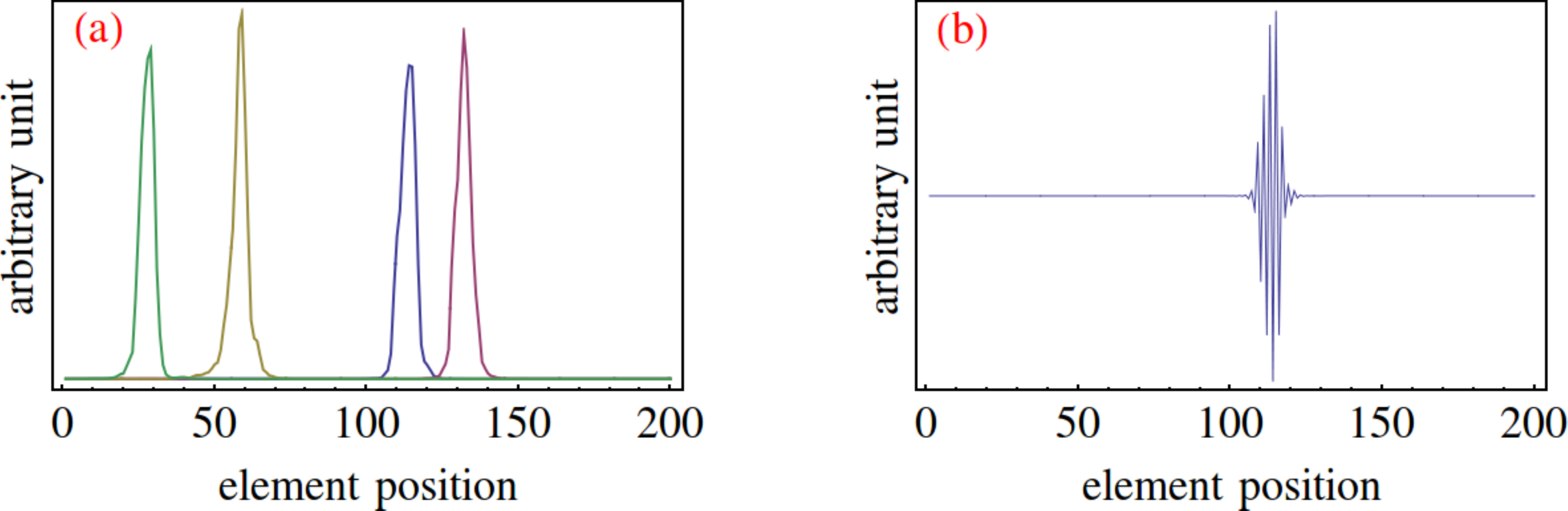}
\caption{Similar to Figure~\ref{fig:orderedMatrix1} and Figure~\ref{fig:randomMatrix1}, except the matrix ${\mathbb M}$ is 
{\bf strongly} randomized according to Eq.~\ref{eq:defMrandom2} and the eigenvectors are {\bf strongly} localized.}
\label{fig:randomMatrix2}
\end{figure}

Not only does the randomness result in localized eigenvectors, but also the relative strength of randomness 
compared with the average off-diagonal values determines the size of the localization (localization length).
This can be verified by increasing the range of the random numbers $r_i$ to $[-0.05,0.05]$. The new, strongly 
randomized ${\mathbb M}$ is formally defined as
\begin{align}
{\mathbb M}_{i,i}=1,\quad {\mathbb M}_{i,i+1}={\mathbb M}_{i+1,i}=0.1+r_i, \quad r_i\in{\rm unif}[-0.05,0.05].
\label{eq:defMrandom2}
\end{align} 
Again, the same eigenvectors are plotted in Figures~\ref{fig:randomMatrix2}{\claop (a)} and~\ref{fig:randomMatrix2}{\claop (b)}. 
The eigenvectors are strongly localized 
over the element-position domain because of the strong randomness in the off-diagonal elements of ${\mathbb M}$.

\begin{figure}[t]
\centering\includegraphics[height=2.in]{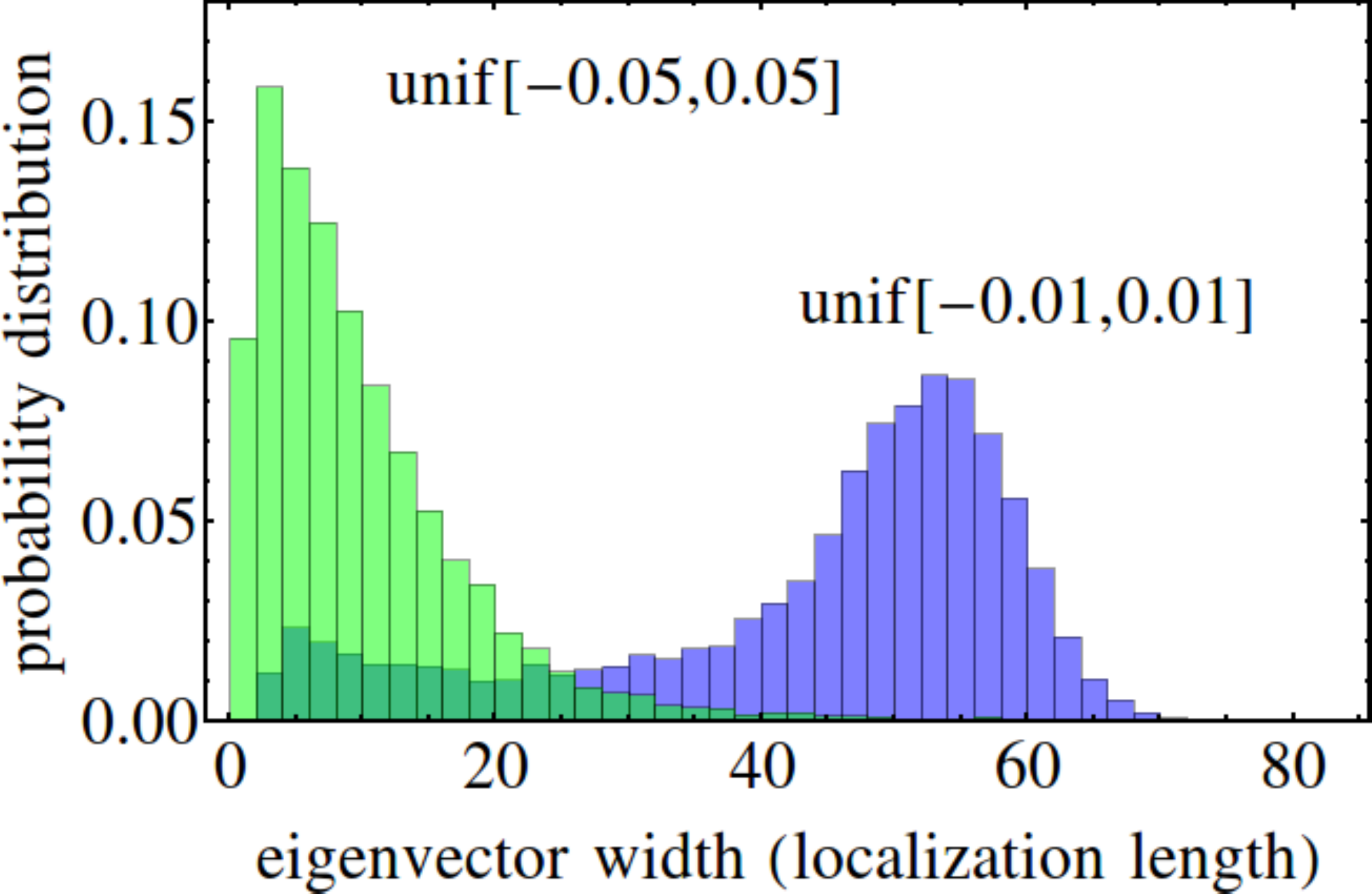}
\caption{The probability distribution of the eigenvector widths for two separate cases
of the weak disorder related to Eq.~\ref{eq:defMrandom1}, and the strong disorder related to
Eq.~\ref{eq:defMrandom2}.}
\label{fig:HistogramRandomMatrix}
\end{figure}

The impact of disorder on the distribution of the width of the eigenvectors on the
element-position domain can be easily visualized in a histogram. In Figure~\ref{fig:HistogramRandomMatrix},
the distribution of the width of the eigenvectors is plotted for the weak disorder of Eq.~\ref{eq:defMrandom1}
and the strong disorder of Eq.~\ref{eq:defMrandom2}. The width is calculated using the second moment method, which is the 
standard deviation around the mean position calculated using the 
absolute-value-squared of the eigenvectors. The width $\sigma_i$ for the eigenvector ${\mathbb V}^{(i)}$ is given by 
\begin{align}
\sigma_i=\left(
\dfrac{\sum^{N}_{j=1} \Big(j-\langle j \rangle\Big)^2~|{\mathbb V}^{(i)}_j|^2}
{\sum^{N}_{j=1} |{\mathbb V}^{(i)}_j|^2}\right)^{1/2}
, \qquad 
\langle j \rangle_i=
\dfrac{\sum^{N}_{j=1} j\ |{\mathbb V}^{(i)}_j|^2}
{\sum^{N}_{j=1} |{\mathbb V}^{(i)}_j|^2}.
\end{align} 

Each probability distribution is presented in a histogram 
and is the result of averaging over 100 independent random simulations. From the distributions in 
Figure~\ref{fig:HistogramRandomMatrix}, it is clear that stronger disorder results in stronger localization of the
eigenvectors; moreover, it is clear that such statements can only be made in a statistical sense. Therefore,
in the case of strong disorder, although the majority of the eigenvectors are more localized, a minority of the 
eigenvectors may actually be less localized than those of the weak disorder.

The exercise presented here shows how off-diagonal disorder results in localization. 
Similar localization behavior can be observed for diagonal disorder, where only the diagonal elements of 
${\mathbb M}$ are randomized, and also for mixed diagonal and off-diagonal disorder. 

\begin{highlight}
\textbf{Highlights:}
\begin{itemize}
\item The eigenvectors of random matrices can be localized in the element position space. 
\item Some modes are very narrow and some are wide, and localization is only meaningful in a statistical sense.
\item By calculating the width of the eigenvectors of a large ensemble of random matrices, it is possible to calculate 
the probability distribution for the width of the eigenvectors.  
\item A stronger level of randomness shifts the eigenvector-width probability distribution to smaller width values, hence a stronger localization.  
\end{itemize}
\end{highlight}
\section{Normal transmission through a random stack of dielectrics}
\label{sec:normalstack}
\begin{figure}[t]
\centering\includegraphics[width=4.5in]{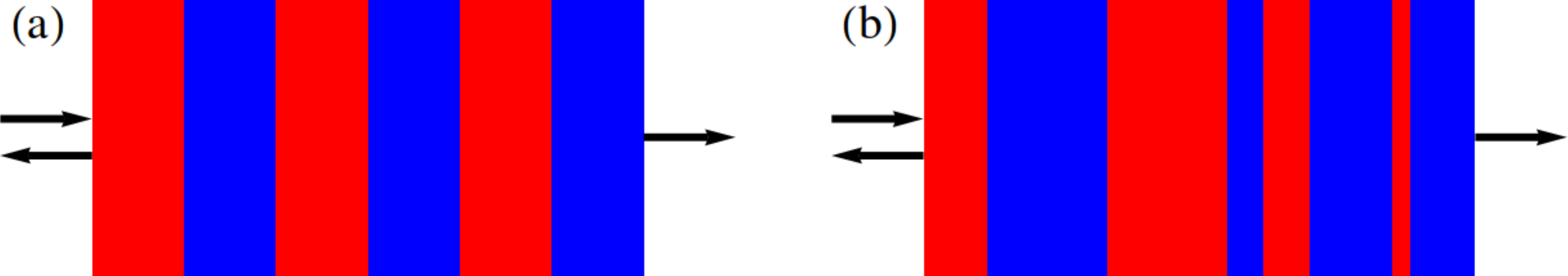}
\caption{{\claop (a)} A periodic array of two different dielectric materials identified with refractive 
indexes $n_1$ and $n_2$ is shown, {\claop (b)} is similar, except the thickness of the layers is chosen 
randomly.}
\label{fig:1D-stack}
\end{figure}

Another interesting example that links randomness to localization is the problem of light
transmission through a random stack of dielectrics. The normalized transmission through a 
stack of dielectrics is shown in Figure~\ref{fig:1D-stack}. The light, which is incident from the left, is 
partially reflected from the stack, while the rest is transmitted through the stack. The dielectrics 
are assumed to be lossless.
Figure~\ref{fig:1D-stack}{\claop (a)} shows a periodic array of two different 
dielectric materials identified with refractive indexes $n_1$ and $n_2$, more commonly
referred to as a Bragg grating. Figure~\ref{fig:1D-stack}{\claop (b)} is
similar, except the thickness of the layers is chosen randomly.
 
\begin{figure}[h]
\centering\includegraphics[width=4.8in]{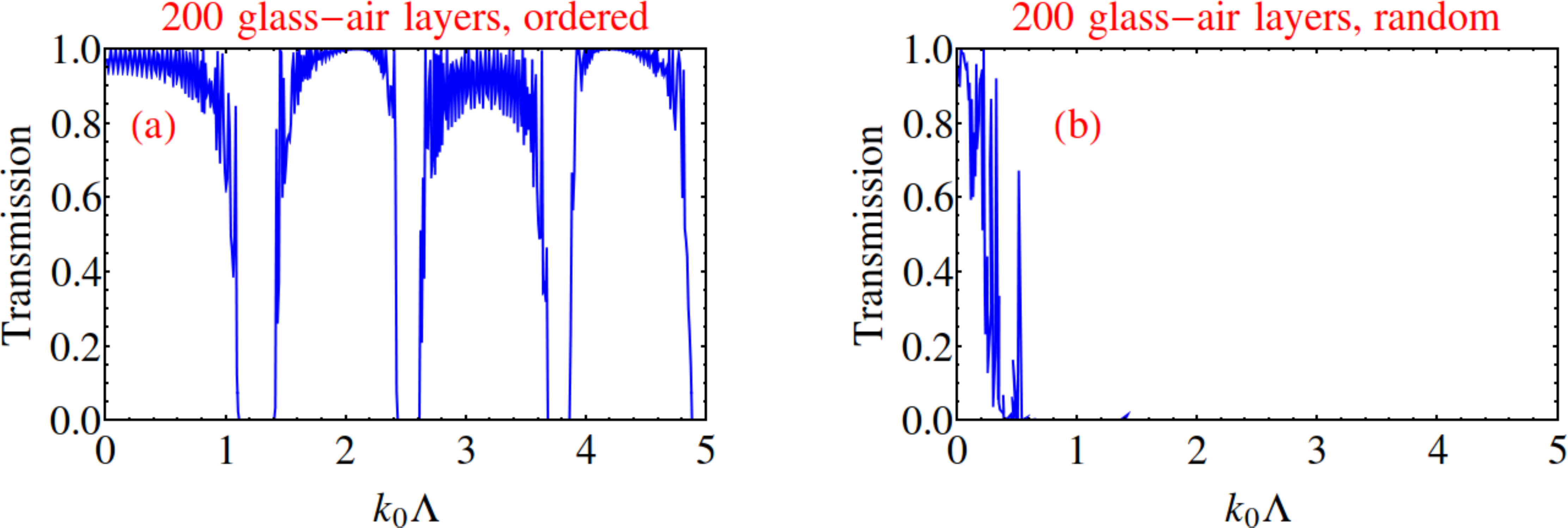}
\caption{The optical transmission through a stack of 200 glass-air layers is plotted versus $k_0\Lambda$ for
{\claop (a)} a periodic layer thickness and {\claop (b)} random layer thickness. $\Lambda$ is the periodicity 
in case {\claop (a)}, while the thickness of each layer in case {\claop (b)} is chosen from a uniform random distribution in the
range $[0,2\Lambda]$.}
\label{fig:1D-PC}
\end{figure}

In Figure~\ref{fig:1D-PC}{\claop (a)}, the relative optical power transmission is plotted as 
a function of the normalized frequency. The thickness of each layer is $\Lambda$ (identical for all layers), 
and $k_0=2\pi/\lambda$ is the wavevector, where 
$\lambda$ is the optical wavelength in vacuum. The stack is made of 200 air-glass layers 
(400 layers total), where $n_1=1.5$ and $n_2=1.0$ is assumed. The usual bandgaps and 
bandpasses are observed in the transmission plot of this periodic Bragg grating.

The setup for Figure~\ref{fig:1D-PC}{\claop (b)} is identical to that of Figure~\ref{fig:1D-PC}{\claop (a)},
except the thickness of each layer is chosen from a uniform random distribution in the
range $[0,2\Lambda]$. Unlike the periodic Bragg grating, transmission through the random 
stack is nearly zero except for very small values of the normalized frequency $k_0\Lambda$.
Therefore, the random stack behaves like a nearly perfect mirror beyond a certain frequency. 

\begin{figure}[h]
\centering\includegraphics[height=1.4in]{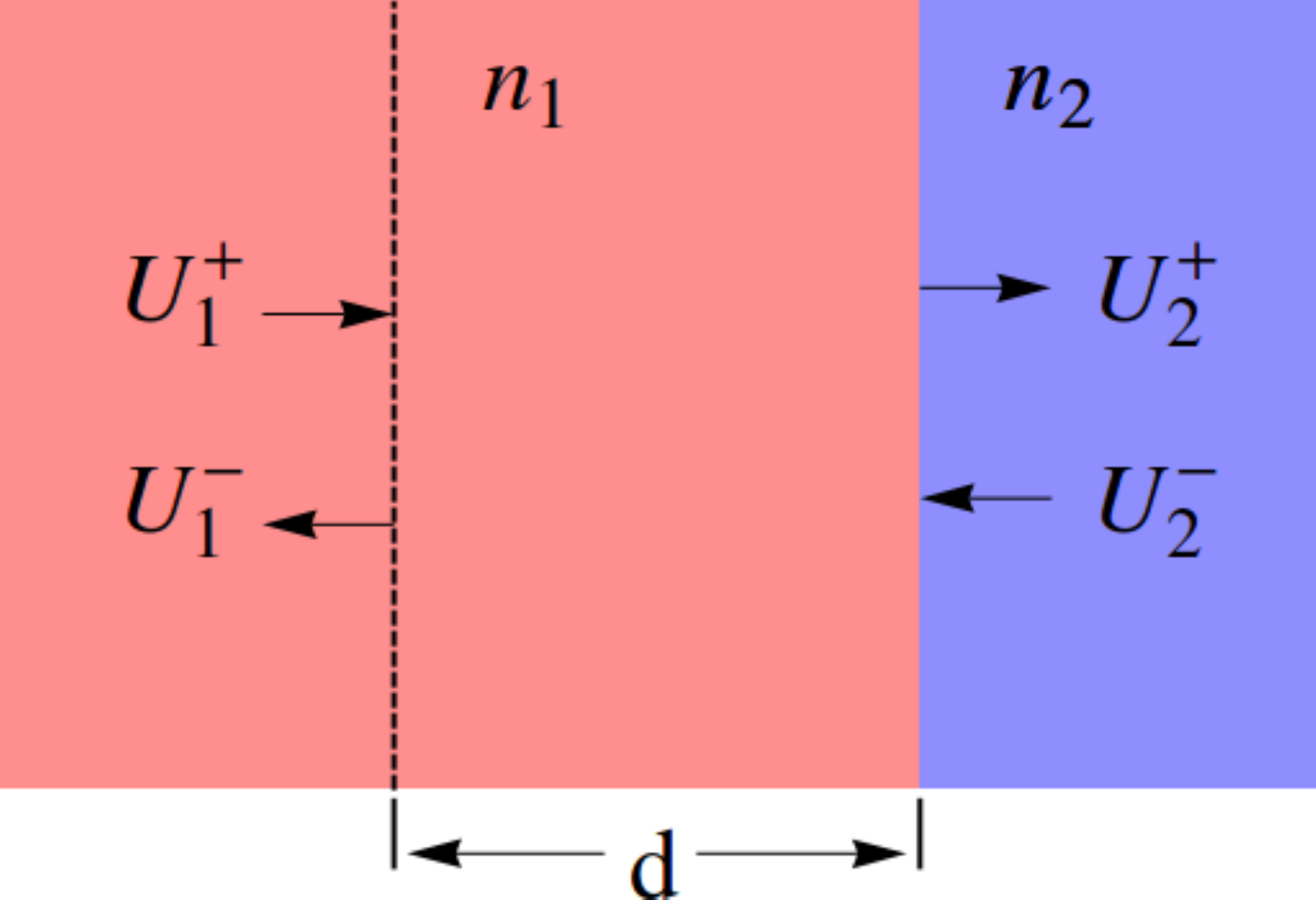}
\caption{A sketch of the field amplitudes, refractive indexes, and geometry related to Eq.~\ref{eq:Mmatrix} and Eq.~\ref{eq:UMU}.}
\label{fig:M-matrix}
\end{figure}

The calculation of the optical transmission in Figure~\ref{fig:1D-PC} is carried out using the 
transmission matrix $M$ defined in Eq.~\ref{eq:Mmatrix} related to Figure~\ref{fig:M-matrix}:
\begin{equation}
M = 
\dfrac{1}{2n_2}
\begin{bmatrix}
(n_2+n_1)e^{i\varphi} & (n_2-n_1)e^{i\varphi} \\[0.3em]
(n_2-n_1)e^{-i\varphi} & (n_2+n_1)e^{-i\varphi}
\end{bmatrix}, \quad \varphi=n_1k_0d.
\label{eq:Mmatrix}
\end{equation}
In Figure~\ref{fig:1D-PC}, the $M$ matrix can be used to relate the right- and left-moving 
components of the optical field in dielectric $n_2$ to those in dielectric $n_1$ according to
\begin{equation}
\begin{bmatrix}
U^+_2  \\[0.3em]
U^-_2 
\end{bmatrix}
=M.
\begin{bmatrix}
U^+_1  \\[0.3em]
U^-_1 
\end{bmatrix}.
\label{eq:UMU}
\end{equation}
The total transmission and reflection can be calculated by cascading all the $M$ matrices for different random
layers of the dielectric stack, as discussed in detail in Ref.~\cite{SalehTeich}.

\begin{figure}[b]
\centering\includegraphics[width=4.8in]{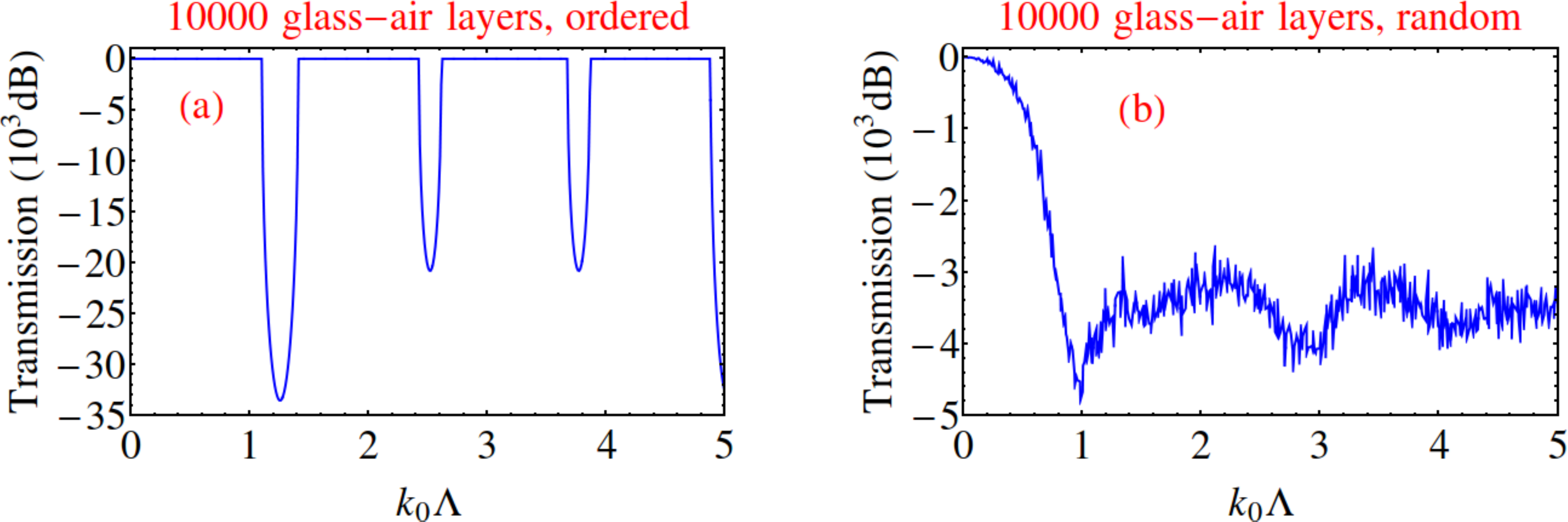}
\caption{Same as Figure~\ref{fig:1D-PC}, except with 10,000 layers.}
\label{fig:1D-PC-10000}
\end{figure}
It is instructive to repeat the exercise of Figure~\ref{fig:1D-PC} with a larger 
number of layers. In Figure~\ref{fig:1D-PC-10000}, the number of layers is increased to
10,000 in each case, and the relative optical power transmission is plotted in dB units.
The logarithmic scaling on the vertical axis shows that the bandgap frequency ranges in the ordered stack of Figure~\ref{fig:1D-PC-10000}{\claop (a)},
although not perfect due to the finite number of layers, attenuate the transmitted light by 
more than 20,000~dB, making the stack a nearly perfect mirror over this frequency range.
The random stack in Figure~\ref{fig:1D-PC-10000}{\claop (b)}, except for low frequency values, is also a very good reflector.
Although the attenuation in transmission at 3,000~dB or more is not as good as the 
bandgap region in the periodic case, it is very broadband and is not limited to bandgap range.

The observations in Figure~\ref{fig:1D-PC-10000} bring about a somewhat philosophical issue.
In practice, any stack of optical dielectrics has some inevitable randomness; therefore, given a sufficiently 
large number of layers, it acts as a perfect mirror beyond a certain frequency. Of course, at very low frequencies,
the wavelength is so large that it hops over the stack and results in large transmission. Therefore,
in practice, any stack of optical dielectrics with a sufficiently large number of layers is a practically 
a broadband bandgap structure. The periodic structure analyzed in Figure~\ref{fig:1D-PC-10000}{\claop (a)} can be viewed
as the limiting case of the random structure, when the randomness goes to zero. When this limit is taken, the bandpasses emerge out of the
broadband bandgap of the more general disordered stack. Therefore, the philosophical point of view is that
the magic of a periodic structure is not in its bandgap, because the bandgap comes naturally in any practical 1D
stack; rather, it is the emergence of the bandpasses that makes periodic structure so special.  
While this may seem like an inconsequential philosophical issue, it is necessary to alert students
who learn about coherent waves in periodic systems in optics or condensed matter physics to ensure
that they develop proper early intuition about the true implications of periodicity. {\em The bandpasses are
at least as glorious as the bandgaps}.
 
The low-frequency behavior, as well as the periodic ``very weak'' resurrection of transmission in Figure~\ref{fig:1D-PC-10000}{\claop (b)}, 
can be intuitively understood based on the work of Berry \textit {et al}.~\cite{Berry}. They have shown that the optical transmission through 
a random stack of $N$ bi-layered transparent plates, each contributing a random phase $\phi$ belonging 
to a uniform distribution $\phi\in{\rm unif}[0,2\pi]$, is given by
\begin{equation}
\tau_{2N} = \exp\Big(-2N\log(1/\tau)\Big),\quad \tau=\dfrac{4n_1n_2}{(n_1+n_2)^2}. 
\label{eq:tauBerry}
\end{equation}
Using $N=10,000$, $n_1=1.5$, and $n_2=1.0$, one obtains $\tau_{2N}\approx -3546~{\rm dB}$, which is in agreement
with the numerical simulation in Figure~\ref{fig:1D-PC-10000}{\claop (b)}.
In the simulations presented in Figures~\ref{fig:1D-PC}~and~\ref{fig:1D-PC-10000},
it was assumed that $(d_1=d_2)\in{\rm unif}[0,2\Lambda]$; therefore, 
\begin{equation}
0 \le \phi=(n_1k_0 d+n_2k_0 d) \le 2(n_1+n_2)k_0\Lambda.
\label{eq:Berry2}
\end{equation}
In order to obtain the near uniform distribution of $\phi\in{\rm unif}[0,2\pi]$, the upper bound in Eq.~\ref{eq:Berry2}
can be set to $2\pi$, where one obtains 
\begin{equation}
{\bar k}_0=k_0\Lambda\approx \pi/(n_1+n_2),
\label{eq:Berry3}
\end{equation}
as the condition for the 
disorder-induced localization, exerting its full power and agreeing with the assumptions of Ref.~\cite{Berry}. 
It should be noted that this value is very close to the value of $k_0\Lambda$
in Figure~\ref{fig:1D-PC-10000}{\claop (b)}, beyond which the attenuation is strong. Therefore, the main reason that the attenuation
is not strong at low $k_0\Lambda$ is that the random phase $\phi$ does not cover the entire range of $[0,2\pi]$.
For $k_0\Lambda>{\bar k}_0$, although the range of the phase $\phi$ covers the entire $[0,2\pi]$, it partially folds over
and makes the distribution of $\phi$ non-uniform over $[0,2\pi]$. 

For example, consider the case of $\phi\in{\rm unif}[0,3\pi]$.
$\phi$ is a phase variable for which $[2\pi,3\pi]\equiv [0,\pi]$; therefore, $\phi$ covers the entire $[0,2\pi]$ range but 
with twice the probability in the $[0,\pi]$ range compared with $[\pi,2\pi]$ range and is no longer uniformly 
distributed over $[0,2\pi]$. This is the main reason behind the partial resurrections of transmission observed in Figure~\ref{fig:1D-PC-10000}{\claop (b)}
for $k_0\Lambda>{\bar k}_0$. 

A final and important point is that the exponential decay of the optical amplitude in the disordered stack is a manifestation of 
the ``coherent'' superposition of partial waves that reflect and transmit at each boundary. The fact that the naive ray theory,
which is based on incoherent transmissions and reflections, fails to predict the Anderson localization behavior observed 
in these examples= defies the intuition at first glance. The interested reader can consult Ref.~\cite{Berry} for a more detailed
discussion on these points. A simple and elegant experiment using a laser pointer and a stack of overhead transparencies is also presented
in Ref.~\cite{Berry}, which can be excellent demonstration in an undergraduate laboratory in optics.
\begin{highlight}
\textbf{Highlights:}
\begin{itemize}
\item Optical transmission through a stack of dielectric layers with differing refractive indexes 
and random thicknesses drops, on the average, exponentially with the number of layers.     
\item This exponential drop is observed for all wavelengths that are comparable or shorter than the 
mean thickness of the dielectric layers. The random stack behaves like an ultra-broadband mirror.
\item The optimum localization is obtained when the cumulative optical phase in each layer is from
a uniformly distributed random number in the range $[0,2\pi]$. If the domain is smaller, or even 
larger such as $[0,3\pi]$, the localization is weaker.
\end{itemize}
\end{highlight}
\section{Oblique transmission through a random stack of dielectrics}
\begin{figure}[b]
\centering\includegraphics[width=4.5in]{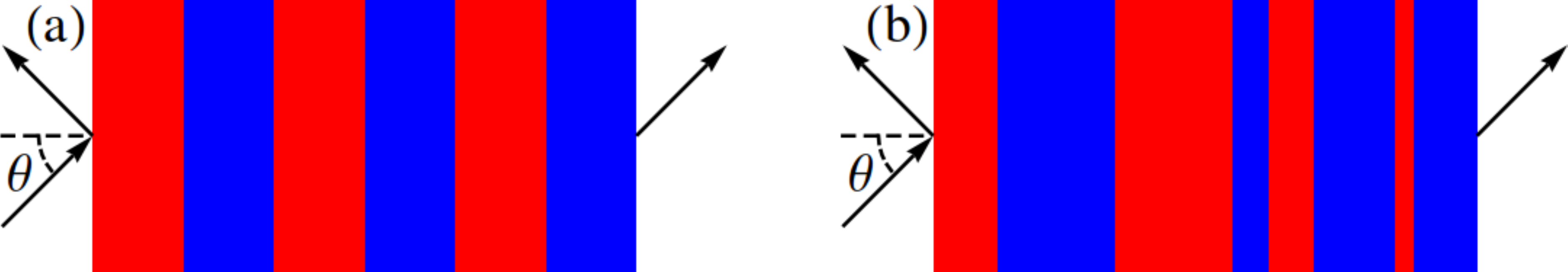}
\caption{{\claop (a)} Transmission and reflection of light incident on a periodic stack of dielectrics. {\claop (b)} Same, except 
the thickness of each dielectric layer is randomly selected.}
\label{fig:1D-stack-angle}
\end{figure}

The example presented in section~\ref{sec:normalstack} can be readily generalized to the case of light 
incident at an angle on the dielectric stack, as illustrated in Figure~\ref{fig:1D-stack-angle}. 
The calculation can be carried out using a simple generalization
of the transmission matrix $M$ defined in Eq.~\ref{eq:Mmatrix} to the case of incidence at an angle (see for example
Ref.~\cite{SalehTeich}). 
\begin{figure}[t]
\centering\includegraphics[width=4.8in]{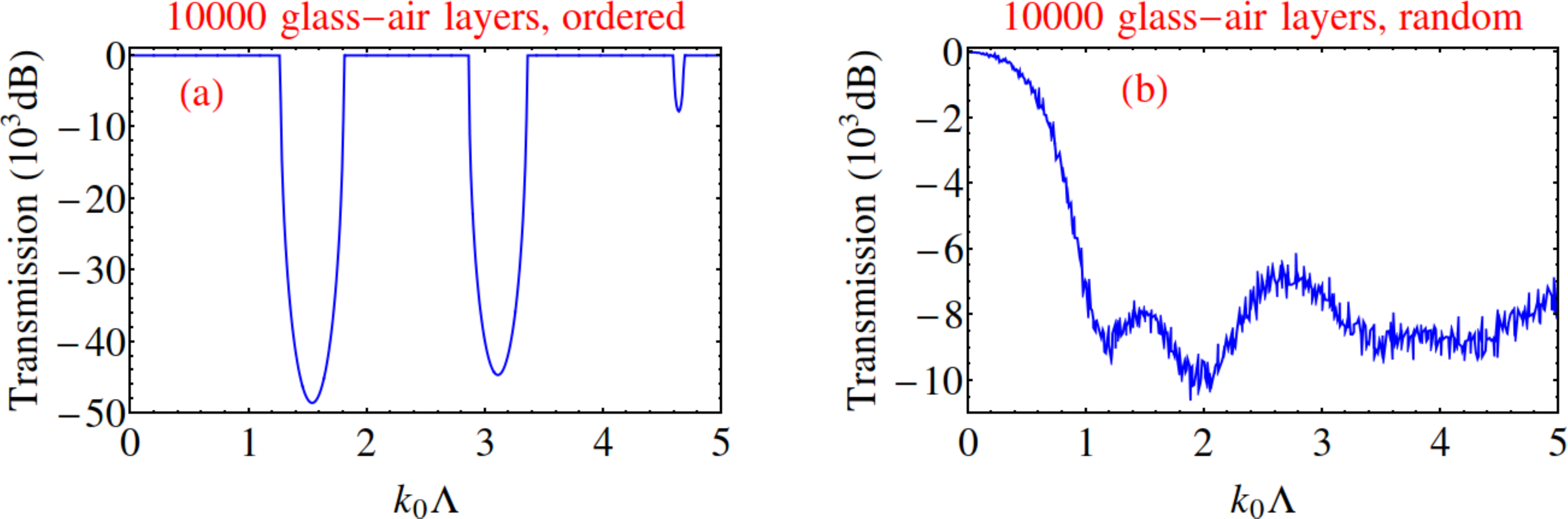}
\caption{Relative optical power transmission is plotted as a function of the
         normalized frequency for oblique incidence at a 45$^\circ$ angle for 
         the TE polarization for {\claop (a)} a periodic dielectric stack and {\claop (b)} a random dielectric stack.}
\label{fig:1D-PC-10000-TE-45degrees}
\end{figure}

In Figure~\ref{fig:1D-PC-10000-TE-45degrees}, the relative optical power transmission is plotted as a function of the
normalized frequency for oblique incidence at a 45$^\circ$ angle (as measured in the air layer) for the TE (transverse electric) 
polarization of light. Figure~\ref{fig:1D-PC-10000-TE-45degrees} should be compared with Figure~\ref{fig:1D-PC-10000}, which was for the case of normal incidence.

In the case of the periodic dielectric stack in Figure~\ref{fig:1D-PC-10000-TE-45degrees}{\claop (a)}, the bandgaps are shifted compared with 
normal incidence in Figure~\ref{fig:1D-PC-10000}(a) and are wider and deeper. 
For the random dielectric stack in Figure~\ref{fig:1D-PC-10000-TE-45degrees}{\claop (b)}, 
the transmission is lower over the entire frequency band compared with Figure~\ref{fig:1D-PC-10000}{\claop (b)}; 
however, besides the lower value and frequency shifts 
of the features in transmission, no substantial qualitative difference exists between oblique incidence at a 45$^\circ$ angle and normal incidence.

\begin{figure}[b]
\centering\includegraphics[width=4.8in]{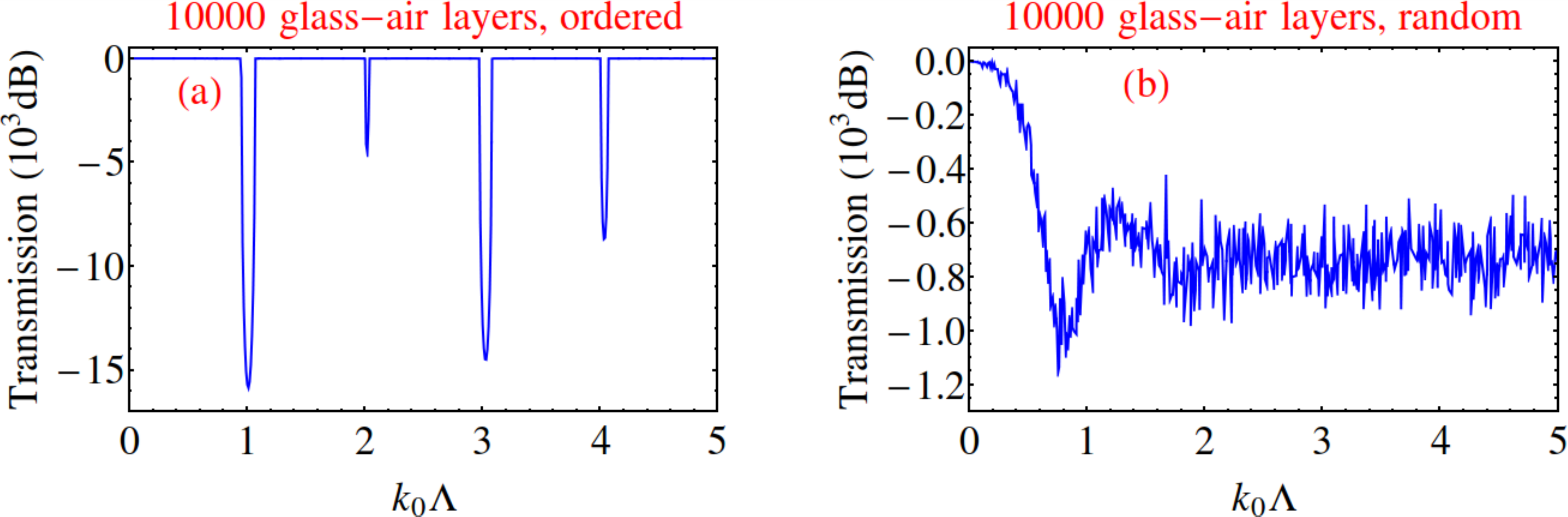}
\caption{Relative optical power transmission is plotted as a function of the
         normalized frequency for oblique incidence at a 45$^\circ$ angle for 
         the TM polarization for {\claop (a)} a periodic dielectric stack and {\claop (b)} a random dielectric stack.}
\label{fig:1D-PC-10000-TM-45degrees}
\end{figure}
The case of a TM (transverse magnetic) polarization is explored in Figure~\ref{fig:1D-PC-10000-TM-45degrees}, where the periodic-stack bandgaps 
in Figure~\ref{fig:1D-PC-10000-TM-45degrees}{\claop (a)} are substantially narrower and more shallow compared with the case of a TE polarization at 45$^\circ$ angle 
in Figure~\ref{fig:1D-PC-10000-TE-45degrees}{\claop (a)} and normal incidence in Figure~\ref{fig:1D-PC-10000}(a). Similarly, the transmission of the 
TM polarization at 45$^\circ$ angle in Figure~\ref{fig:1D-PC-10000-TM-45degrees}{\claop (b)} through a random stack is stronger in the case of a TM polarization
compared with the TE polarization in Figure~\ref{fig:1D-PC-10000-TE-45degrees}{\claop (b)} and the case of normal incidence in 
Figure~\ref{fig:1D-PC-10000}{\claop (b)}.

\begin{figure}[t]
\centering\includegraphics[width=4.8in]{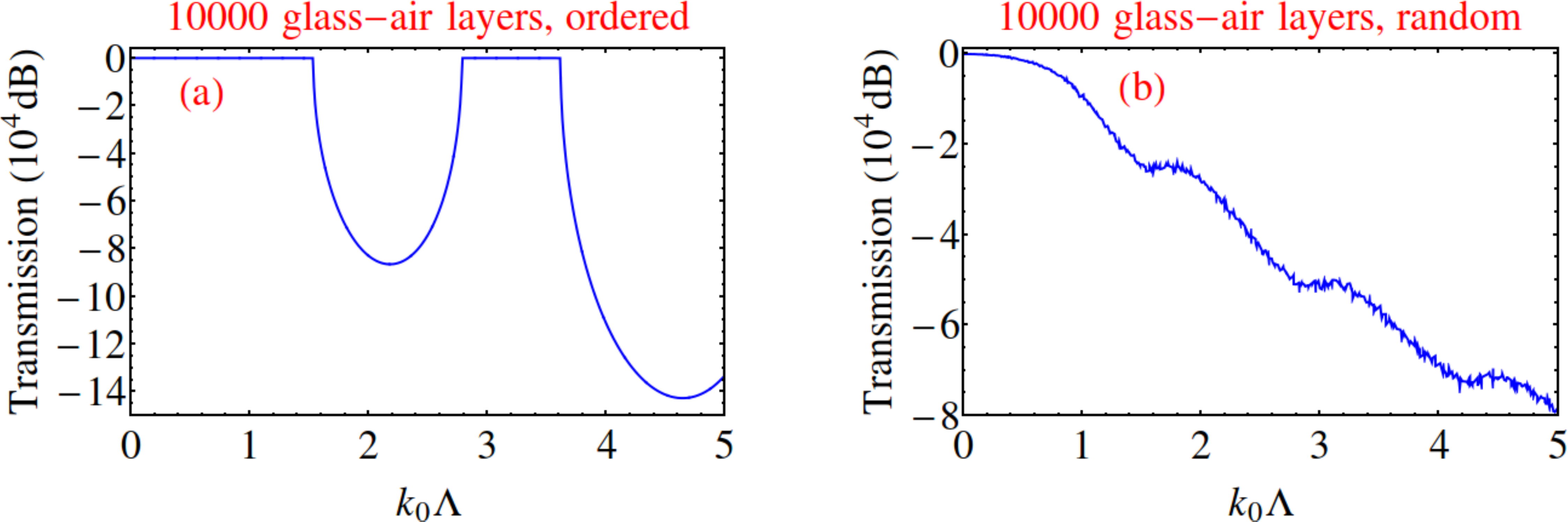}
\caption{Relative optical power transmission is plotted as a function of the
         normalized frequency for oblique incidence at an 85$^\circ$ angle for 
         the TE polarization for a {\claop (a)} periodic dielectric stack and a {\claop (b)} random dielectric stack.}
\label{fig:1D-PC-10000-TE-85degrees}
\end{figure}
For the larger incidence angle of 85$^\circ$, the difference between transmission through a periodic dielectric stack
and a random dielectric stack is more pronounced; and so is the difference between TE and TM poalizations.  
The bandgaps are wider and deeper (note the $10^4$ dB label in the vertical scale) 
for the TE polarization incident on a periodic dielectric stack in Figure~\ref{fig:1D-PC-10000-TE-85degrees}{\claop (a)} compared with the case of 45$^\circ$ 
and the case of normal incidence. 
The gaps are also shifted in normalized frequency and are more widely separated. Similar behavior
is observed for the case of the TE polarization incident on a random dielectric stack in Figure~\ref{fig:1D-PC-10000-TE-85degrees}{\claop (b)}, where the
full strength of the localization begins at a higher normalized frequency than what is shown in Figure~\ref{fig:1D-PC-10000-TE-85degrees}{\claop (b)}; 
this behavior agrees with the general shift of the spectral features to the higher normalized frequency observed in the case of 
a periodic dielectric stack in Figure~\ref{fig:1D-PC-10000-TE-85degrees}{\claop (a)}.

\begin{figure}[b]
\centering\includegraphics[width=4.8in]{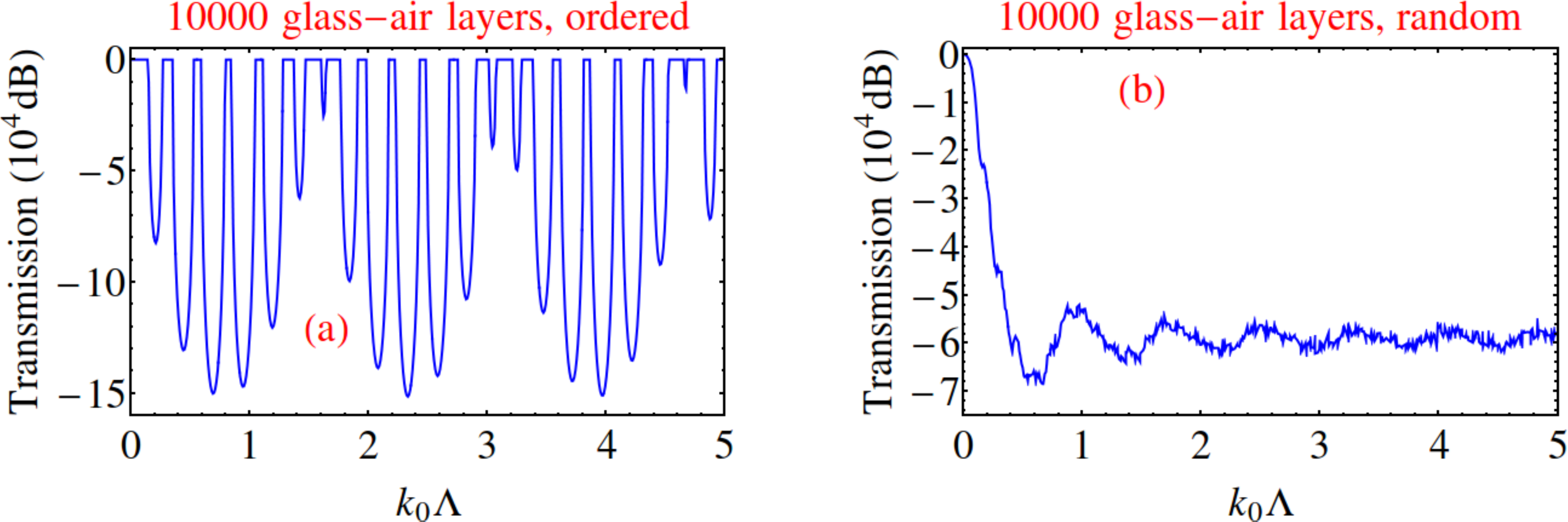}
\caption{Relative optical power transmission is plotted as a function of the
         normalized frequency for oblique incidence at an 85$^\circ$ angle for 
         the TM polarization for {\claop (a)} a periodic dielectric stack and {\claop (b)} a random dielectric stack.}
\label{fig:1D-PC-10000-TM-85degrees}
\end{figure}
For the TM polarization incident on a periodic dielectric stack in Figure~\ref{fig:1D-PC-10000-TM-85degrees}{\claop (a)}, many narrow and closely separated 
gaps appear in the studied normalized frequency range. The gaps are deeper compared with the case of 45$^\circ$ and the case of normal incidence.
Similarly, the spectral features shift to lower values in the case of the TM polarization incident on a random dielectric stack in 
Figure~\ref{fig:1D-PC-10000-TM-85degrees}{\claop (b)}; therefore, the full strength of the localization begins at a lower normalized frequency compared
with TE polarization and smaller incidence angles (for both polarizations).

\subsection*{TM polarization and Brewster's angle}
\begin{figure}[h]
\centering\includegraphics[width=4.8in]{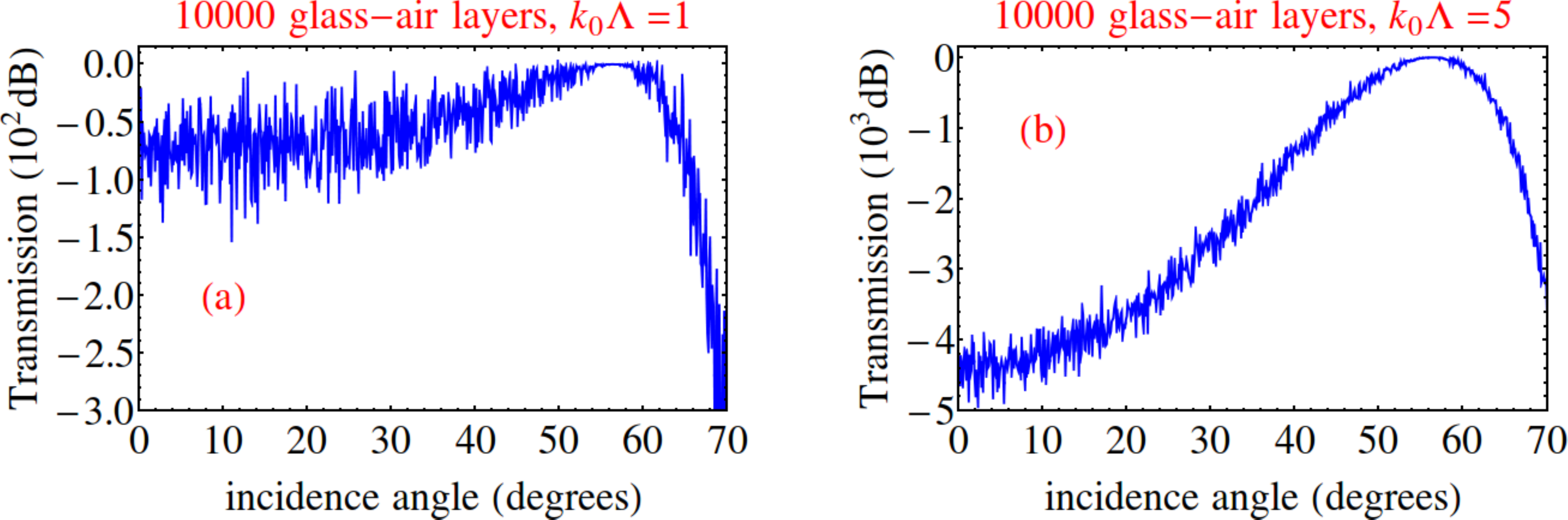}
\caption{Relative optical power transmission is plotted as a function of the
         incidence angle of TM polarized light on a random dielectric stack for
         {\claop (a)} $k_0\Lambda=1$, and {\claop (b)} $k_0\Lambda=5$. Perfect transmission is observed 
         at Brewster's angle $\theta_B=56.3^\circ$.}
\label{fig:1D-PC-10000-TM-vs-angles}
\end{figure}

In the above examples, it is observed that the TM polarization behaves differently from the TE polarization. 
The root of this distinction is in the different boundary conditions for the electric field in the TE and TM polarizations at the 
interface of each layer, and the difference is most 
apparent at Brewster's angle, while the TM polarization shows perfect transmission. 
In Figure~\ref{fig:1D-PC-10000-TM-vs-angles}, the relative optical power transmission is plotted as a function of the
incidence angle of the TM polarized light on a random dielectric stack for {\claop (a)} $k_0\Lambda=1$, and {\claop (b)} $k_0\Lambda=5$.
Perfect transmission is observed at Brewster's angle $\theta_B=56.3^\circ$ through 10,000 glass-air layers of random thickness.
Of course, like other simulations so far, intrinsic attenuation of glass is neglected in these simulations. The transmission
is lower (localization is stronger) for $k_0\Lambda=5$ in Figure~\ref{fig:1D-PC-10000-TM-vs-angles}{\claop (b)} compared with the case of
$k_0\Lambda=1$ in Figure~\ref{fig:1D-PC-10000-TM-vs-angles}{\claop (a)}, as expected from previous simulations and related arguments.

\begin{figure}[b]
\centering\includegraphics[width=4.8in]{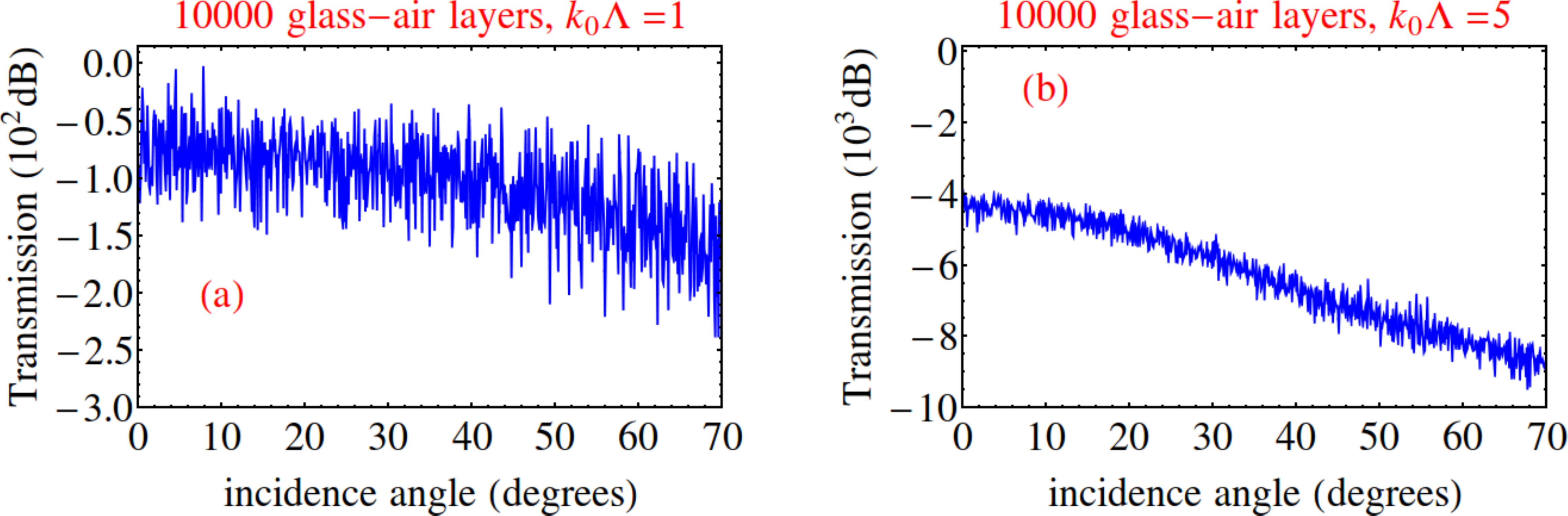}
\caption{Relative optical power transmission is plotted as a function of the
         incidence angle of TE polarized light on a random dielectric stack for
         {\claop (a)} $k_0\Lambda=1$, and {\claop (b)} $k_0\Lambda=5$.}
\label{fig:1D-PC-10000-TE-vs-angles}
\end{figure}
For comparison, in Figure~\ref{fig:1D-PC-10000-TE-vs-angles}, the relative optical power transmission is plotted as a function of the
incidence angle of TE polarized light on a random dielectric stack for {\claop (a)} $k_0\Lambda=1$, and {\claop (b)} $k_0\Lambda=5$.
Compared with Figure~\ref{fig:1D-PC-10000-TM-vs-angles}, the absence of perfect transmission at Brewster's angle is notable.
It can also be concluded that below Brewster's angle, transmission is weaker in the TM polarization compared with the TE.
This is expected, because at normal incidence TE and TM polarizations have the same transmission; increasing the angle 
results in a monotonic increase in transmission for the TM polarization all the way up to 100\% at Brewster's angle, while
the transmission for the TE polarization monotonically decreases. Beyond Brewster's angle, the TE polarization continues its slow
decline; however, the TM polarization goes through a steep decline. Depending on the value of $k_0\Lambda$, at large angles, 
the TM polarization will have a lower transmission compared with TE (compare Figure~\ref{fig:1D-PC-10000-TM-vs-angles}{\claop (a)} with
Figure~\ref{fig:1D-PC-10000-TE-vs-angles}{\claop (a)}), or vice versa (compare Figure~\ref{fig:1D-PC-10000-TM-vs-angles}{\claop (b)} with
Figure~\ref{fig:1D-PC-10000-TE-vs-angles}{\claop (b)}).
The interested reader is encouraged to consult a detailed account of the polarization dependence of an obliquely incident light
on layered media in Ref.~\cite{Bliokh}.
\subsection*{Random dielectric stack as a broadband waveguide}
\begin{figure}[h]
\centering
\includegraphics[width=4in]{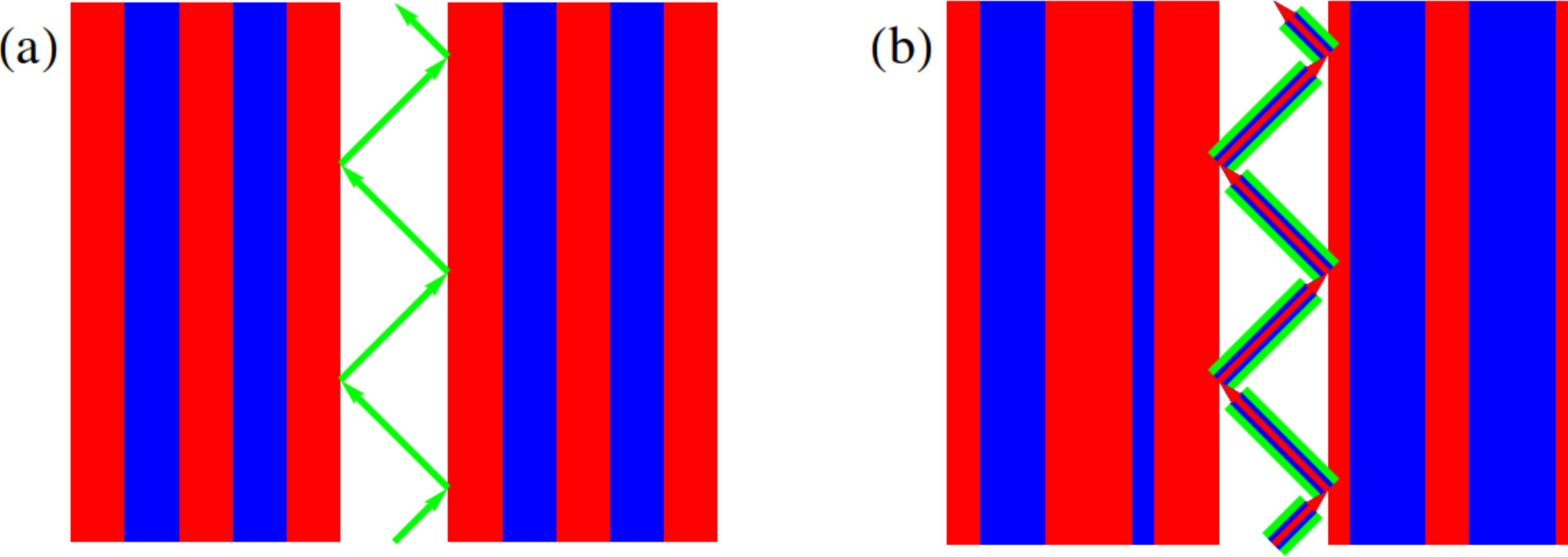}
\caption{A 1D dielectric waveguide using {\claop (a)} a periodic dielectric 
stack, which can have a very low leakage over a narrow range of wavelengths in 
the bandgaps corresponding to the specific incident wavelength and incident angle; and {\claop (b)} 
a random dielectric stack, which can operate over a broad range of wavelengths and
angles, but with slightly more leakage compared with a periodic waveguide
of the same number of layers optimized to operate in the center of the bandgap.}
\label{fig:1D-waveguide-angle}
\end{figure}

Based on the detailed discussions of the oblique incidence e.g., the results in    
Figure~\ref{fig:1D-PC-10000-TM-vs-angles} and Figure~\ref{fig:1D-PC-10000-TE-vs-angles}, 
it is conceivable to make a 1D waveguide in the form shown in Figure~\ref{fig:1D-waveguide-angle}. 
For a periodic dielectric stack, a waveguide can efficiently operate over a narrow range of 
wavelengths in the bandgaps corresponding to the specific incident wavelength~\cite{PCBook,Yeh,Fink}. 
The operating bandwidth can be increased by considering a chirped grating as shown in Ref.~\cite{Ghosh0}.
Alternatively, a design based on a random dielectric stack is very broadband; the trade off
is that the leakage is slightly higher than that of a bandgap design for a periodic 
dielectric stack. Therefore, if ultra-broadband operation is desired, a random stack is
likely the solution. 

\begin{highlight}
\textbf{Highlights:}
\begin{itemize}
\item Considerable differences are observed between the transmission of the TE and TM polarizations, when a random dielectric
stack is illuminated at an angle. This hints at the possibility that the vectorial nature of the electromagnetic
field may play an important role in setting the localization behavior in certain situations. 
\item The strongest difference between the TE and TM polarizations can be observed for oblique incidence at
Brewster's angle, at which the TM polarization is entirely transmitted through the random dielectric stack.
\item Below Brewster's angle, the TE polarization usually has a lower transmission; however, above Brewster's angle, 
the situation depends on the ratio of the size of the optical wavelength to the average lattice size. 
\item The concept of a broadband waveguide is introduced, where the random dielectric stacks are used as broadband reflecting
walls for the waveguide.   
\end{itemize}
\end{highlight}
\section{Transverse Anderson localization of light in one transverse dimension}
\begin{figure}[htb]
\centering\includegraphics[width=4.in]{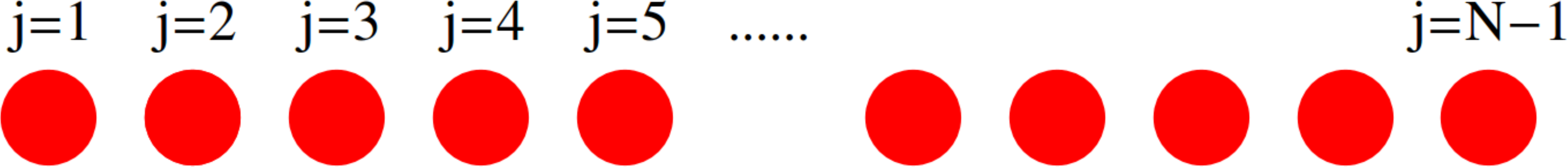}
\caption{
A one dimensional array of $N$ identical single-mode optical fibers, referred to as a coupled waveguide array.
}
\label{fig:1D-array-fiber}
\end{figure}

Consider the one dimensional array of $N$ identical single-mode 
optical fibers sketched in Figure~\ref{fig:1D-array-fiber}. The propagation constant of each fiber is 
$\beta_0$ and the direction of the propagation of the optical wave is assumed to be 
into the page. Each fiber is weakly coupled to its nearest neighbor, and
the strength of the coupling is determined by the separation between the fibers. 
The coupled mode equations for the propagation of the optical field through 
this optical fiber array can be expressed as~\cite{SalehTeich}
\begin{align}
\Big(i\dfrac{\partial}{\partial z}+\beta_0\Big) A_j(z)
+ c^+_{j} A_{j+1}(z) + c^-_{j} A_{j-1}(z) = 0,\quad j=1,\cdots,N.
\label{eq:coupledmode1}
\end{align} 
$A_j$ is the amplitude of the optical field in the $j$th fiber,
$c^+_{j}$ ($c^-_{j}$) is the coupling strength of the $j$th fiber
to its right neighbor $A_{j+1}$ (left neighbor $A_{j-1}$), 
and we assume symmetric coupling, so $c^+_{j}=c^-_{j+1}$. Of course, 
we need to assume $c^-_{1}=0$ and $c^+_{N}=0$, because the 
1st ($N$th) fiber does not have a neighbor to its left (right).

\begin{figure}[b]
\centering
\includegraphics[width=5.in]{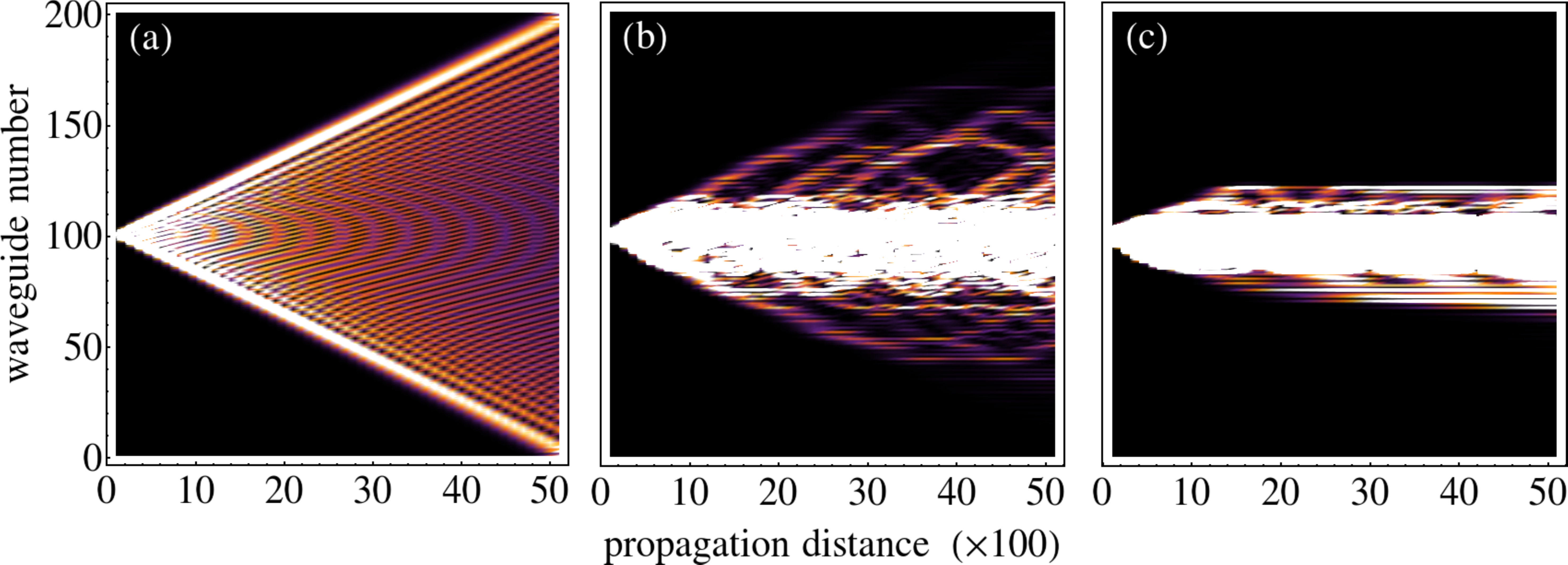}
\caption{Propagation through a waveguide coupled array for the case of {\claop (a)} a periodic array, 
{\claop (b)} a disordered array, and {\claop (c)} a highly disordered array. Higher level of disorder results in
a more localized propagation.}
\label{fig:randomCoupledWaveguides1D}
\end{figure}

Here, we consider the simple case, where the input light is only coupled 
to the middle fiber at $z=0$. As the optical field propagates through
the middle fiber, it couples to its neighboring fibers; those will
couple to their neighbors as well, and this cascading event results 
in a discrete diffraction pattern. This scenario is shown in Figure~\ref{fig:randomCoupledWaveguides1D}{\claop (a)},
where the intensity of the propagating light is plotted 
as a function of the propagation distance and the waveguide number. The parameters used in this simulation are: 
$\beta_0 = 6$, $c_0=0.01$, $N=201$, and $0 \le z \le 5000$. The input boundary 
condition is set to $A_j(z=0)=\delta_{j,101}$, as mentioned above.

The situation is quite different if the coupling between the waveguides 
is randomized. For example, consider perturbing the array slightly, so
that the average coupling remains $c_0$, but individual couplings 
vary with a uniform distribution according to $c^+_{j}=c_0+r_j$, where 
$r_j\in{\rm unif}[-0.006,0.006]$. In Figure~\ref{fig:randomCoupledWaveguides1D}{\claop (b)}, we observe that the light 
in the middle waveguide cannot spread as efficiently to its neighbors
and the diffraction slows down as the beam propagates farther in $z$.
This behavior is certainly induced by the disorder and randomness 
introduced in the cross coupling of fibers. Therefore, it is reasonable 
to expect that more randomness should result in slower diffraction.
This is verified in Figure~\ref{fig:randomCoupledWaveguides1D}{\claop (c)} by repeating the experiment for
$r_j\in{\rm unif}[-0.01,0.01]$. 

Let's take a closer look at the optical field pattern in Figure~\ref{fig:randomCoupledWaveguides1D}{\claop (c)}. 
The propagating light initially follows a diffracting pattern, but after a certain propagation 
distance the diffraction halts and the total width of the beam remains more or 
less the same for the rest of the propagation. It appears that the fibers near 
the edges of the array always remain dark. This behavior is referred to as
transverse Anderson localization. In fact, similar behavior can be observed 
in Figure~\ref{fig:randomCoupledWaveguides1D}{\claop (b)} after a sufficiently long propagation distance;
however, the transition to a stable width occurs at a longer propagation distance and
the beam width is larger because of the smaller amount of disorder. 

The light propagation in a one-dimensional disordered coupled fiber array is transversely localized 
for any amount of disorder; however, if the disorder is too small, the stable localized beam width may 
be larger than the transverse size of the structure and the localization effect cannot be observed in 
practice.
 
The coupled fiber array of Figure~\ref{fig:1D-array-fiber} described by Eq.~\ref{eq:coupledmode1} can be 
analyzed alternatively in the language of normal modes. The advantage of the normal mode description
is that the equations describing light propagation in a disordered coupled fiber array can be directly 
mapped to a random matrix discussed in section~\ref{sec:randommatrix}: the transverse Anderson localization
is nothing but the localization of the eigenvectors in the element-position domain index $j$ of $A_j$.
In order to see this, we define the vector ${\mathbb A}$ as
\begin{align}
{\mathbb A}=(A_1,A_2,A_3,\cdots,A_{N-1},A_N),
\label{eq:AvecDefine}
\end{align} 
and rewrite Eq.~\ref{eq:coupledmode1}
\begin{align}
i\dfrac{\partial}{\partial z}{\mathbb A}+{\mathbb B}.{\mathbb A}=0,
\label{eq:coupledmode2}
\end{align} 
where all diagonal elements of the symmetric tridiagonal matrix 
${\mathbb B}$ are equal to $\beta_0$ and the off-diagonal elements
are $c^+_{j}=c_0+r_j$. The boundary condition is set to
${\mathbb A}_j(z=0)=\delta_{j,101}$, because the input light is only 
coupled to the middle fiber.

The real tridiagonal matrix ${\mathbb B}$ is symmetric; therefore, all of its eigenvectors
are real and mutually orthogonal and its eigenvalues are real. We identify the $i$th eigenvalue
as ${\bar \beta}_i$ and the corresponding eigenvector as ${\mathbb V}^{(i)}$,
and ${\mathbb V}^{(i)}_j$ represents the $j$th element of the ${\mathbb V}^{(i)}$ eigenvector. 
Furthermore, without loss of generality, we assume that the eigenvectors are properly scaled such that they
are orthonormal, i.e., ${\mathbb V}^{(i)}.{\mathbb V}^{(j)}={\mathbb \delta}^{ij}$. 

Using the eigenvectors, we can construct a real orthogonal matrix ${\mathbb Q}$ such that 
${\mathbb Q}_{ij}={\mathbb V}^{(i)}_j$, where ${\mathbb Q}.{\mathbb Q}^T={\mathbb I}$,  and carry out an orthogonal transformation to 
rewrite Eq.~\ref{eq:coupledmode2} as
\begin{align}
i\dfrac{\partial}{\partial z}{\mathbb {\bar A}}+{\mathbb {\bar B}}.{\mathbb {\bar A}}=0,\quad {\rm where} \quad  
{\mathbb {\bar A}}={\mathbb Q}.{\mathbb A},\ \ 
{\mathbb {\bar B}}={\mathbb Q}.{\mathbb B}.{\mathbb Q^T}.
\label{eq:coupledmode3}
\end{align} 
The orthogonal transformation of the matrix ${\mathbb B}$ results in a diagonal
matrix whose elements are the eigenvalues of ${\mathbb B}$, i.e., ${\mathbb {\bar B}}_{ij}={\bar \beta}_i\delta_{ij}$.

Using the above information, we can rewrite Eq.~\ref{eq:coupledmode3} as
\begin{align}
i\dfrac{\partial}{\partial z}{\bar A}_j(z)+{\bar \beta}_j {\bar A}_j(z)=0,\quad j=1,\cdots,N,
\label{eq:coupledmode4}
\end{align} 
where we have defined
\begin{align}
{\mathbb {\bar A}}=({\bar A}_1,{\bar A}_2,{\bar A}_3,\cdots,{\bar A}_{N-1},{\bar A}_N).
\label{eq:AbarvecDefine}
\end{align} 
The solutions to the $N$ independent first-order differential equations~\ref{eq:coupledmode4} for the unidirectional propagation of light 
can be simply written as
\begin{align}
{\bar A}_j(z)={\bar A}_j(0)\exp[-i{\bar \beta}_j z],\quad j=1,\cdots,N.
\label{eq:coupledsol1}
\end{align} 
Using the orthogonal transformation and a few lines of simple algebra, Eq.~\ref{eq:coupledsol1} can be expressed as
\begin{subequations} 
\begin{align}
{A}_j(z)&=\sum^N_{k=1}\ \sum^N_{l=1}\ Q_{k,j}Q_{k,l}A_l(0)\exp[-i{\bar \beta}_k z],\quad j=1,\cdots,N.\\
        &=\sum^N_{k=1}\ Q_{k,j}Q_{k,101}\exp[-i{\bar \beta}_k z],\\
        &=\sum^N_{k=1}\ {\mathbb V}^{(k)}_j{\mathbb V}^{(k)}_{101}\exp[-i{\bar \beta}_k z].
\label{eq:coupledsol2}
\end{align} 
\end{subequations} 

Equation~\ref{eq:coupledsol2} brings the problem to its final form, using which we can now discuss the 
transverse Anderson localization in terms of the random matrix ${\mathbb B}$ in the language of
Section~\ref{sec:randommatrix}. Recall that we showed in Figure~\ref{fig:randomMatrix1} and the corresponding 
discussions that the eigenvectors of a symmetric tridiagonal matrix with random off-diagonal elements are 
localized over the element-position domain. Therefore, only a few elements of each ${\mathbb V}^{(k)}$ in 
Eq.~\ref{eq:coupledsol2} are non-zero. We also observed in Figure~\ref{fig:randomMatrix1} that the localization 
of the eigenvectors occurs at different locations in the  element-position domain; therefore, only very few 
eigenvectors ${\mathbb V}^{(k)}$ have non-zero elements in the $j=101$ position. As a result, the sum 
in Eq.~\ref{eq:coupledsol2} is practically limited to only a few eigenvectors (modes) ${\mathbb V}^{(k)}$
that have non-zero $j=101$ elements and the other non-zero elements of these few eigenvectors are all concentrated
around $j=101$ due to localization. Therefore, $A_j(z)$ will be equal to zero for all $z$ if $j$ is very different
from $101$.

From the above discussions we can conclude that for a stronger disorder, the localization becomes stronger
as seen in Figure~\ref{fig:randomMatrix1}, and $A_j(z)$ will remain zero unless $j$ is even closer to $101$.
The picture portrayed here is consistent with Figure~\ref{fig:randomCoupledWaveguides1D}, especially 
in describing the difference between Figure~\ref{fig:randomCoupledWaveguides1D}{\claop (b)} and Figure~\ref{fig:randomCoupledWaveguides1D}{\claop (c)}.

Lets recap these observations in the more common modal language.
The initial beam only couples to those guided modes that are localized 
in the vicinity of the center fiber. Near the entrance, the optical excitation
closely resembles the in-coupling beam, as a result of a coherent sum
of the excited modes. Each excited mode propagates with a different
phase velocity determined by the propagation constant ${\bar \beta}_k$; therefore
the detailed balance between the excitation amplitudes of the guided modes that is 
responsible for the narrow excitation at the entrance $A_j(z=0)=\delta_{j,101}$ is 
broken as the relative phases between the modes change when the beam propagates along the 
fiber array. As the beam propagates and the detailed balance is further broken,
the beam expands; however, it can never expand beyond the size of its constituent modes,
which are only a few, each being very localized. Therefore, the expansion is eventually 
halted, as observed in Figure~\ref{fig:randomCoupledWaveguides1D}{\claop (b)} and Figure~\ref{fig:randomCoupledWaveguides1D}{\claop (c)}.

Using the modal language we can explain a few other interesting phenomena. One question that
is often asked is what happens if the initial excitation is wider than the typical width of an individual mode?
The answer is that a wider beam excites a few more modes; therefore, the propagating light 
eventually localizes to a slightly wider beam. Another interesting fact is that because
in Eq.~\ref{eq:coupledsol2} only a few terms in the sum play a role, after a sufficiently long
propagation distance $z_r$, the beam refocuses back to its initial intensity distribution 
(self-imaging)~\cite{SalmanModal,SalmanImage}.
This revival distance can be calculated by noting that at $z_r$,
$({\bar \beta}_k-{\bar \beta}_{101}) z_r = 2\pi m_k$ must be satisfied for all the relevant terms in the sum,
where each $m_k$ is an integer.
The intensity pattern at $z_r$ will be identical to that at $z=0$. In practice, if the number of excited modes 
is sufficiently large, the revival distance can be much longer than the waveguide and the refocusing
is not observed.

It must be noted that a similar localization behavior is obtained if instead of randomizing the coupling 
coefficients between the fibers in the array~\cite{Martin}, the propagation constants of individual fibers are 
slightly randomized~\cite{Ghosh3}. This was briefly discussed in Section~\ref{sec:randommatrix} as diagonal localization.
\subsection*{Mode shapes}
\label{Sec:modeshapes}
\begin{figure}[htpb]
\centering
\includegraphics[width=2.5in]{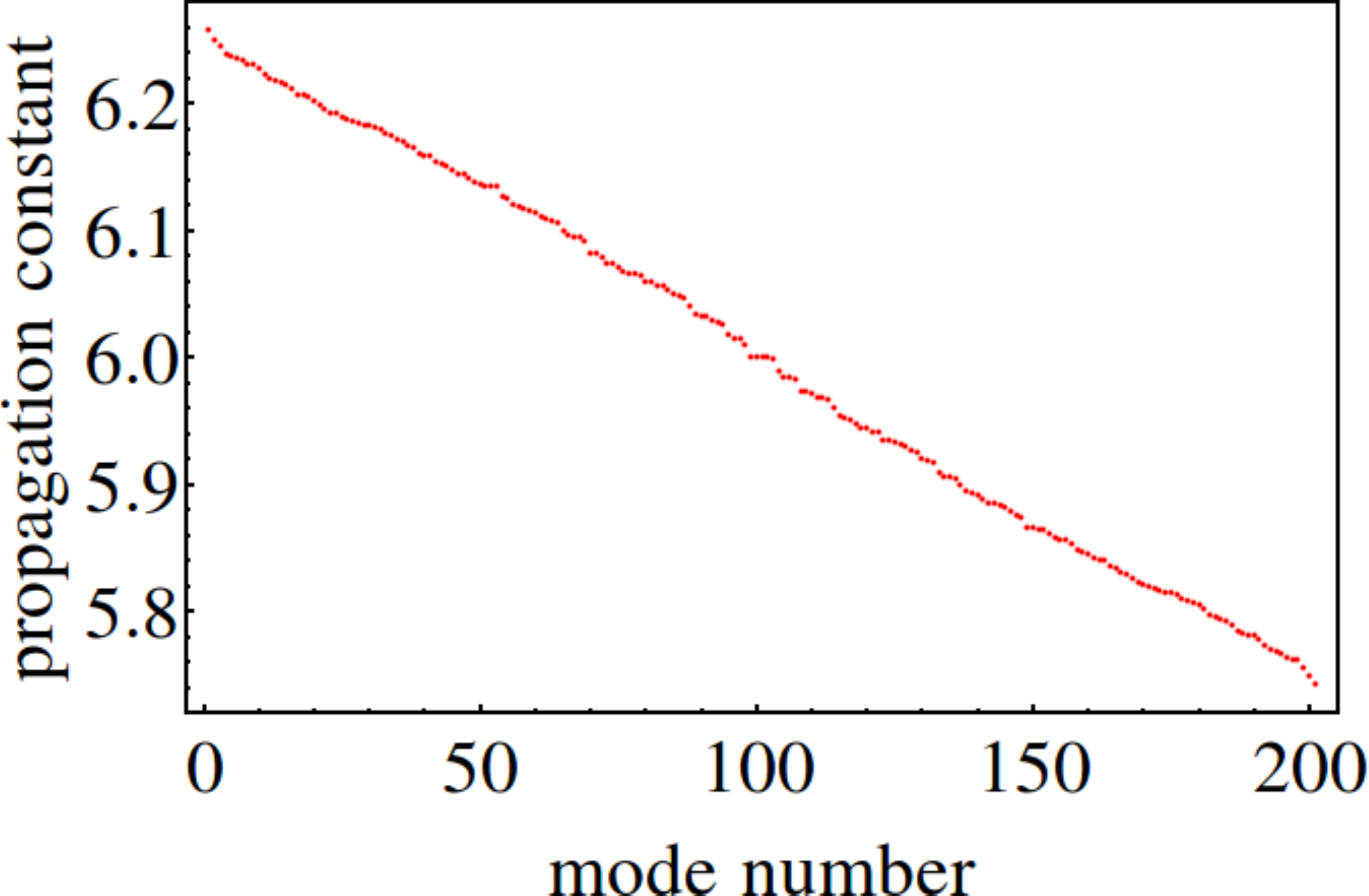}
\caption{Propagation constants of a random coupled waveguide array, where the modes are 
numbered from $1$ to $N=201$, sorted in a descending order of the value of the propagation constant.}
\label{fig:1D-waveguide-prop-const}
\end{figure}

In Figure~\ref{fig:1D-waveguide-prop-const} we show the distribution of the propagation constants 
of the eigenmodes of the coupled waveguide array of Figure~\ref{fig:1D-array-fiber}; i.e., 
${\bar \beta}_j$ in Eq.~\ref{eq:coupledsol1}, for $j=1,\cdots,201$. 
The parameters used for the disorder relate to the case discussed in 
Figure~\ref{fig:randomCoupledWaveguides1D}{\claop (b)}, where we used $\beta_0 = 6$, $c_0=0.01$, $N=201$, and
$c^+_{j}=c_0+r_j$, with $r_j\in{\rm unif}[-0.006,0.006]$. The calculated modes are numbered from $j=1$ to $j=N$, 
sorted in a descending order of the value of the propagation constant. As expected, the calculated
propagation constants are in the vicinity of $\beta_0 = 6$. The range of variation of ${\bar \beta}_j$
is determined by the range of $r_j$ (${\rm unif}[-0.006,0.006]$), while the exact values vary for each different
random waveguide.

\begin{figure}[htpb]
\centering
\includegraphics[width=\textwidth]{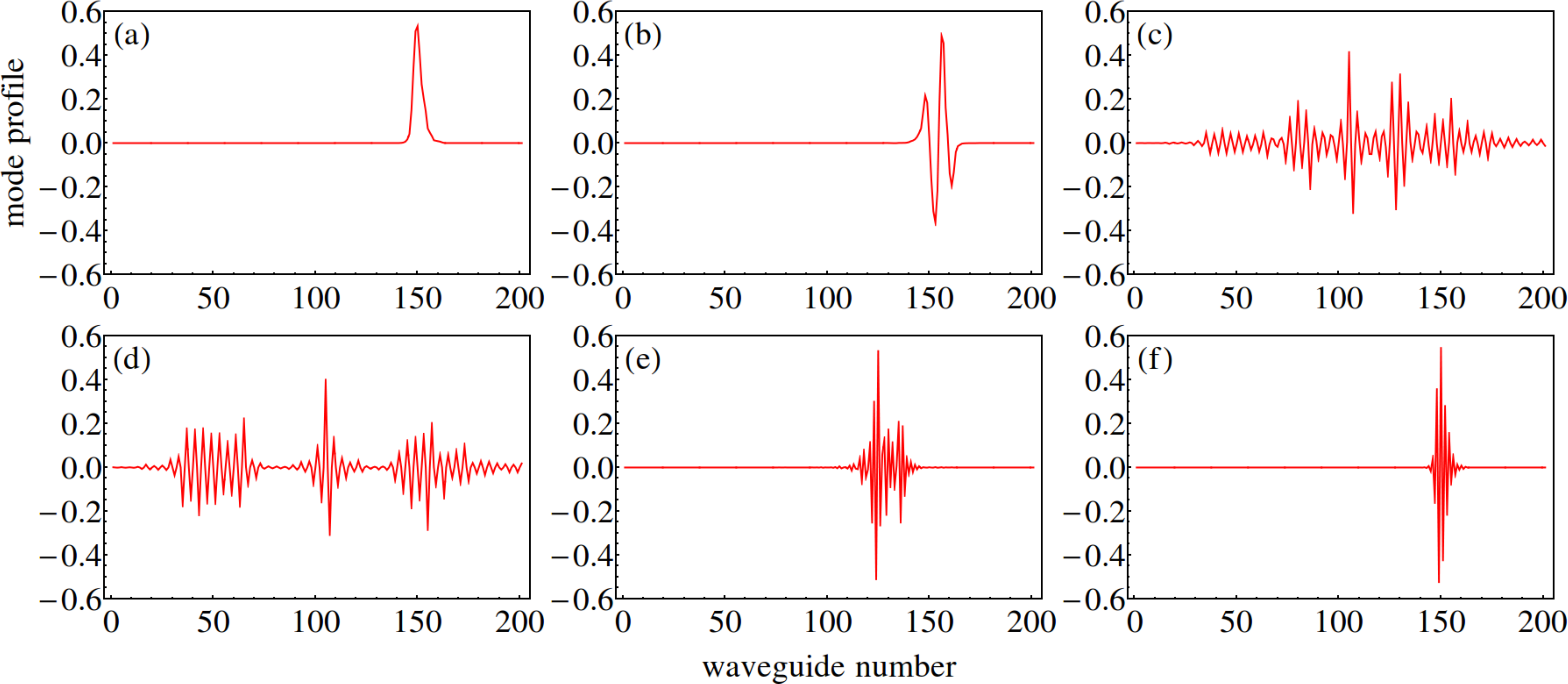}
\caption{Mode profiles of the random coupled waveguide array of Figure~\ref{fig:1D-array-fiber} are plotted. Subfigures
{\claop(a)}, {\claop(b)}, {\claop(c)}, {\claop(d)}, {\claop(e)}, {\claop(f)}, correspond to mode numbers 1, 25, 99, 101, 170, 201, 
respectively. The modes whose propagation constants belong to the region near the top edge of the band in Figure~\ref{fig:1D-waveguide-prop-const} are
highly localized with no or few oscillations as in {\claop(a)} and {\claop(b)}. The modes with propagation constants near the middle of the band 
spread over many waveguides and oscillate  as in {\claop(c)} and {\claop(d)}. The modes with propagation constants near the bottom of the band 
are highly localized and oscillate so rapidly that the sign of the mode profile flips between adjacent waveguides as in {\claop(c)} and {\claop(d)}.}
\label{fig:1D-mode-profiles}
\end{figure}

In Figure~\ref{fig:1D-mode-profiles} we plot the shape of a few modes of the waveguide. Of course, 
the coupled waveguide array supports a total of $N=201$ modes. In Figure~\ref{fig:1D-mode-profiles} each mode
is normalized to unity for easier comparison. The horizontal axis is the waveguide number in Figure~\ref{fig:1D-array-fiber}.
Figures~\ref{fig:1D-mode-profiles}{\claop(a)}, {\claop(b)}, {\claop(c)}, {\claop(d)}, {\claop(e)}, {\claop(f)} correspond to
mode numbers 1, 25, 99, 101, 170, 201, respectively. Recall that the mode numbers here correspond to the
mode numbers used in Figure~\ref{fig:1D-waveguide-prop-const}. Therefore, Figure~\ref{fig:1D-mode-profiles}{\claop(a)}  
is the mode with the largest propagation constant in Figure~\ref{fig:1D-waveguide-prop-const}, which is the
top edge of the propagation constant band. 
The modes that correspond to the region in the vicinity of the top edge 
of the band are highly localized with very few oscillations. As the propagation constant is increased e.g., for
mode number 25 the mode profile oscillates a few times as in Figure~\ref{fig:1D-mode-profiles}{\claop(b)}.
The modes near the middle of the band in the vicinity of $j=100$ generally oscillate rapidly and spread
over many waveguides, as can be seen for modes number 99 and 101 in Figures~\ref{fig:1D-mode-profiles}{\claop(c)} and 
{\claop(d)}, respectively. These modes become more localized if the disorder is increased.
Once the mode number gets closer to the bottom edge of the band, the oscillations become
so rapid that the sign of the mode profile flips between adjacent waveguides, and the modes localize again.
This behavior can be clearly seen for modes number 170 and 201 in Figures~\ref{fig:1D-mode-profiles}{\claop(e)} and 
{\claop(f)}, respectively. 

\begin{figure}[htpb]
\centering
\includegraphics[width=4in]{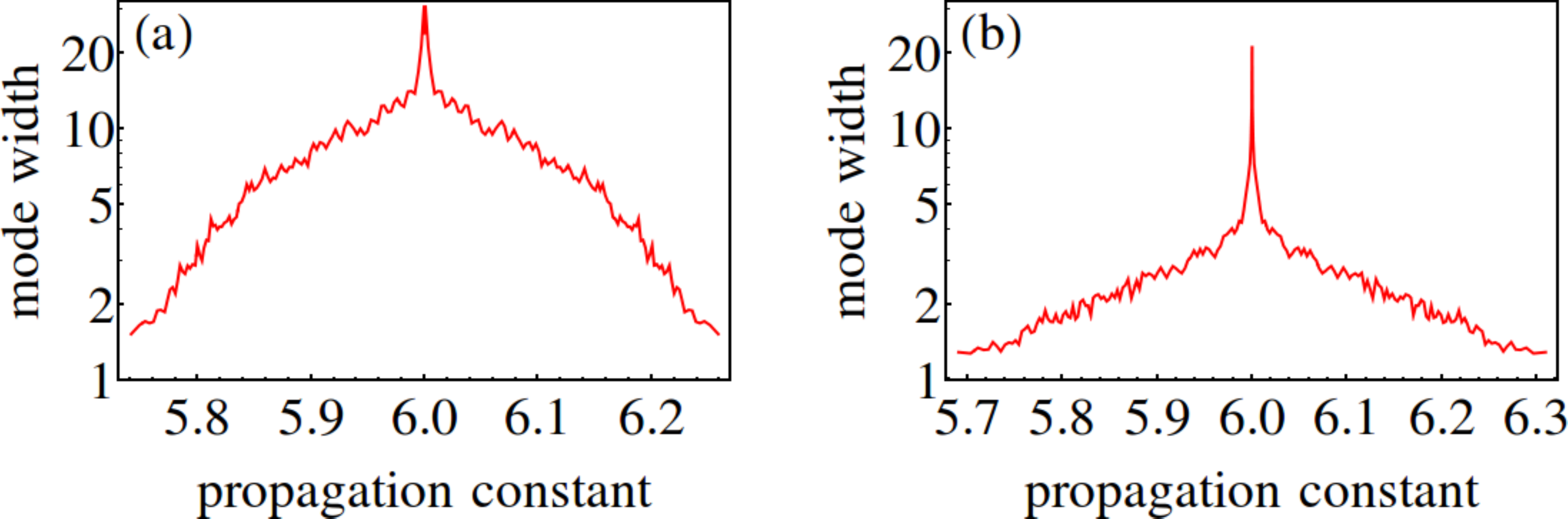}
\caption{The mode width is plotted against the propagation constant of each mode. The results are averaged 
over 1000 simulations for {\claop (a)} $r_j\in{\rm unif}[-0.006,0.006]$ and {\claop (b)} $r_j\in{\rm unif}[-0.01,0.01]$.}
\label{fig:1D-waveguide-prop-const-mode-widths}
\end{figure}

The mode width is plotted against the propagation constant of each mode in Figure~\ref{fig:1D-waveguide-prop-const-mode-widths},
averaged over 1000 simulations for {\claop (a)} $r_j\in{\rm unif}[-0.006,0.006]$ and {\claop (b)} $r_j\in{\rm unif}[-0.01,0.01]$.
It is clear that the band edges result in the narrowest modes, while the modes near the middle of the band are more spread out. Also, 
the stronger disorder in Figure~\ref{fig:1D-waveguide-prop-const-mode-widths}{\claop (b)} results in smaller mode widths compared with
the weaker disorder case in Figure~\ref{fig:1D-waveguide-prop-const-mode-widths}{\claop (a)}.

The disorder in the model studied here is off-diagonal. Although the effects of diagonal and off-diagonal disorder
are generally quite similar, there exist subtle differences between the two types of disorder. For example,
it has been argued that while all modes are exponentially localized for diagonal disorder, this may not
be strictly true for off-diagonal disorder~\cite{Theodorou,Stone,Soukoulis1}. Specifically, the modes at the
center of the band in a 1D off-diagonally disordered lattice
fall off asymptotically as $\exp(-\lambda \sqrt{|j-j_0|})$ instead of the simple exponential
form $\exp(-\lambda |j-j_0|)$, where $|j-j_0|$ represents the distance from the center of the mode at $j_0$.
The $\exp(-\lambda \sqrt{|j-j_0|})$ form is also responsible for the divergence of the mode width 
in the middle of the band in both subfigures of Fig.~\ref{fig:1D-waveguide-prop-const-mode-widths}. We emphasize that even 
in the middle of band, the mode is localized, although the weaker localization form results in a
divergent mode width. There are also differences between diagonal and off-diagonal disorder in 2D that will 
be pointed out later in section~\ref{sec:AAmechanism}.

\subsection*{Beam width in position and momentum space}
\label{Sec:momentum}
\begin{figure}[htpb]
\centering
\includegraphics[width=5.in]{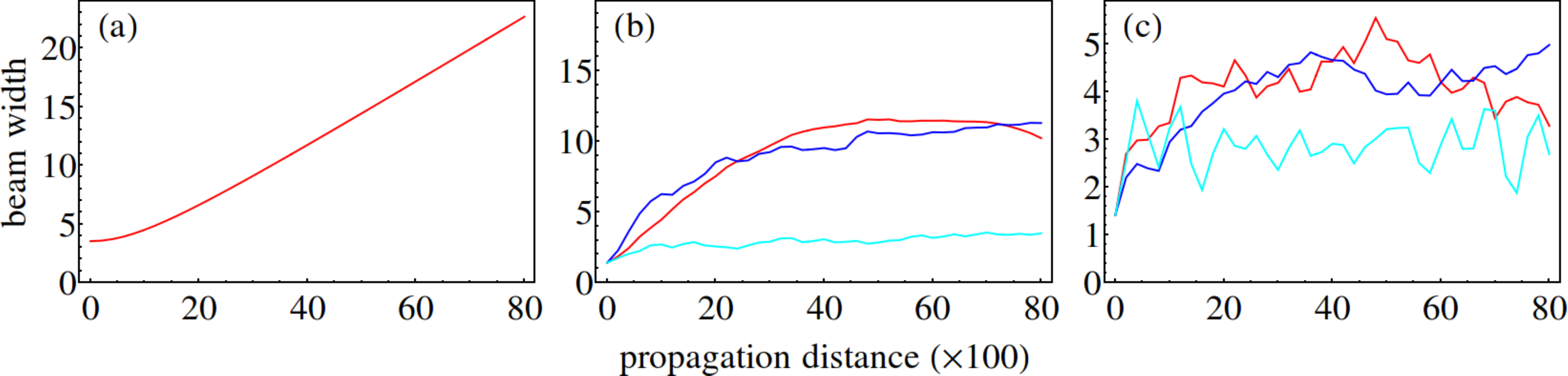}
\caption{This figure shows the expansion of the {\em x-space} beam width in the {\claop (a)}
disorder-free lattice, {\claop (b)} weakly disordered lattice, and 
{\claop (c)} strongly disordered lattice. Three sample realization
of the random waveguide are shown for each disorder level. The disorder-free
lattice shows ballistic expansion, while disorder-induced localization is apparent
in disordered samples.
}
\label{fig:beamExpansion-sampleX-Waveguides1D}
\vspace*{3mm}

\includegraphics[width=5.in]{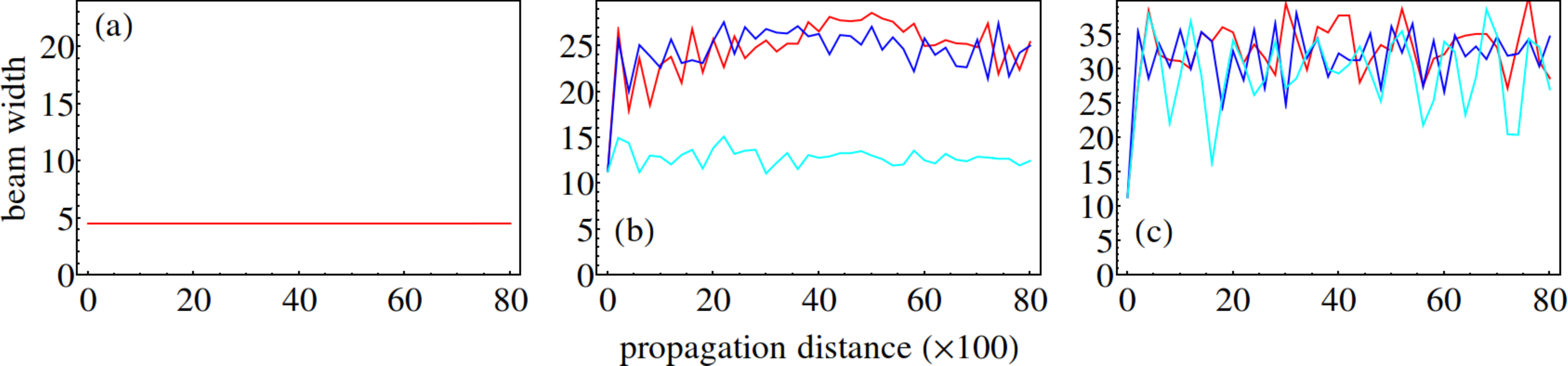}
\caption{Same as Figure~\ref{fig:beamExpansion-sampleX-Waveguides1D}, except the beam width is calculated in the {\em k-space}.
The {\em k-space} beam width for diffractive propagation in the disorder-free periodic lattice of {\claop (a)} remains unchanged. 
}
\label{fig:beamExpansion-sampleF-Waveguides1D}
\vspace*{3mm}

\includegraphics[width=5.in]{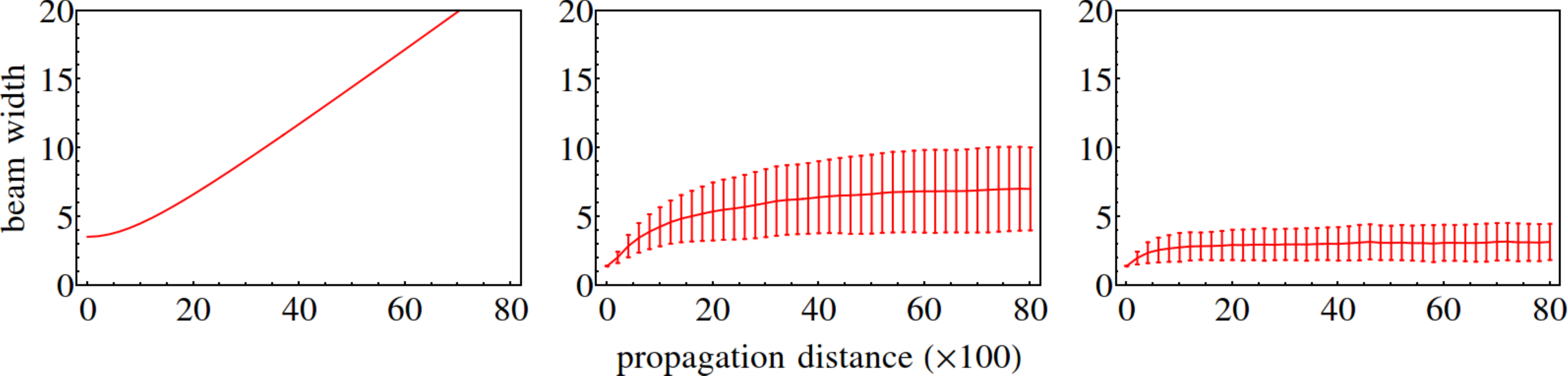}
\caption{Same as Figure~\ref{fig:beamExpansion-sampleX-Waveguides1D}, 
except the {\em x-space} beam width is averaged over 100 independent statistical realizations 
of the disordered waveguides. The error bars signify the one standard deviation 
for the beam width over the 100 samples.}
\label{fig:beamExpansion-XSPACE-Waveguides1D}
\vspace*{3mm}

\includegraphics[width=5.in]{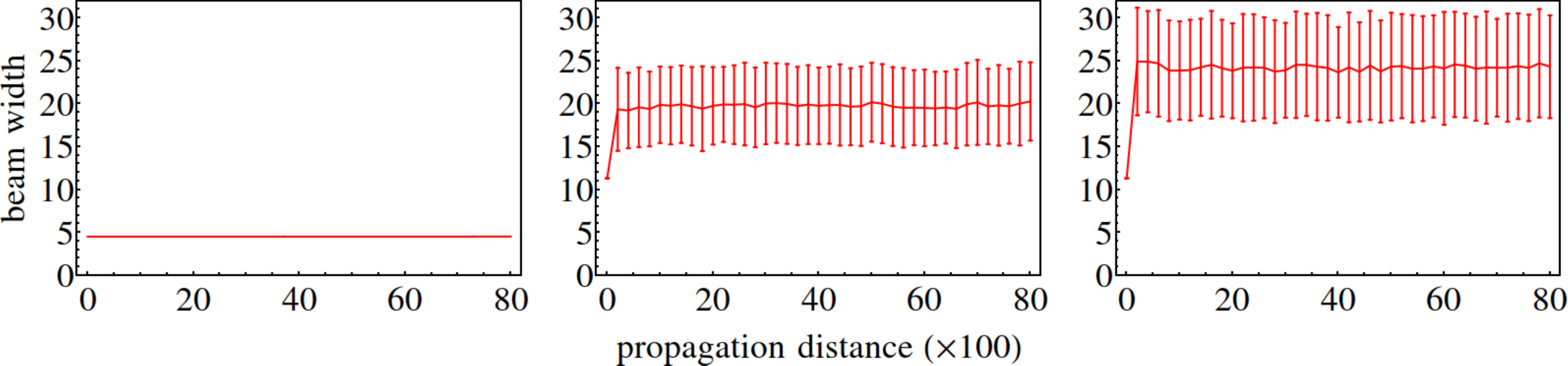}
\caption{Same as Figure~\ref{fig:beamExpansion-sampleF-Waveguides1D}, 
except the {\em k-space} beam width is averaged over 100 independent statistical realizations 
of the disordered waveguides. The error bars signify the one standard deviation 
for the beam width over the 100 samples.}
\label{fig:beamExpansion-FSPACE-Waveguides1D}
\end{figure}

In Figures~\ref{fig:beamExpansion-sampleX-Waveguides1D},~\ref{fig:beamExpansion-sampleF-Waveguides1D},~\ref{fig:beamExpansion-XSPACE-Waveguides1D},~\ref{fig:beamExpansion-FSPACE-Waveguides1D}, we
explore in some detail the expansion and localization of the optical beam that is coupled
to the waveguide array of Figure~\ref{fig:1D-array-fiber}, as a function of the longitudinal coordinate.
The vertical axis is the beam width, which is determined by the 
second moment method of Eq.~\ref{eq:xi}. The parameters used for the disorder relate to the 
cases already discussed in Figure~\ref{fig:randomCoupledWaveguides1D}. For these figures 
we have used $\beta_0 = 6$, $c_0=0.01$, and $N=201$. Any figure labeled with
{\claop(a)} corresponds to the disorder-free periodic array;
labeled with {\claop(b)} corresponds to the weakly disordered case of $c^+_{j}=c_0+r_j$, where 
$r_j\in{\rm unif}[-0.006,0.006]$; and 
labeled with {\claop(c)} corresponds to the strongly disordered case of $r_j\in{\rm unif}[-0.01,0.01]$.
The only minor differences are that $0 \le z \le 8000$, and the boundary condition is set to 
\begin{equation}
A_j(z=0)=\exp[-\dfrac{(j - 101)^2}{4{\cal W}^2_0}],\quad j=1,\cdots,N, \quad {\cal W}_0=\sqrt{2}.
\label{eq:exponentialInput}
\end{equation}
We note that the beam width formula of Eq.~\ref{eq:xi} (presented later in this article) 
also gives the value of $\sqrt{2}$ for the beam width at $z=0$.

Figure~\ref{fig:beamExpansion-sampleX-Waveguides1D}{\claop (a)} shows the expansion of the beam width in the 
disorder-free lattice. The expansion is ballistic, i.e., the width grows linearly with propagation 
distance $z$ (at large $z$). In Figure~\ref{fig:beamExpansion-sampleX-Waveguides1D}{\claop (b)} we plot 
the expansion of the beam width for three sample realizations of the random waveguide for the weakly disordered
case, as explained above. The tendency to localize can be seen in each of these samples, but one can immediately
see the random nature of the process, where different random realizations of the coupled waveguide system 
result in different rates of initial expansion and final beam localization width. 
Lastly, the expansion of the beam width for three sample realizations of the random waveguide for the strongly 
disordered case is shown in Figure~\ref{fig:beamExpansion-sampleX-Waveguides1D}{\claop (c)}. Note the different
vertical scale in these three subfigures. It is clear that despite the random variation the disorder 
slows and eventually halts the expansion, and a stronger disorder results in a smaller eventual localization 
width {\em on average}.

It is interesting to study the beam expansion in the Fourier $k${\em -space} (momentum space) as well. A well-known
characteristic of the ballistic expansion of the beam is that the beam width in the $k${\em -space} remains invariant
under propagation, as shown in Figure~\ref{fig:beamExpansion-sampleF-Waveguides1D}{\claop (a)}. The $k${\em -space} expansion
of the cases plotted in Figures~\ref{fig:beamExpansion-sampleX-Waveguides1D}{\claop (b), (c)} are shown in
Figures~\ref{fig:beamExpansion-sampleF-Waveguides1D}{\claop (b), (c)}. Interestingly, the beam widths in the $k${\em -space}
conform well to our intuition that narrower beams in $x${\em -space} are wider in $k${\em -space}; 
of course, this simplistic intuition is not always true, especially in the presence of spatial chirp.
 
As was noted above, the statistical nature of Anderson localization means that the beam expansion rate and the 
eventual localization width varies depending on the particular ``random'' realization of the disordered waveguide.
Therefore, no two propagations will be identical. However, some general conclusions can still be 
drawn, in a statistical sense, about the average beam width, as well as the variation around the average. 
In Figure~\ref{fig:beamExpansion-XSPACE-Waveguides1D} we show the average beam width over 100 independent
statistical realizations of the disordered waveguide, plotted as a function of the longitudinal propagation
coordinate $z$. Figure~\ref{fig:beamExpansion-XSPACE-Waveguides1D}{\claop (a)} is the disorder-free ballistic propagation
and is plotted again for comparison with Figures~\ref{fig:beamExpansion-XSPACE-Waveguides1D}{\claop (b)} and {\claop (c)},
which correspond to the weakly and strongly disordered cases, respectively. The vertical scales in these subfigures 
are all chosen identically, for easier comparison. It is clear that, on average, the localized beam width is narrower
for the case of a stronger disorder, as expected. The error bars in Figures~\ref{fig:beamExpansion-XSPACE-Waveguides1D}{\claop (b)} and {\claop (c)}
signify one standard deviation for the beam width around the mean over the 100 independent random samples. 
Strong disorder results, not only in a smaller beam width, but also in a smaller variation around the average beam width. 
Therefore, a stronger disorder is equivalent to a better predictability: when the disorder is strong, 
the beam width is smaller and is almost the same 
in all random realizations of the disordered waveguide.  

Finally, in Figure~\ref{fig:beamExpansion-FSPACE-Waveguides1D} we plot the corresponding average beam width
in $k${\em -space}, where the error bars are now defined as the one standard deviations of the beam width in 
$k${\em -space}. The interested reader is urged to consult Refs.~\cite{Lahini1,Izrailev,Nakhmedov,SegevNaturePhotonicsReview}
for a more detailed account of the beam expansion in 1D and 2D disordered lattices e.g., it is argued that
in a 2D disordered waveguide the expansion starts as ballistic then goes through a diffusive phase, and eventually
the localization takes over. However, in 1D the diffusive transport regime is absent and the expansion
turns from ballistic directly to localized.
\begin{highlight}
\textbf{Highlights:}
\begin{itemize}
\item The light propagating through a coupled array of identical optical waveguides remains confined to only a few waveguides 
(transversely localized), if the waveguide-to-waveguide couplings are random.
\item Similar localization behavior can be observed if the individual waveguides are randomized, regardless of whether
the waveguide-to-waveguide couplings are random or not. 
\item The stronger the disorder, the more localized the beam is.   
\end{itemize}
\end{highlight}
\section{Transverse Anderson localization of light in two transverse dimensions}
\label{sec:AAmechanism}
\begin{figure}[h]
\centering
\includegraphics[height=2.in]{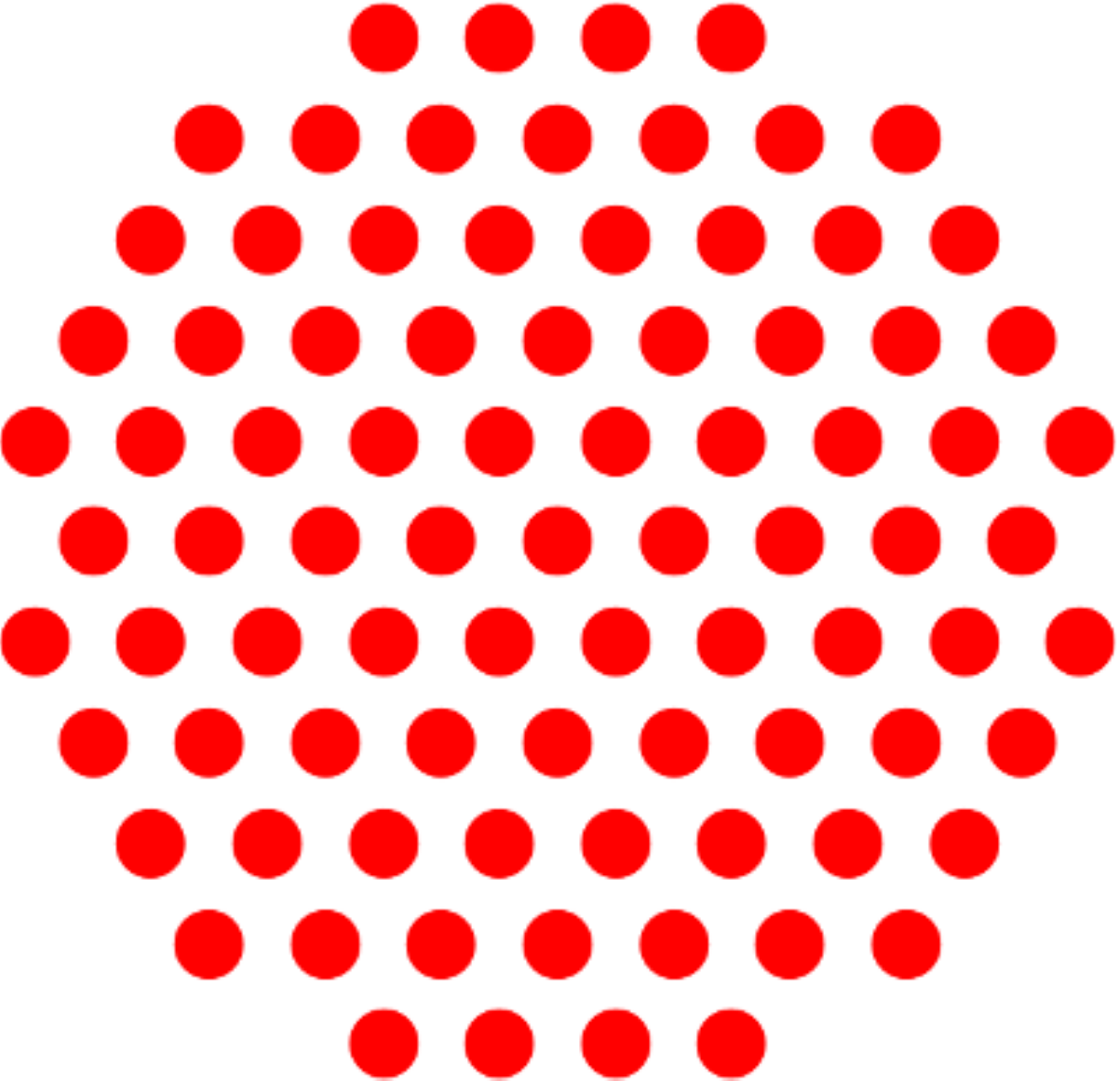}
\caption{A coupled array of optical fibers are placed on a hexagonal lattice. Actual simulations in this section are
performed on a hexagonal lattice of $N=817$ fibers.}
\label{fig:2D-array-fiber}
\end{figure}

Transverse Anderson localization in two transverse dimensions can be explored in a
similar fashion to 1D transverse localization discussed in the previous section.
Consider the 2D array of optical fibers placed in a hexagonal lattice in 
Figure~\ref{fig:2D-array-fiber}. The hexagonal lattice is special because each fiber
has six nearest neighbors separated by an equal distance $\Lambda$, and the next-to-nearest 
separations are $\sqrt{3}\Lambda$. In contrast, a fiber in a square lattice
has four nearest neighbors and the next-to-nearest separations are $\sqrt{2}\Lambda$.
Therefore, the next-to-nearest couplings are weaker in a hexagonal lattice and
the simplifying assumption of the nearest neighbor coupling is more reliable.
Similar to Eq.~\ref{eq:coupledmode1}, the coupled mode equations for the propagation of 
the optical field through this optical fiber lattice can be expressed as
\begin{align}
\Big(i\dfrac{\partial}{\partial z}+\beta_0\Big) A_j(z)
+ \sum_{k\ \in{\ \rm NN}} c_{jk} A_{k}(z)=0,
\label{eq:coupledmode2D}
\end{align} 
where the sum is only on the nearest neighbors, and the coupling matrix 
is symmetric. 

As we discussed in the previous section, the transverse Anderson localization 
can be explored by studying the distributions of the elements of eigenvectors 
of the ${\mathbb B}$ matrix over the element position space. Here, ${\mathbb B}$ is
the effective propagation constant matrix of the coupled fiber system and is defined
as ${\mathbb B}_{ij}=\beta_0\delta_{ij}+c_{jk}$, noting that $c_{jk}$ are non-zero
only when $j$ and $k$ are the nearest neighbor fibers. The fiber coupling constants 
are assumed to be random in the general form of $c_{jk}=c_0+r_{jk}$, where $c_0$
is a fixed value of $r_{jk}$ and belongs to a random distribution. Here, we take 
$\beta_0 = 6$ and $c_0=0.1$ on a hexagonal lattice of $N=817$ fibers. 

\begin{figure}[htp]
\centering
\includegraphics[width=5.in]{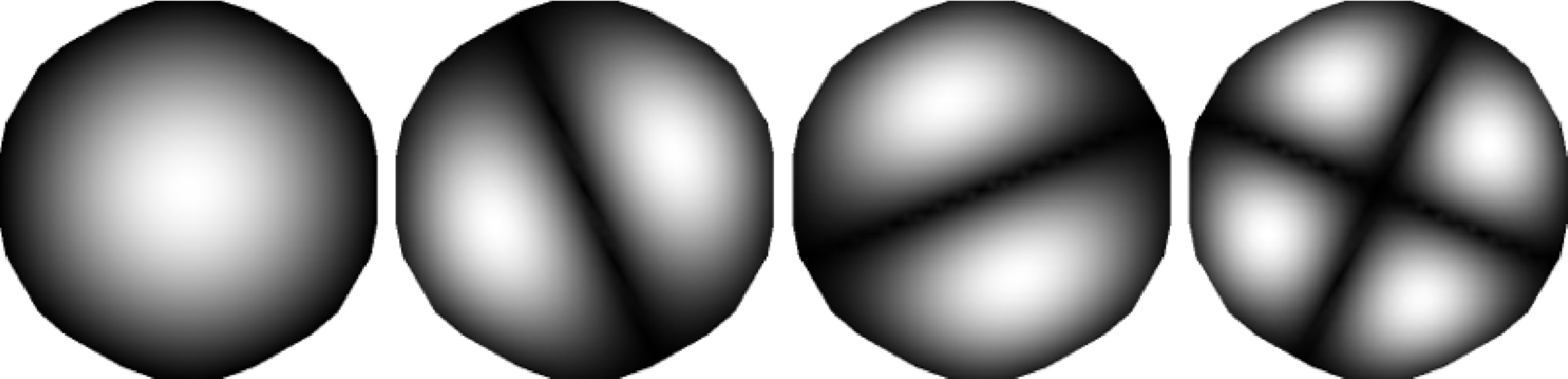}
\caption{Each subfigure shows the density plot of the elements of an eigenvector of the effective 
propagation constant matrix ${\mathbb B}$ on the element-position domain of the hexagonal lattice. 
The density plot shows only the absolute value of the elements of each eigenvector. In the language of
coupled mode theory, each density plot signifies the intensity distribution of a supermode of the
entire coupled fiber lattice. This Figure relates to the non-random deterministic situation, where
all fibers are identical and all coupling strengths to the nearest neighbors are equal. Note that
the eigenvectors spread over nearly the entire lattice.}
\label{fig:hex-array-uniform}
\vspace*{3mm}

\includegraphics[width=5.in]{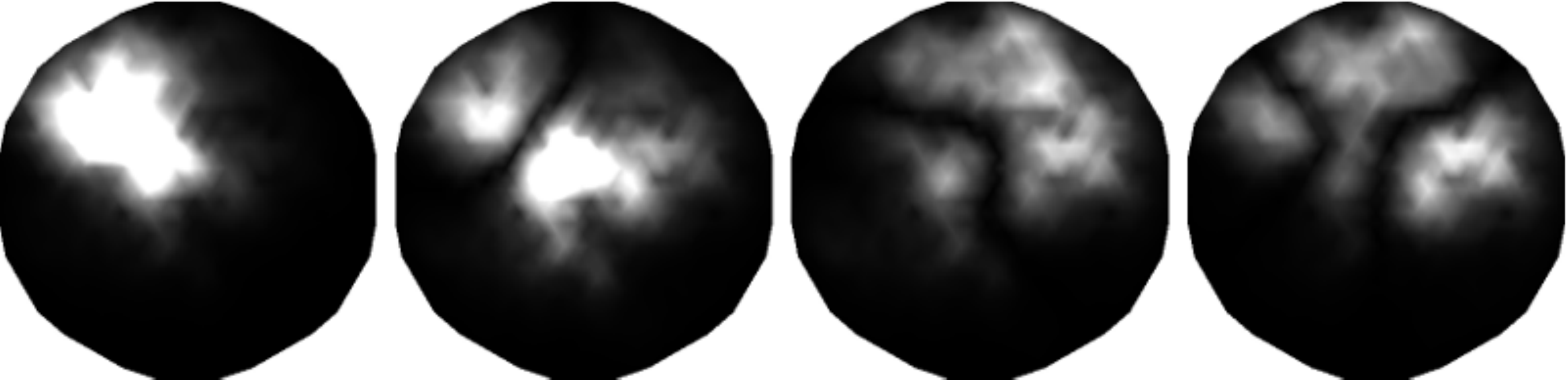}
\caption{This figure is similar to Figure~\ref{fig:hex-array-uniform}, except the nearest neighbor 
coupling strengths are randomized (off-diagonal disorder). Note that each eigenvector is localized in a certain region on the 
lattice, as expected from the transverse Anderson localization.}
\label{fig:hex-array-random-offdiagonal}
\vspace*{3mm}

\includegraphics[width=5.in]{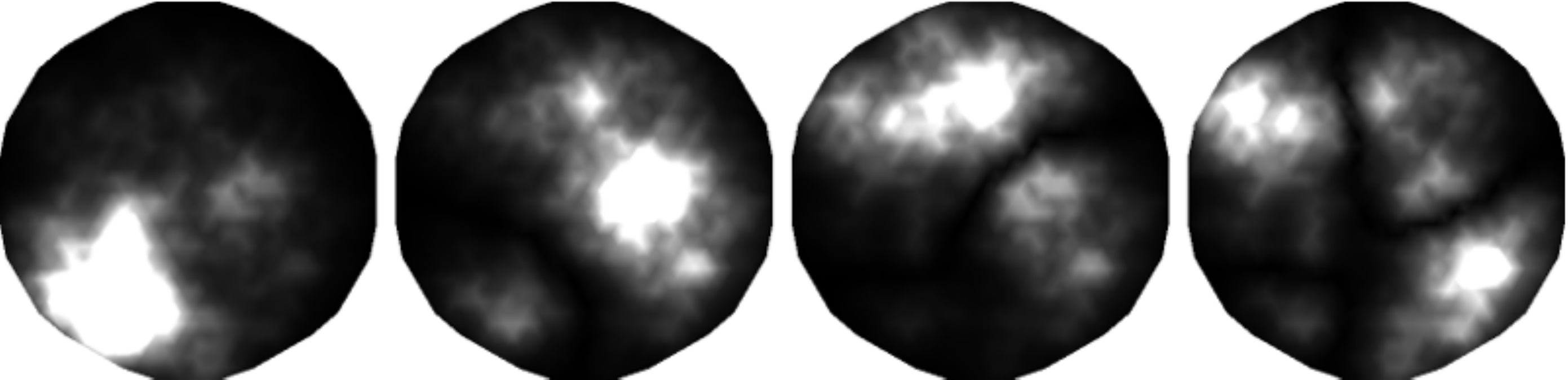}
\caption{This figure is similar to Figure~\ref{fig:hex-array-uniform}, except the propagation constants of the individual fibers
are randomized (diagonal disorder). Note that each eigenvector is localized in a certain region on the 
lattice similar to Figure~\ref{fig:hex-array-random-offdiagonal}, as expected from the transverse Anderson localization.}
\label{fig:hex-array-random-diagonal}
\end{figure}

In Figure~\ref{fig:hex-array-uniform} we consider the non-random deterministic 
case with $r_{jk}=0$. The magnitude of the four eigenvectors of the ${\mathbb B}$ matrix
with the largest eigenvalues are shown as density plots on the element-position domain
of the hexagonal lattice. For the reader familiar with coupled mode theory, the calculated
eigenvectors are the supermodes of the entire waveguide structure~\cite{MafiModeShape}.
The left most figure shows an azimuthally symmetric distribution
of non-zero elements that decrease monotonically from the center of the lattice toward
the boundary. Other eigenvectors have different distributions and symmetry properties,
but they all nearly fill the entire element-position domain, i.e., the non-zero elements 
of the eigenvectors spread over nearly the entire lattice.       

Let's consider what happens when the coupling constants are randomized according to
$r_{jk}\in{\rm unif}[-0.1,0.1]$ in Figure~\ref{fig:hex-array-random-offdiagonal}. Similar to the above, 
the magnitude of four eigenvectors of the ${\mathbb B}$ matrix are shown as density plots on the element-position 
domain of the hexagonal lattice. All figures show that the non-zero elements of the eigenvectors are localized
in certain regions on the lattice, as expected from the transverse Anderson localization. 
Of course, the localization region for each eigenvector is different.

As mentioned before, random variations of the diagonal elements of the ${\mathbb B}$ matrix can also induce
localization of the eigenvectors in the element-position domain. This is shown in Figure~\ref{fig:hex-array-random-diagonal},
where the off-diagonal elements are assumed to be constant values of $c_0=0.1$ and the diagonal elements (propagation constants
of the individual optical fibers) are randomized according to ${\mathbb B}_{ii}=\beta_0+\delta\beta_{i}$, where
$\delta\beta_{i}\in{\rm unif}[-0.2,0.2]$. Therefore, the effect of the diagonal disorder is similar to that of the off-diagonal
disorders, at least qualitatively. Of course, transverse Anderson localization is similarly observed if both diagonal and off-diagonal 
disorder are implemented simultaneously. Transverse Anderson localization in a 2D lattice of optical fibers with random variations 
of the parameters was first proposed by S.~S.~Abdullaev and F.~Kh.~Abdullaev of the Heat Physics Department of Uzbekistan Academy 
of Sciences in 1980~\cite{Abdullaev}.

{
Earlier in this tutorial review, we briefly highlighted the subtle differences between diagonal and off-diagonal disorders 
in a 1D disordered lattice. The weaker localization in the middle of the band for off-diagonally disordered lattices persists 
in 2D as well. For example, it has been argued in Ref.~\cite{Soukoulis2} that in the middle of the band
for a 2D disordered square lattice, the wave
amplitude falls off with distance as $R^{-\lambda}$ where $R$ is the distance from the center of the mode and $\lambda$
is an exponent that depends on the amount of disorder. The power-law localization is also attributed to the geometry 
of the square lattice (and the logarithmic singularity of the density of states in the middle of the band). Therefore,
for the triangular lattice that has been studied in this section (which does not have a logarithmic singularity of the 
density of states in the middle of the band), all modes are exponentially localized.  
}
\begin{highlight}
\textbf{Highlights:}
\begin{itemize}
\item Transverse Anderson localization is observed for a 2D disordered coupled waveguide array, just as in 1D.
\end{itemize}
\end{highlight}
\section{Transverse Anderson localization of light: the RLV mechanism}
\label{sec:RLV}
The 2D transverse Anderson localization of light that was proposed by 
Abdullaev and Abdullaev and discussed in the previous section strongly resembles
the original proposal by P.~W.~Anderson. We will refer to this as the AA mechanism
for the rest of this article, not to be confused with the Aubry-Andre model~\cite{Aubry}. 
In either case, the starting point is
a lattice-periodic potential where each site on the lattice is characterized by
a ``bound-state'' energy ($\beta_0$ for fibers) and off-diagonal 
couplings ($c_{jk}$ for fibers). In the absence of randomness, the solutions to the
wave equations are Bloch-periodic solutions that extend over the entire lattice. If
sufficient randomness is introduced in the energy and/or couplings, the solutions
of the wave equation no longer extend over the entire lattice; rather, they localize to
certain regions on the lattice.  

\begin{figure}[htp]
\centering\includegraphics[width=5.in]{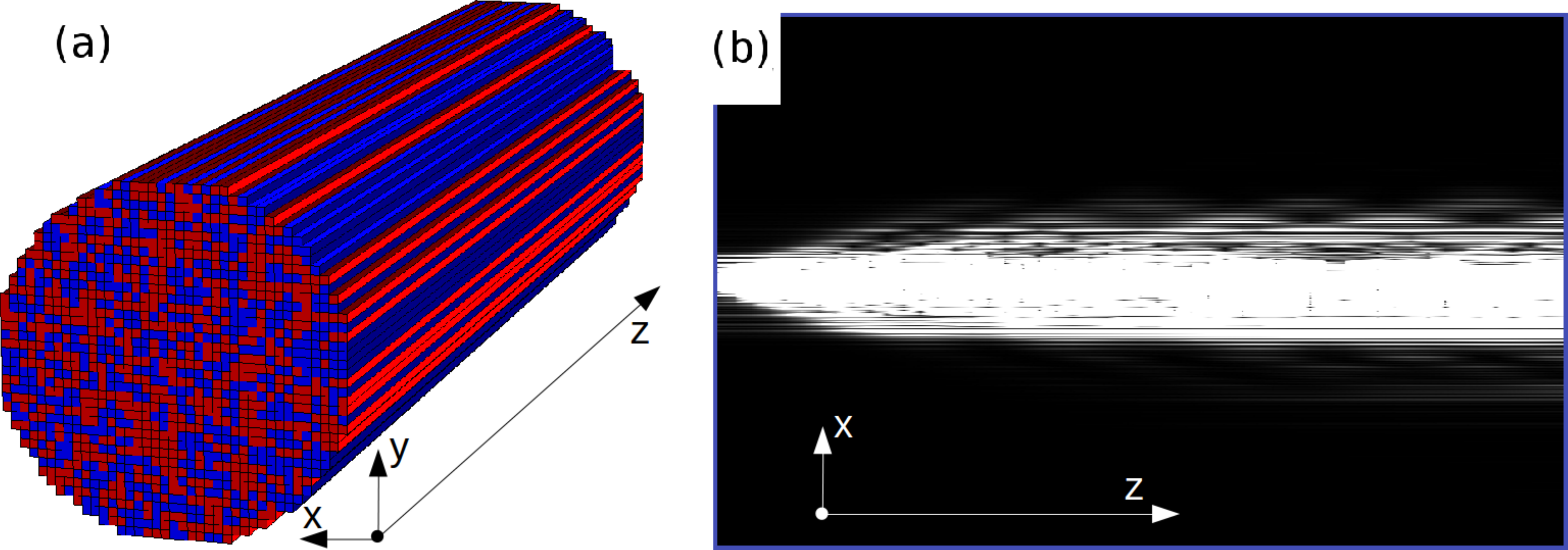}
  \caption{{\claop (a)} Sketch of the transversely random and longitudinally invariant dielectric medium for 
               the observation of the transverse Anderson localization. {\claop (b)} Cross section of 
               a Gaussian beam that is coupled to the disordered waveguide. The intensity distribution 
               shows that the beam goes through an initial expansion and eventually localizes to a 
               stable width, as expected from the transverse Anderson localization.}
    \label{fig:beam-prop-sketch}
\end{figure}

An alternative method to obtain the 2D transverse Anderson localization of light was 
independently proposed by de Raedt, Lagendijk, and de Vries in 1989~\cite{DeRaedt}.
Recall that for the AA mechanism, the disorder was introduced on top of an existing 
ordered lattice.
However, in the suggested method by de Raedt, Lagendijk, and de Vries (RLV mechanism), 
the randomness is not superimposed on 
top of an existing ordered lattice; rather, the underlying potential is completely 
random~\cite{DeRaedt}. 
De~Raedt \textit {et al}.~\cite{DeRaedt} proposed that this
scheme can be realized in a quasi-2D optical system in a dielectric with a transversely 
random and longitudinally uniform refractive index profile. This is sketched in  
Figure~\ref{fig:beam-prop-sketch}{\claop a}, where a long dielectric ``waveguide'' has a random 
refractive index profile that is invariant along the fiber. De~Raedt \textit {et al}. 
suggested a binary random system, where each pixel is randomly chosen to have a refractive 
index of $n_1$ or $n_2$ with equal probabilities, where $n_1$ and $n_2$ are marked by
red and blue color, respectively in Figure~\ref{fig:beam-prop-sketch}{\claop a}.

Using extensive numerical simulations, De~Raedt \textit {et al}. showed that for 
a properly designed random fiber e.g. with $n_1=1.0$, $n_2=1.5$; and $d\approx \lambda$,
where $d$ is the width of each pixel and $\lambda$ is the wavelength of light, they obtain
quasi-2D localization in the transverse plane of the waveguide. An optical field that is 
launched in the longitudinal direction initially expands until it reaches a terminal 
localization radius, after which the freely propagating beam fluctuates around a relatively 
stable radius. In Figure~\ref{fig:beam-prop-sketch}{\claop b}, a slice of the intensity
distribution in the $x$-$z$ plane is shown for a representative localized propagation through a 
random binary waveguide, where $x$ is a transverse coordinate and $z$ is the longitudinal 
coordinate. 
\begin{highlight}
\textbf{Highlights:}
\begin{itemize}
\item One way to observe transverse Anderson localization is to use a 2D disordered coupled waveguide 
array. We refer to this as the AA mechanism because it was originally proposed by
Abdullaev and Abdullaev.
\item Alternatively, transverse Anderson localization can be observed in a optical fiber-like 
medium with a fully random refractive index profile. We refer to this as the RLV mechanism
because it was originally proposed by de Raedt, Lagendijk, and de Vries. The RLV mechanism 
is distinct from the AA mechanism by the absence the underlying lattice.
\end{itemize}
\end{highlight}
\section{Experimental observation of localization: the AA mechanism}
\label{sec:SegevAA}
The first experimental observation of the transverse Anderson localization of light was carried 
out in Segev's group in 2007~\cite{Schwartz}. The experiment was more in line with the AA mechanism
discussed earlier in section~\ref{sec:AAmechanism}, where the disorder was superimposed
on top of an existing triangular lattice of waveguides. The ordered lattice and the superimposed disorder
were formed, by means of the optical induction technique~\cite{Efremidis}, in a photo-refractive crystal 
(SBN:60) using a laser beam at 514~nm wavelength.

In order to write the underlying ordered lattice in the photo-refractive crystal, 
Schwartz \textit {et al}.~\cite{Schwartz} used the interference pattern generated by three 
symmetrically arranged lasers and obtained a fully periodic hexagonal interference pattern 
with periodicity of 11.2~\textmu m inside the 10~mm-long crystal. The disorder is generated 
from a speckled beam that is invariant in the longitudinal direction, as required for the 
observation of transverse Anderson localization. The speckled beam was formed from a
Bessel beam, created by passing a Gaussian laser beam through an axicon, and then passed 
through a 4-f imaging system, where a diffuser was placed at the joint Fourier plane between 
the two lenses. The 4-f imaging system with the diffuser transformed the Bessel beam into a 
broad and speckled beam that was coherently combined with the original lattice-forming beam
to create the desired refractive index fluctuations on top of the ordered lattice in the 
photo-refractive crystal. 

In order to investigate the localization, Schwartz \textit {et al}. used another
probe beam at 514~nm wavelength. The beam width was 10.5~\textmu m full-width at half-maximum (FWHM)
and was always launched into the crystal at the same location. In the absence of disorder,
they observed diffraction patterns after 10~mm of propagation. They observed clear transverse 
localization of the probe beam when the disorder level was increased to more than 30\%.
In order to obtain an appropriate ensemble to calculate the mean localization radius (localization
length), the diffuser was rotated by a step-motor, such that the laser beam passed through a 
different location on the diffuser in each step, where a new intensity measurement was taken. 

For the large disorder level of 45\%, Schwartz \textit {et al}. fitted the properly averaged intensity
distribution of the localized beam, over 100 independent experiment, to an exponential of the form
$\exp(-2|r|/\xi)$, and obtained $\xi$=64~\textmu m for the localization length of the Anderson localized 
beam. The experiment carried out by Schwartz \textit {et al}. was quite interesting, as they were able to
vary the disorder level by controlling the intensity level of the disorder-inducing lasers,
and also obtained an ensemble of independent measurements for the statistical analysis of the localization 
phenomenon by rotating the diffuser. However, the variations of the refractive index of random sites 
in Ref.~\cite{Schwartz} were on the order of $10^{-4}$. As we will discuss later in greater detail,
the radius of the localized beam (localization length) depends on the magnitude of the refractive index
fluctuations. In order to obtain a more localized beam for device applications e.g., comparable
to the beam radius in a conventional optical fiber, it is necessary to increase the magnitude of the 
index fluctuations~\cite{ElDardiry}. Moreover, when the magnitude of the index fluctuation is larger, the sample-to-sample
variation in the beam radius becomes smaller; therefore, each element of the ensemble closely
resembles the average. This self-averaging behavior alleviates the need for averaging over a large 
ensemble, and for a sufficiently large magnitude of the index fluctuation the average is almost identical to
each element. We will discuss a recent realization of the transverse Anderson localization in the 
presence of a large index contrast in the next section.
\begin{highlight}
\textbf{Highlights:}
\begin{itemize}
\item The first observation of transverse Anderson localization was reported by Schwartz \textit {et al}
in an AA-mechanism setup.
\item The underlying lattice and the overlaying disorder were optically induced in a photo-refractive crystal.
A separate beam was used to probe the localization behavior. 
\item The setup allowed them to vary the amount of disorder on demand.   
\end{itemize}
\end{highlight}
\section{Experimental observation of localization: the RLV mechanism}
\label{sec:MafiRLV}
\begin{figure}[htbp]
\centering\includegraphics[width=5.in]{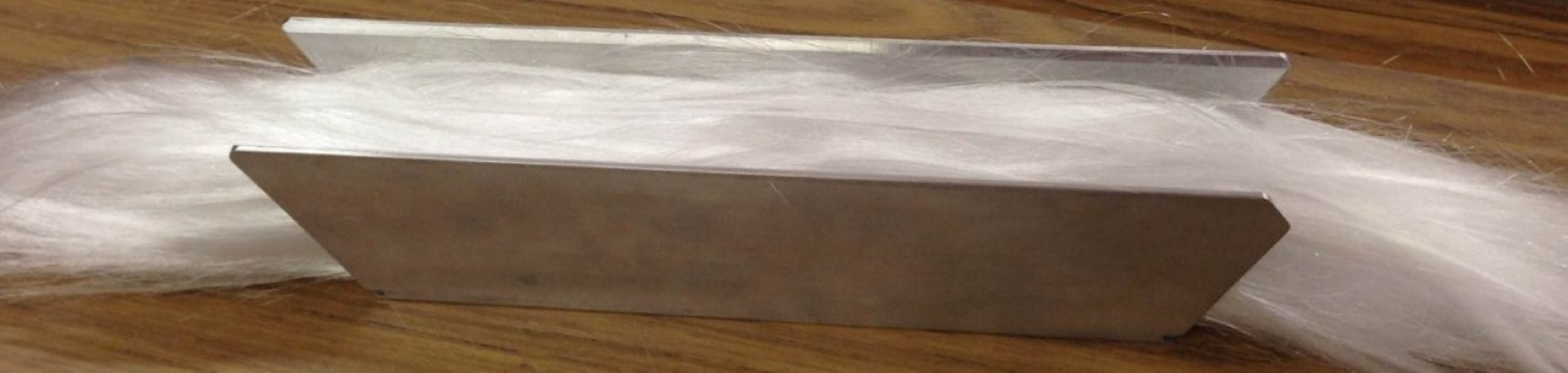}
  \caption{Random mixture of the PS and PMMA fiber strands.}
    \label{fig:fiber-strands}
\end{figure}

In section~\ref{sec:RLV} we discussed the RLV mechanism for the transverse Anderson localization of light. 
The quasi-2D system proposed by De~Raedt \textit {et al}.~\cite{DeRaedt} strongly resembles an optical fiber.
In 2012, Karbasi \textit {et al}.~\cite{SalmanOL} designed and fabricated an optical fiber that functioned based
on the RLV mechanism, i.e., transverse Anderson localized in the presence of an entirely random refractive 
index profile. In order to obtain the random and pixelated refractive index profile of the RLV mechanism,
Karbasi \textit {et al}.~\cite{SalmanOL} used 40,000~pieces of a low index polymer fiber polymethyl methacrylate (PMMA) 
with refractive index of 1.49 and 40,000 pieces of a high index polymer fiber polystyrene (PS) with refractive index 
of 1.59. Each fiber was 8 inches long with an approximate diameter of 200~\textmu m. The fibers were randomly mixed,
assembled to a square preform as shown in Figure~\ref{fig:fiber-strands}, fused together, and redrawn to a fiber 
with a nearly square profile and approximate side width of 250~\textmu m~\cite{SalmanJOVE}.

We recall that one of the requirements for transverse Anderson localization is the longitudinal invariance.
Some of the randomly mixed optical fibers may have crossed over each other during the assembly and
redraw process; however, the large draw ratio of the fiber guarantees that the refractive index profile remains 
relatively unchanged along the fiber. Moreover, small perturbations are likely not going to disturb the 
transverse localization noticeably.

\begin{figure}[htp]
\centering\includegraphics[width=5in]{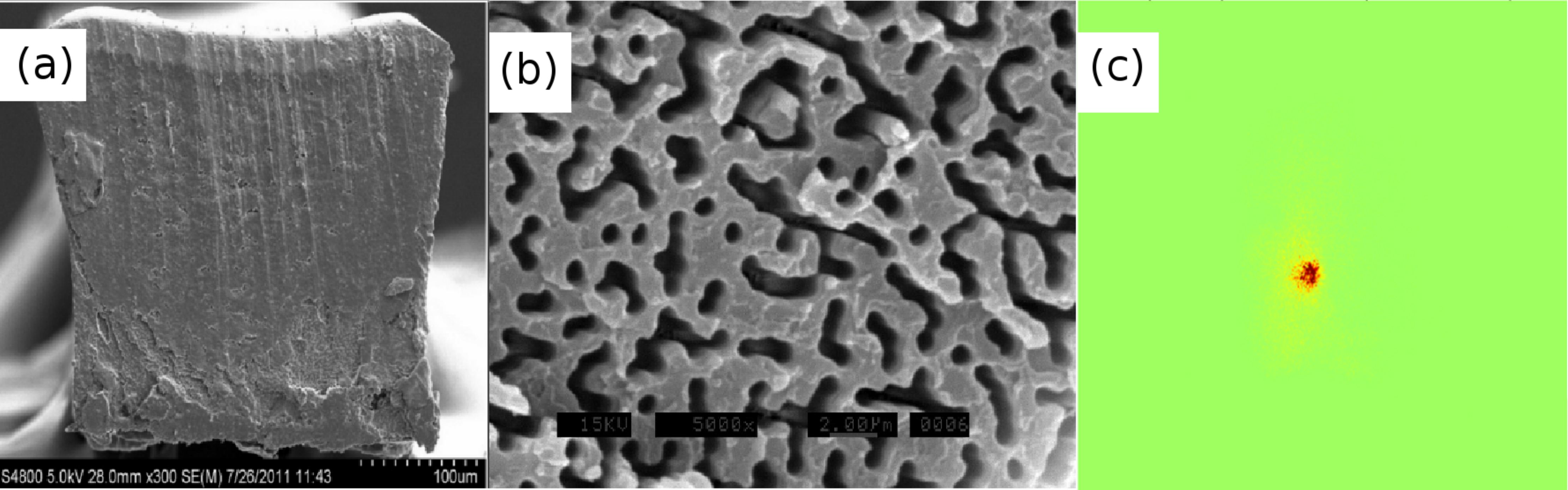}
  \caption{{\claop (a)} cross section of pALOF with a nearly square profile and an approximate side width 
               of 250~\textmu m; {\claop (b)} zoomed-in SEM image of a 24~\textmu m wide region on the tip 
               of pALOF exposed to a solvent to differentiate between PMMA and PS 
               polymer components, where feature sizes are around 0.9~\textmu m and darker regions are PMMA;
               and {\claop (c)} experimental measurement of the near-field intensity profile of the localized beam 
               after 60~cm of propagation through pALOF. The total side width of subfigure~c is 250~\textmu m, so it can
               be directly compared with subfigure~a. Adapted with permission, copyright 2012, Optical Society of America~\cite{SalmanOL}.}
    \label{fig:pALOF-profile-localization}
\end{figure}

Figure~\ref{fig:pALOF-profile-localization} shows the scanning electron microscope (SEM) image of a polished polymer 
Anderson localized optical fiber (pALOF). Figure~\ref{fig:pALOF-profile-localization}{\claop a } is the image of the cross section
of the optical fiber with an approximate side width of 250~\textmu m, where the high and low index regions
are not distinguishable in this figure. Figure~\ref{fig:pALOF-profile-localization}{\claop b} 
is a zoomed-in SEM image of a 24~\textmu m wide region on the tip of pALOF exposed to 70\% ethanol
solvent to to dissolve the PMMA, so that the PMMA region can be differentiated by the darker color.
The sizes of the random features (pixels) in Figure~\ref{fig:pALOF-profile-localization}{\claop b} are around 0.9~\textmu m. 

In order to investigate the guidance and localization properties of the pALOF, light from a single mode optical fiber was
directly launched (nearly butt-coupled) into the pALOF. The output near-field image was collected by a 40X objective and
was projected onto a CCD camera. Figure~\ref{fig:pALOF-profile-localization}c shows an experimental measurement of the 
near-field intensity profile of the localized beam after 60~cm of propagation through pALOF, where the wavelength of light 
is 633~nm. The localization was observed to be strong: when the input beam was scanned across the input facet, the output
beam clearly followed the transverse position of the incoming beam~\cite{SalmanOPEX}.

We encourage the interested reader to consult Ref.~\cite{Boguslawski} for another method to obtain transverse 
Anderson localization based on the RLV mechanism. The optically induced randomized potential is created by a computer 
controlled spatial light modulator (SLM) in a photorefractive crystal.
\begin{highlight}
\textbf{Highlights:}
\begin{itemize}
\item Experimental observation of transverse Anderson localization for the RLV mechanism was reported by 
Karbasi \textit {et al} in a disordered optical fiber medium.
\item  The disordered fiber was drawn from a random preform. The preform consisted of 80,000 strands of 
optical fibers with different refractive indexes that were randomly mixed and fused together. 
\item When the input beam was scanned across the input facet, the output
beam clearly followed the transverse position of the incoming beam
\end{itemize}
\end{highlight}
\section{Detailed analysis of the RLV localization scheme}
An attractive feature of the pALOF is the large index contrast of 0.1 that is helpful in reducing the localization 
radius (localization length) of the beam. The measured value for the localization radius was reported
as $\xi_{\rm avg}\sim$31~\textmu at 633~nm wavelength. The measured radius was only slightly larger than the calculated value; however, the
measured standard deviation of $\sigma_\xi\sim$14~\textmu m was considerably larger than the variations calculated numerically
and reported in Figure~3 of Ref.~\cite{SalmanOL}. The measured standard deviation for the beam radius was taken from 
100 separate measurements of the beam profile radius with 20 different fiber samples and 5 different locations across
the fiber. This discrepancy between the theoretical and experimental variation of the beam radius was mainly attributed 
to the quality of the polishing of the pALOF tip as the surface quality of a polished polymer optical fiber tip is 
generally lower than what is routinely achievable for glass optical fibers.
\subsection*{Numerical modeling of wave propagation in a disordered fiber}
\label{Sec:numerics}
The simulation of light propagation in a disordered fiber is carried out 
by numerically solving the wave propagation equation Eq.~\ref{eq:bpm}
using the finite difference beam propagation method (FD-BPM)~\cite{SalmanOPEX,Schwartz,Huang}.
\begin{equation}
\label{eq:bpm}
i\dfrac{\partial A}{\partial z}+
\dfrac{1}{2n_0k_0}\left[\nabla^2_T A+k_0^2\left(n^2-n^2_0\right)A\right]=0.
\end{equation}
Eq.~\ref{eq:bpm} is the paraxial approximation to the Helmholtz equation,
where $A({\bf r})$ is the slowly-varying envelope of the primarily transverse
electric field
$E({\bf r},t)={\rm Re}\left[A({\bf r})\exp(i n_0k_0 z-i\omega t)\right]$
centered around frequency $\omega$ and $k_0=2\pi/\lambda$.
$n(x,y)$ is the (random) refractive index of the optical fiber, which is a function
of the
transverse coordinates, and $n_0$ is average refractive index of the fiber.
The forward propagation scheme is
implemented using the fourth order Runge-Kutta method~\cite{Butcher}.

The stability condition for the fourth order Runge-Kutta method limits the size of the
steps in the longitudinal direction as $d z\le \alpha n_0k_0 dx^2$, where $dx=dy$ is
assumed to be the size of the transverse grid in the finite-difference numerical
scheme, and $\alpha=1/\sqrt{2}$ in a uniform medium. In the simulations in
Ref.~\cite{SalmanOPEX}, $\alpha=0.02$ is chosen in Ref.~\cite{SalmanOPEX} to ensure stability and no power dissipation 
for reliable long distance propagation. While transparent boundary condition~\cite{Hadley} is 
implemented, the size of the simulation domain is taken to be large enough to ensure that the
total power in the simulation region remains unchanged along the fiber for Anderson localized beams.

A typical reliable simulation of transverse Anderson localization e.g. for pALOF, requires a transverse area 
in the range of $\sim 10^5~\lambda^2$. The random refractive index pixel size is on the order of a wavelength,
and each pixel must be resolved by roughly 10 points in each direction in the finite difference scheme; therefore,
$\sim 10^7$ points are required in the transverse domain with $dx\sim\lambda/10$. This results in the fourth order 
Runge-Kutta stability criterion of the form $d z\le \alpha \lambda/10$, which is $d z\approx \lambda/500$ for
$\alpha=0.02$. For $\lambda\approx$500~nm,  $d z\approx$1~nm, and for a typical propagation distance of 1~cm required 
for reliable localization, $\sim 10^7$ steps in the longitudinal direction are required, which is computationally intensive.

The effective beam radius (localization length) is calculated by the variance method
\cite{DeRaedt} as
\begin{equation}
\xi(z)=\sqrt{\langle A({\bf r})|({\bf R}-\bar{{\bf R}})^2 |A({\bf r})\rangle},
\label{eq:xi}
\end{equation}
where the angle brackets denote integration over transverse $x-y$ coordinates.
${\bf R} = (x, y)$ is the transverse position vector and $\bar{{\bf R}}$ is the vector pointing to
the center of the beam, defined as the mean intensity position by
$\bar{\bf R} = \langle A({\bf r})|{\bf R}|A({\bf r})\rangle$. The optical field is assumed to be normalized according to
$\langle A({\bf r})|A({\bf r})\rangle = 1/2$. 

We note that an estimate of the localization length is sometimes given using a quantity called the Inverse Participation Ratio (IPR)
~\cite{Schwartz}. IPR is defined as
\begin{equation}
{\rm IPR}=\dfrac{\int~I^2(x,y;z)~dx dy}{\left(\int~I(x,y;z)~dx dy\right)^2},
\label{eq:IPR}
\end{equation}
where $I$ is the optical intensity defined in the $x-y$ transverse plane, at the longitudinal coordinate $z$.
 IPR has units of inverse area and the average effective width is defined as $\omega_{\rm eff}=\langle P\rangle^{-1/2}$,
where $\langle \cdots\rangle$ represents the statistical averaging over the ensemble. We note that IPR as defined by 
Eq.~\ref{eq:IPR} is originally rooted in the definition of the mode effective area in nonlinear optics (see e.g.~\cite{AgrawalBook}).
Equation~\ref{eq:xi} used by De~Raedt \textit {et al}.~\cite{DeRaedt} is likely a better representation of the localization 
phenomena and more true to the random-walk nature of the scattered wave. Therefore, we prefer to use the second moment method 
of Eq.~\ref{eq:xi}, rather than the Inverse Participation Ratio to calculate the localization length.
\subsection*{Small localized beam radii with low variations are desired}
The experimental realization of the RLV scheme by Karbasi \textit {et al}.~\cite{SalmanOL} showed that the transversely 
disordered optical fibers can be embraced as a completely new class of optical fibers that guide light, not in a 
conventional core-cladding setting, but by means of the Anderson localization, where any location across the transverse 
profile of the fiber can be used to guide light. As we will see later in image transport applications of this disordered
optical fiber, a small beam localization radius is likely its main desired attribute. A reduced variance is also desired,
because as a device it is not helpful to have variations in the beam radius that depend on the transverse location 
of the beam in a random and unpredictable way. In other words, although the underlying waveguiding mechanism is based on 
disorder and randomness, the main measurable device attribute, the localized beam radius, should be predictable.  
These considerations lead one to ask what can be done, in the design of these disordered fiber in order to minimize the localization
radius and also reduce the sample-to-sample variations in the localization radius. 
\subsection*{Design parameters upon which the beam radius depends}
The localized beam radius can depend on: the input characteristics of the optical beam such as the wavelength 
and the in-coupling beam diameter; the dimensionful fiber characteristics such as the transverse dimensions of 
the fiber, the width of each pixel; and the dimensionless fiber characteristics such as the refractive index of
the components, especially the index difference, and the geometrical ratio of the components used in the fiber
characterized by the fill-fraction.

Given the intuition we established based on the modal picture, we learned that if the initial excitation width 
is smaller than the radius of a typical mode of the disordered waveguide (localization radius or localization length) ,
the propagating light eventually localizes to a radius that is comparable to the localization length; and if the initial 
excitation is wider than the typical width of an individual mode, the propagating light eventually localizes to a slightly 
wider beam than the original size. Therefore, the design program to optimize the disordered waveguide for best localization 
must be independent of the input beam profile and must focus on the transverse dimensions of the localized beams~\cite{Ghosh1,Ghosh2,MarcoSharp}. 

Another important property of the disordered fiber is the transverse size of the fiber. If the fiber cross section is 
not large enough, the boundaries of the fiber strongly affect the localization radius of the modes. The impact of the 
boundary on the localization radius of the modes that reside near the boundary is inevitable, but in an ideal
case the fiber cross section must be large enough so that the modes in the interior region are shielded from the
boundary to prevent excessive scattering loss; therefore, the physical characteristics of the majority or 
nearly all of the interior modes are solely determined by the transverse Anderson localization mechanism.

In the ideal case, when the disordered fiber cross section is much larger than the light wavelength, feature
size (width of each index pixel), and the localization radius of the typical modes, the total transverse size of
the fiber becomes irrelevant. In that case, the only two dimensionful fiber design parameters are the light wavelength 
and the width of each pixel. Considering the fact that Maxwell's equations are scale invariant, if the wavelength and 
feature size (fluctuating index pixel size) are scaled by some factor, the localized beam radius of 
each mode is scaled by the same factor. 

\begin{figure}[htbp]
\centering\includegraphics[width=2.5in]{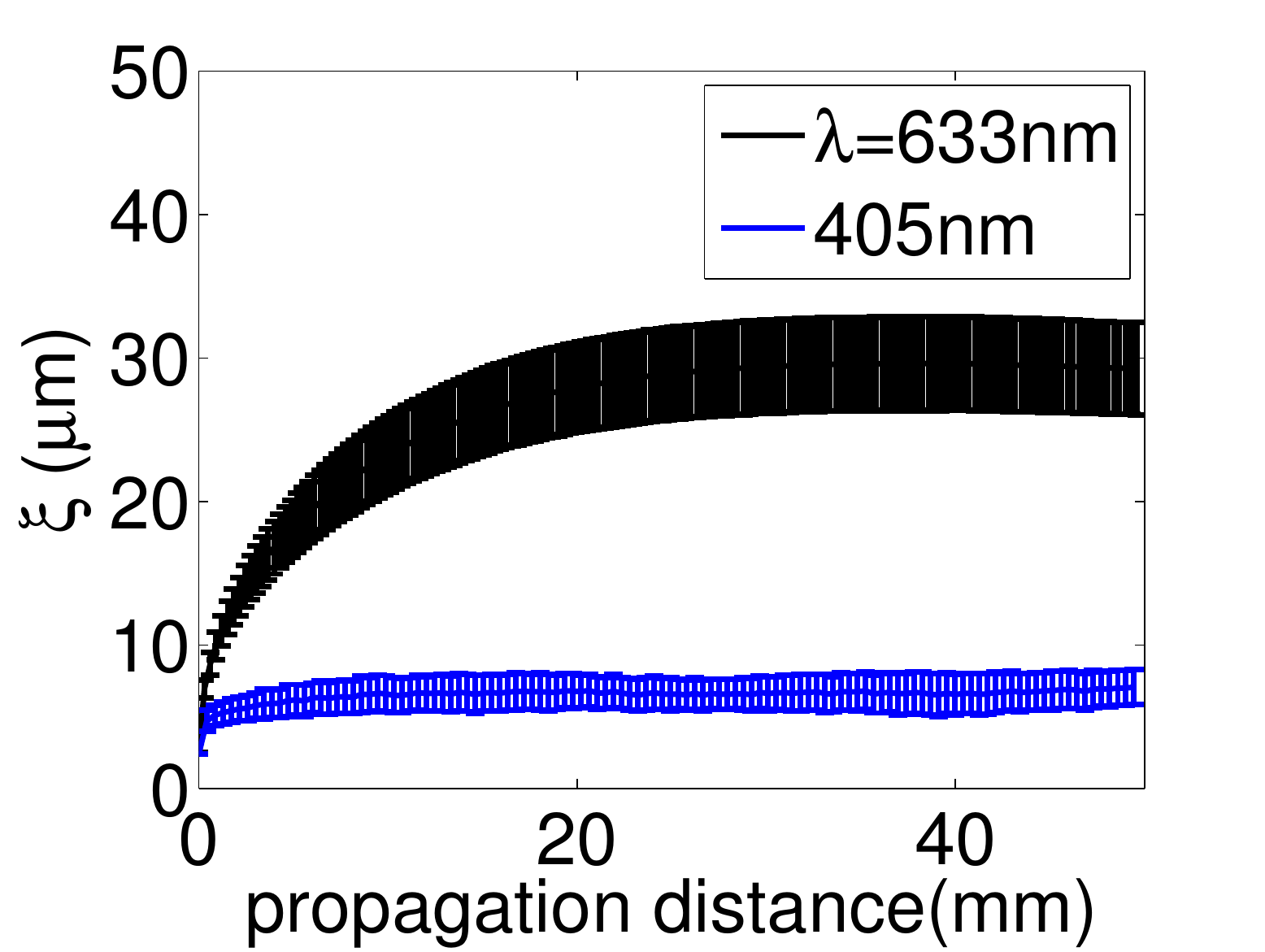}
  \caption{The localized beam radius is smaller at 405~nm wavelength compared with 633~nm wavelength, for the
           pALOF design with an approximate 0.9~\textmu m feature size. Adapted with permission, copyright 2012, Optical Society of America~\cite{SalmanOPEX}.}
    \label{fig:W633nmVS405}
\end{figure}

The feature size of the pALOF studied in Refs.~\cite{SalmanOL,SalmanOPEX} are approximately 0.9~\textmu m.
Experimental measurements and numerical simulations have confirmed that the localization radius is much smaller 
at 405~nm wavelength than at 633~nm wavelength, as shown in Figure~\ref{fig:W633nmVS405}. It is speculated that the optimum feature size for strong 
localization using the RLV scheme is around $2\lambda$. This statement is merely an educated guess and
the optimum value of the feature size relative to the wavelength has yet to be found. Intuitively, if the feature 
size is too small compared with the wavelength, the optical field would hop over the random sites and merely average
the refractive index, with very weak scattering. On the other hand, if the feature size is too large compared with 
the wavelength, the mean free path for wave scattering will be large, resulting in a large localization radius. 
Therefore, there must be an optimum value of the feature size for a given wavelength. Finding the optimum value is 
computationally very challenging.

\begin{figure}[htbp]
\centering\includegraphics[width=2.5in]{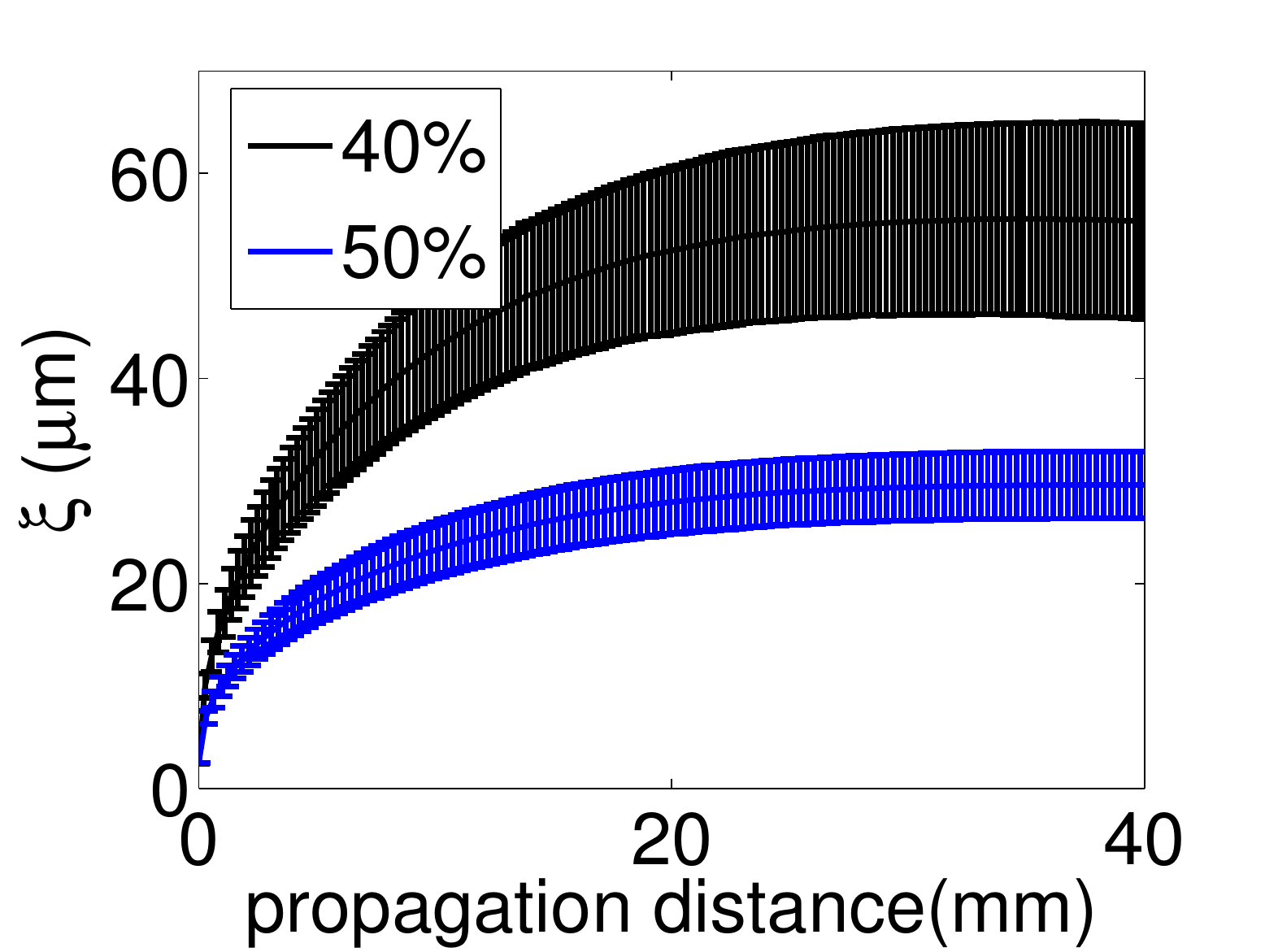}
  \caption{The evolution of the effective beam radius is plotted versus the propagation distance is shown 
for different values of the fill-fraction of $p = 40\%$, and $p = 50\%$, where the latter provides a lower effective 
beam radius and localization length. Adapted with permission, copyright 2012, Optical Society of America~\cite{SalmanOPEX}.}
    \label{fig:40p50p}
\end{figure}

Another design parameter that affects the localized beam radius in the binary disordered RLV scheme is the
portion of each random refractive index component, characterized by the fill-fraction $p$. In Ref.~\cite{SalmanOPEX},
$p$ is defined as the portion of the low-index material in the higher index host medium. Using an intuitively
plausible argument that the maximum transverse scattering is obtained when there is an equal amount of low-index 
and high-index material in the disordered matrix, and also a few instances of simulations relevant to the
experimental regime of interest, Karbasi \textit {et al}.~\cite{SalmanOPEX} have argued that $p=50\%$ should be
regarded as the ideal design target. A sample simulation is shown in Figure~\ref{fig:40p50p}, where the evolution of 
the effective beam radius versus the propagation distance is shown for different values of the fill-fraction of 
$p = 40\%$ and $p = 50\%$; and the latter provides a lower effective beam radius and localization length.

\begin{figure}[htbp]
\centering\includegraphics[width=2.5in]{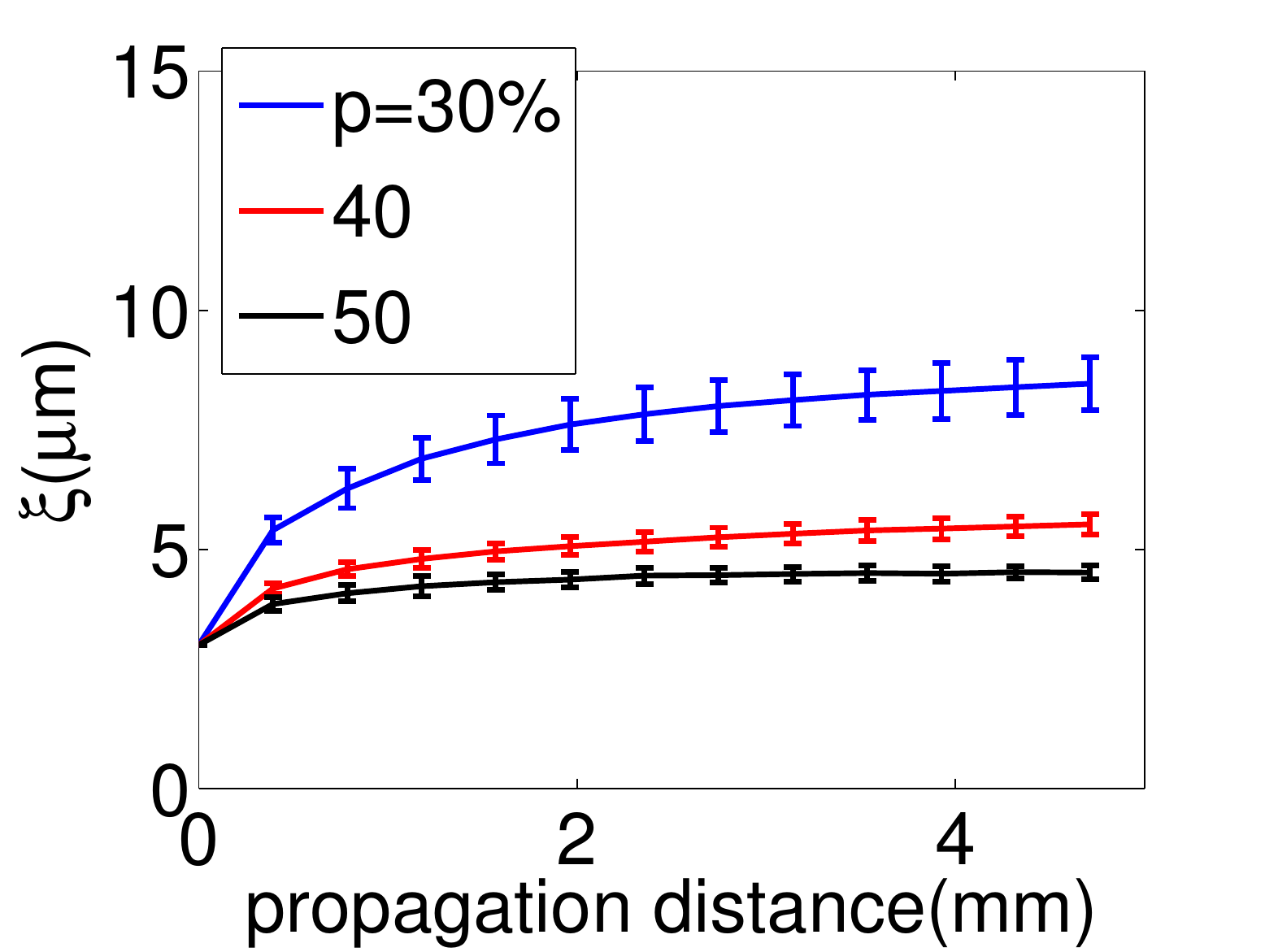}
  \caption{Effective beam radius vs propagation distance for different values of fill-fraction, $p$, in glass 
           disordered optical fibers with random air holes. The index difference of 0.5 between the random
           sites results in a very small localization radius. Adapted with permission, copyright 2012, Optical Society of America~\cite{SalmanOPEX}.}    
     \label{fig:FillFractionGlass}
\end{figure}

The last, but definitely not the least important design parameter to reduce the localization radius is the 
index difference between the random components of the disordered fiber. It is generally believed that 
a stronger localization (smaller average mode radius) is obtained if the refractive index difference between 
the random constituents is increased. This is confirmed in the 1D model of the transverse Anderson localization
of Ref.~\cite{SalmanModal}. However, it is also shown that the dependence of the average mode radius on the
index difference asymptotically saturates beyond a certain value of the index difference; therefore, the 
payback in the reduction of the localization length may be quite small beyond a certain threshold index difference.
While these issues are well studied in the 1D model of Ref.~~\cite{SalmanModal}, the 2D analysis remains to be
done. For the 2D geometry, Karbasi \textit {et al}.~\cite{SalmanOPEX} compared the localization radius in a
glass-air disordered fiber with pALOF. The refractive index of sites in the glass-air disordered fiber 
are randomly picked as $n_2 = 1.5$ and $n_1 = 1.0$ for the glass host and random air-hole sites, respectively.
The index difference of 0.5 in the glass-air disordered fiber compared with 0.1 in pALOF results in a considerable 
reduction in the localization radius. In Figure~\ref{fig:FillFractionGlass},
the calculated beam radius versus propagation distance for different values
of fill-fraction are plotted, and thelocalized beam radius is clearly smaller than that provided by the index difference
of 0.1 for pALOF.

Finally, we have repeatedly observed that whenever the scattering strength is increased to reduce the localization 
length (localized beam radius), whether by selecting the right wavelength, or by bringing the fill-factor 
close to 50\%, or by increasing the index difference, the statistical variation of the beam radius
decreases. Therefore, we have the luxury of solving two problems with one solution (a more animal friendly expression
than {\em killing two birds with one stone!}). 
\subsection*{Spatial beam multiplexing}
\label{sec:bm}
The possibility of spatial beam multiplexing in pALOF was studied in Ref.~\cite{SalmanMB} in 2013.
It was shown, both numerically and experimentally, that a pALOF can be used to simultaneously 
transport multiple beams~\cite{Fini}. 
\begin{figure}[htbp]
\centering\includegraphics[width=4.5in]{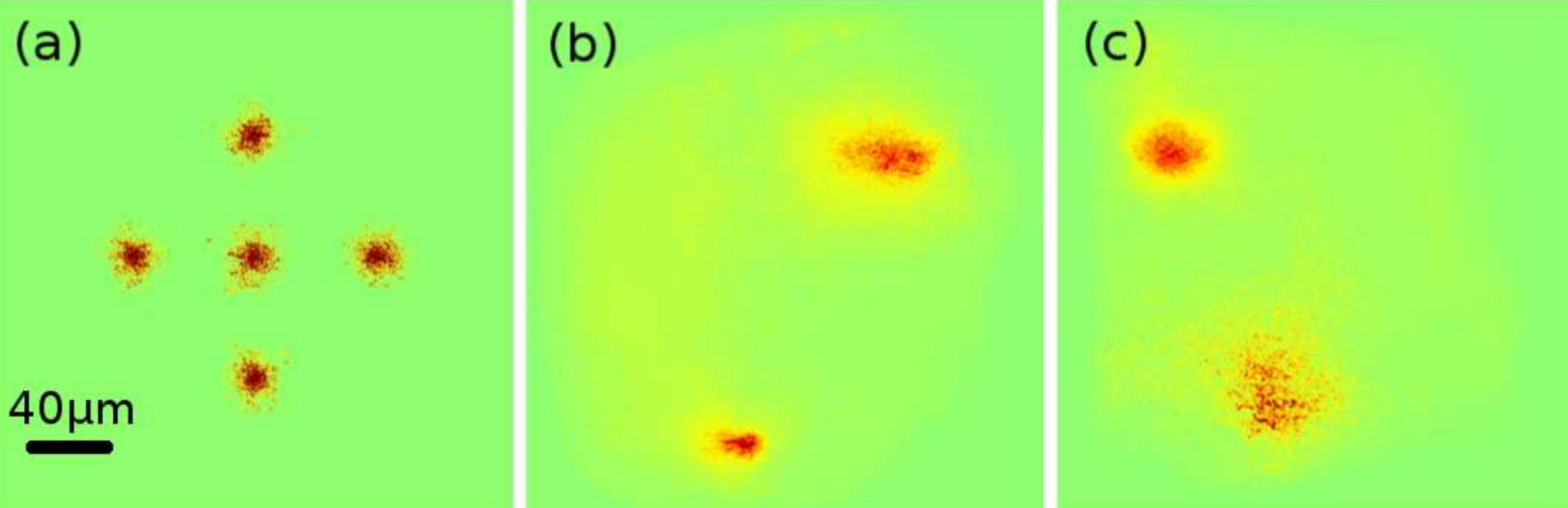}
  \caption{Multiple-beam propagation in a 5~cm-long pALOF 
           {\claop (a)} simulation for five beams;
           {\claop (b)} experiment for two beams; and 
           {\claop (c)} experiment for two beams with different wavelengths. 
           All beams are at 405~nm wavelength, except the bottom-middle beam in subfigure {\claop (c)}, which is at 633nm wavelength.
     Adapted with permission, copyright 2013, Optical Society of America~\cite{SalmanMB}.}    
     \label{fig:beam-multiplex}
\end{figure} 
In Figure~\ref{fig:beam-multiplex}{\claop (a)} beam multiplexing in pALOF is studied numerically at 405~nm wavelength,
where the intensity profile of a multiplexed beam is plotted after 5~cm of propagation along the fiber; the four 
exterior beams are launched at a distance of 70~\textmu m from the central beam.
The output beams do not show any appreciable drift and remain in the same transverse location across the fiber.

Experimental verification of beam multiplexing are shown in Figure~\ref{fig:beam-multiplex}{\claop (b)}, 
where both beams are 405~nm wavelength, and in Figure~\ref{fig:beam-multiplex}{\claop (c)}, 
where the upper beam is at 405~nm wavelength and the lower beam is at 633~nm wavelength. The localization 
is clearly stronger at 405~nm due to the choice of the design parameters of pALOF, as discussed before.
\subsection*{Macro-bending loss}
In section~\ref{sec:bm} the possibility of beam multiplexing in an disordered fiber was discussed. Given the
unconventional nature of the beam confinement in a disordered optical fiber, one must worry about the 
possibility that the spatially multiplexed beams drift across the fiber when the
fiber is subjected to substantial macro-bending. This issue was studied in detail in Ref.~\cite{SalmanMB}.
Using numerical simulations, it was shown that transverse Anderson localization is very robust and can
withstand, at least in theory, a degree of macro-bending that is beyond what is acceptable for conventional
fibers. To explore the macro-bending experimentally, a 10~cm section of a 15~cm-long pALOF was wrapped 16 times 
around a mandrel with an approximate radius of 1mm, where no appreciable loss of walk-off effect was observed in the 
localized beam.
\section{Image transport through the disordered fiber}
\label{sec:imaging}
\begin{figure}[htp]
\centering
        \includegraphics[height=2.in]{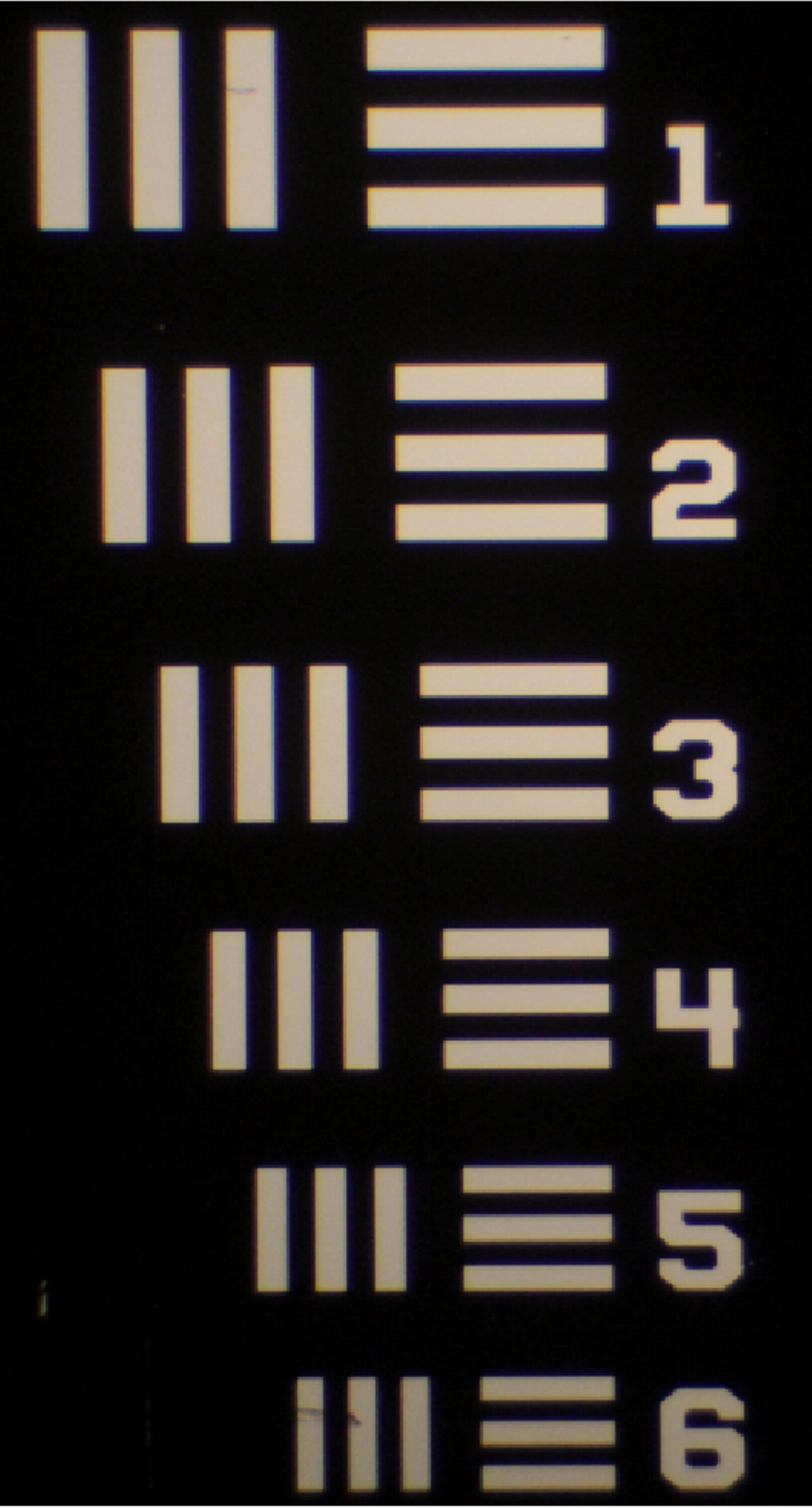}
	\includegraphics[height=2.in]{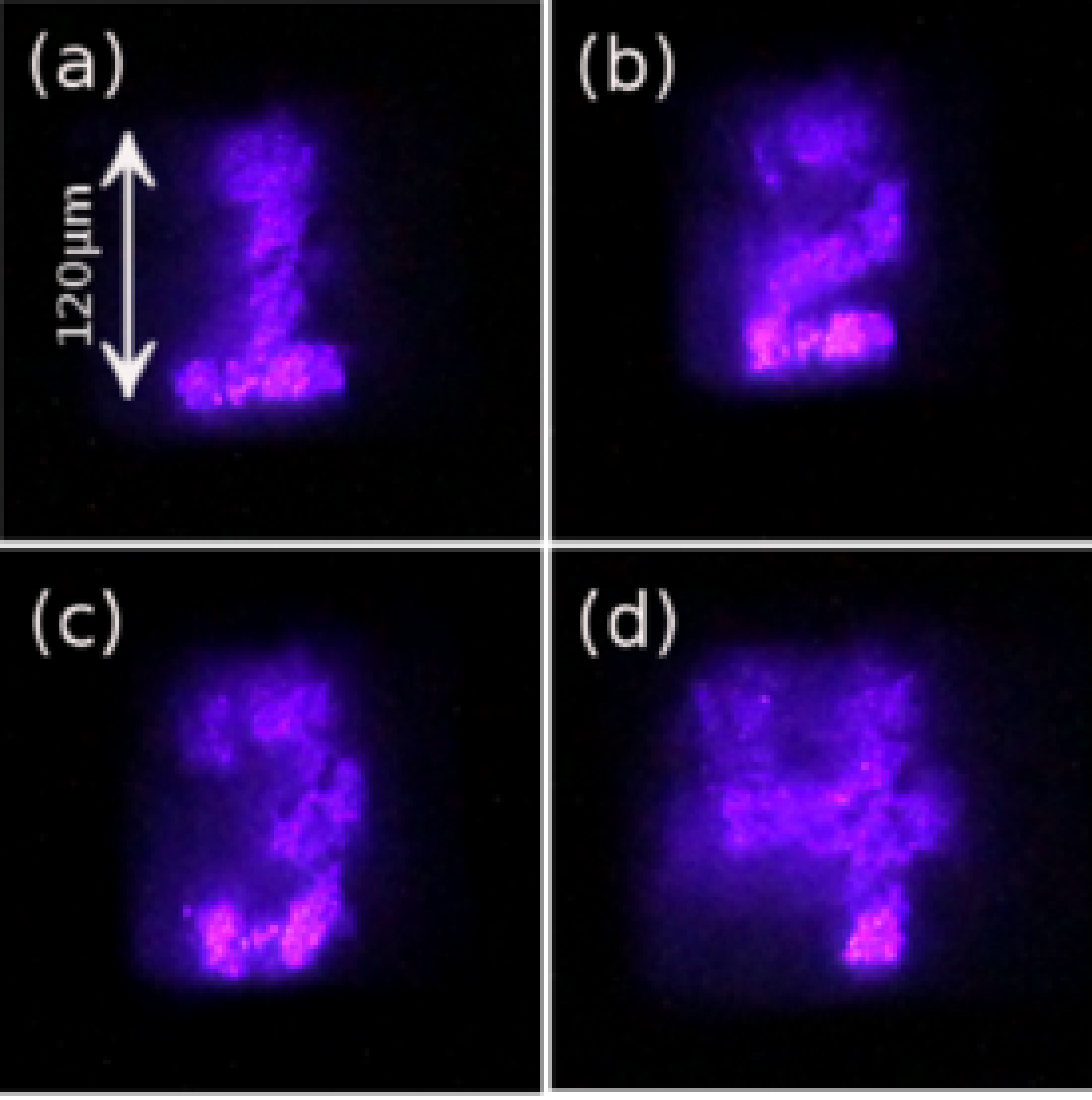}
	\includegraphics[height=2.in]{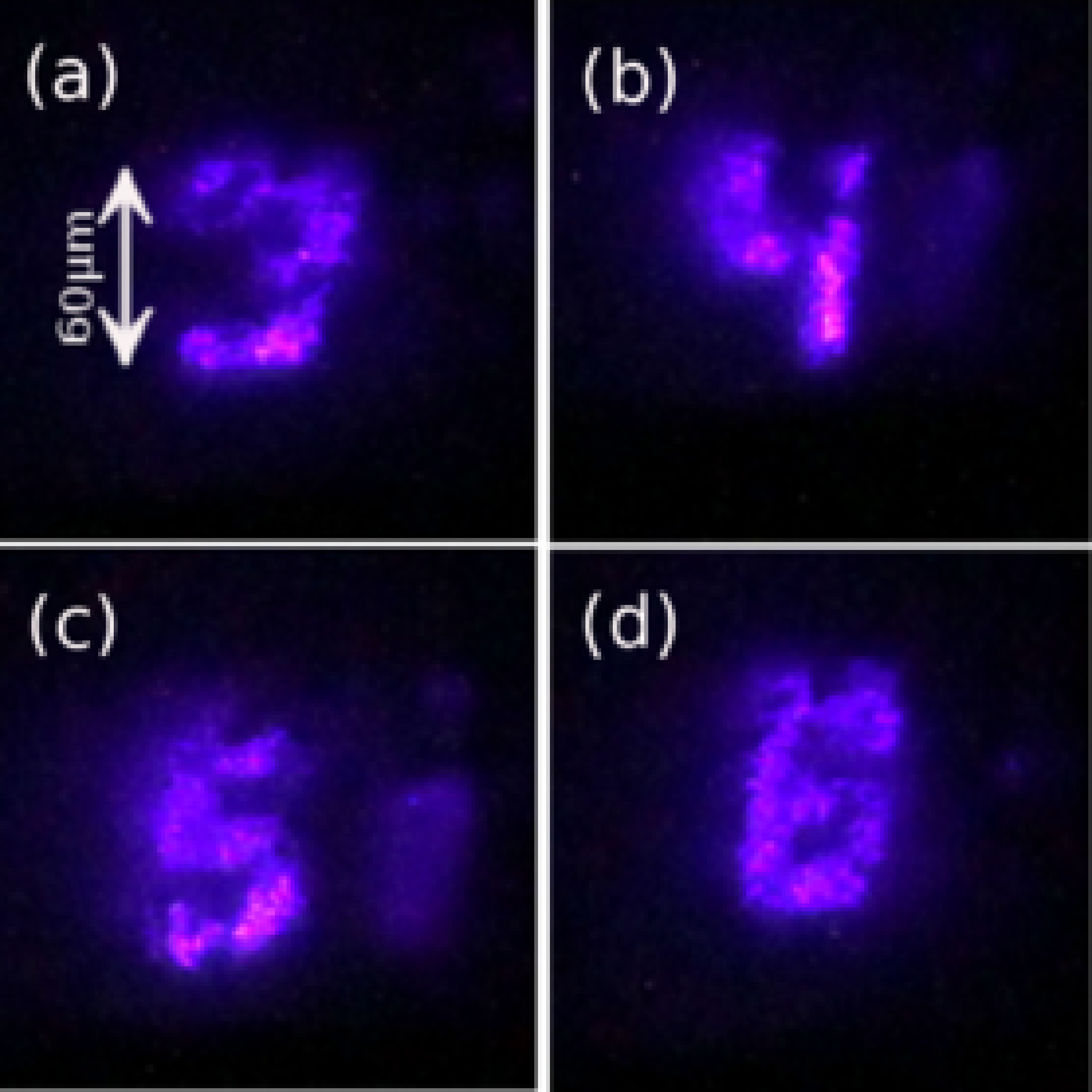}
    \caption{\small\em(left) Elements of a group on 1951 U.S. Air Force test target (1951-AFTT). Transported images of different numbers 
             through a disordered optical fiber: (middle) subfigures {\claop (a)}-{\claop (d)} are related to the group 3 on the test target and
                                               (right) subfigures {\claop (a)}-{\claop (d)} are related to the group 5 on the test target.
    Adapted with permission, copyright 2014, Nature Communications~\cite{SalmanNature}.}
	\label{fig:Experimental405}
\end{figure}
Motivated by the successful demonstration of beam multiplexing, Karbasi \textit {et al}. 
used pALOF for endoscopic fiber-optic imaging. To their pleasant surprise, the image transport quality was 
comparable to or better than some of the best commercially available multicore imaging fibers, with less 
pixelation and higher contrast~\cite{SalmanNature}. Figure~\ref{fig:Experimental405} shows some of the transported images in the 
form of numbers from a section of the 1951 U.S. Air Force resolution test chart through pALOF. The test-target, in the form of a 
stencil in which numbers and lines were carved, was butt-coupled to the hand-polished input facet of pALOF and was illuminated 
by white light. The near-field output was projected onto a CCD camera with a 40$\times$ microscope objective. 

The minimum resolution of the images is determined by the width of the point spread function of the disordered optical 
fiber imaging, which was calculated to be smaller than 10~\textmu m at 405~nm wavelength~\cite{SalmanOPEX}.
In practice, the imaging resolution in pALOF is limited by the quality of the cleave and polishing of the fiber. 
The fiber surface quality is partially responsible for the distortions 
in the transported images in Figure~\ref{fig:Experimental405}. The high quality image transport in the 
proof-of-concept experiment has been achieved without any optimization in the design.

Highly multicore optical fibers, similar to those introduced in section~\ref{sec:AAmechanism},
have been used for direct transportation of images in various 
configurations~\cite{Hopkins,Han,ChrisXu,ChrisXu2}.
The highly multicore fiber, often referred to as the ``coherent fiber bundle,'' is commonly used in 
medical and industrial endoscopy~\cite{Hopkins,Fujikura,Schott}.

For an Anderson localized fiber, a higher amount of disorder and a larger level of 
fluctuation in the refractive index 
provides stronger beam localization, resulting in an improved image resolution. A similar statement can be said for 
multicore imaging fiber as well: the coherent core-to-core coupling is detrimental and blurs the image; therefore,
core-to-core coupling must be suppressed by varying the size of the cores so that neighboring cores cannot 
couple resonantly. Also, in a multicore imaging fiber, the cores must be as close as possible to create a less
pixelated image, while being very different in size so that they cannot couple efficiently even when they are close. 
One can view the disordered Anderson localized fiber as taking the two limits of the high-packing of the cells and the large variation 
of the sizes to an extreme, so that the individuality of each core is completely lost: {\em all neighboring sites are strongly
coupled, but the extreme randomness prevents light leakage and blurring.}  

\begin{figure}[htp]
  \begin{center}
    \includegraphics[width=85mm]{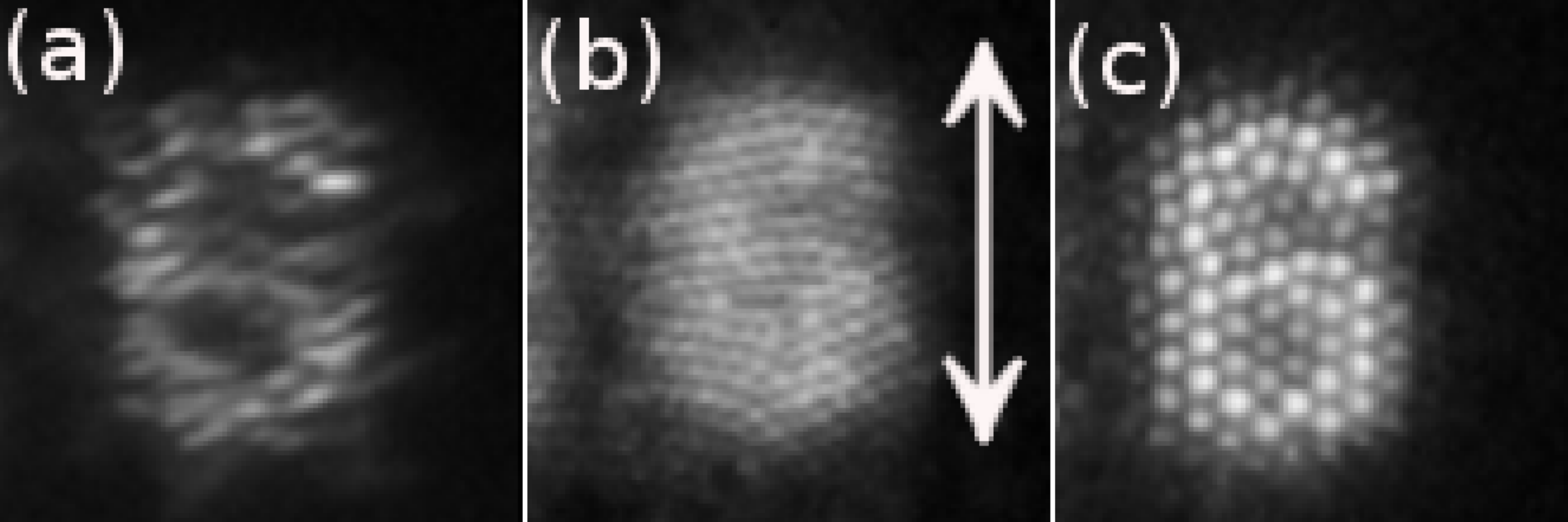}
  \end{center}
    \caption{Transported images through the disordered fiber and the commercial image fibers. Images related to group 5 of the 1951-AFTT test chart 
             in (a) pALOF, (b) FIGH-10-350S image fiber and (c) FIGH-10-500N image fiber (experimental measurements).
             The scale bar in (b) is 30~$\mu$m-long and the same scale bar can be used for (a) and (c). Each fiber is approximately 5~cm long.
    Adapted with permission, copyright 2014, Nature Communications~\cite{SalmanNature}.}
    \label{fig:Fig-5}
\end{figure} 
The imaging performance of the ``unoptimized'' pALOF compares with some of the best commercially available multicore imaging
optical fibers, as is further confirmed in Figure~\ref{fig:Fig-5}. The transported images over 5~cm of the number ``6'' from
group 5 of the 1951-AFTT test chart are compared between pALOF in Figure~\ref{fig:Fig-5}{\claop (a)},
Fujikura FIGH-10-350S in Figure~\ref{fig:Fig-5}{\claop (b)}, and Fujikura FIGH-10-500N in Figure~\ref{fig:Fig-5}{\claop (c)}.
The image quality of the transported image through the pALOF is clearly better than FIGH-10-350S and is comparable with FIGH-10-500N.
The feature sizes in Figure~\ref{fig:Fig-5} are on the order of 10-20~\textmu m. The
Rayleigh range for this level of resolution is approximately 0.8-3~mm, which is substantially
shorter than the typical propagation length in these imaging fibers. Therefore, the imaging results are non-trivial and cannot be
obtained using bulk propagation or conventional multimode fibers.
\begin{highlight}
\textbf{Highlights:}
\begin{itemize}
\item In an Anderson localized disordered optical fiber, a smaller localized beam radius is obtained via increasing the differences between the refractive
indexes of the random dielectric constituents.
\item The feature size should be $\sim 2\lambda$, and a fill-fraction of 50\% is preferred.
\item A stronger localization is generally accompanied by a smaller variation in the localized beam radius, resulting in 
a more uniform and predictable beam radius.
\item Beam multiplexing and high-quality image transport are featured as device-level applications of the transverse Anderson localization in a disordered optical fiber. 
\end{itemize}
\end{highlight}
\section{Transverse Anderson localization in a disordered silica optical fiber}
The first observation of Anderson localization in a silica fiber was reported in Ref.~\cite{SalmanOMEX}. 
The main motivation for using a glass-air structure has been the larger index contrast that results in a 
smaller beam diameter (better image transport resolution), as well as a lower sample-to-sample variation in 
the value of the beam diameter (better image uniformity)~\cite{SalmanOMEX}.
The reported glass-air disordered fiber was drawn at Clemson University. The preform was made from 
``satin quartz'' (Heraeus Quartz), which is a porous artisan glass. By drawing the preform, the 
airholes (bubbles) in the glass are stretched to form the hollow air-rods required for transverse 
Anderson localization. The large draw ratio sufficiently preserves the longitudinal invariance, without 
significant disturbance over typical lengths used in the experiments.  
 
\begin{figure}[htp]
\centering\includegraphics[width=4in]{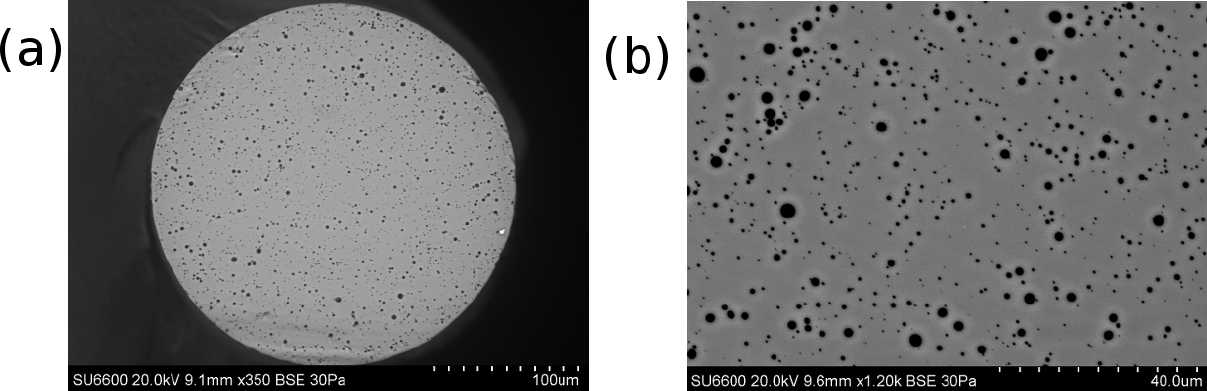}
  \caption{{\claop (a)} SEM image of the glass optical fiber with random airholes reported in 
               Ref.~\cite{SalmanOMEX}; and {\claop (b)} zoomed-in SEM image of the same fiber.}
    \label{fig:glass-profile}
\end{figure} 
The cross-sectional SEM image of the disordered glass-air optical fiber is shown in Figure~\ref{fig:glass-profile}{a},
and a zoomed-in SEM image is shown in Figure~\ref{fig:glass-profile}{b}. The SEM images provide
a good estimate of the refractive index profile of the fiber; the light gray
background matrix is glass and the black dots represent the random airholes. The diameter of the disordered glass
fiber is measured to be 250~\textmu m. The diameters of the airholes vary between 0.2~\textmu m and 5.5 ~\textmu m.
Unfortunately, the airhole fill-fraction was shown to be as low as 2\% in the central regions, which was far below the ideal
value of 50\%, so strong localization was not observed near the center. However, Anderson localization was observed near
the boundary, where the airhole fill-fraction is 8\%.

More recently, transverse Anderson localization has been reported by scientists from Corning Incorporated in random airline 
fibers (RALF)~\cite{MingJun}. The preform for RALF is fabricated using the outside vapor deposition (OVD) process.
The consolidation process is done in the presence of 100\% nitrogen gass, during which nitrogen is trapped in the blank
to form glass with randomly distributed air bubbles. When the preform is drawn, the random air bubbles stretch to form
random airlines. RALFs with 150, 250 and 350 \textmu m diameters were fabricated, where the averaged air line 
diameters were 177, 247 and 387 nm in these fibers, respectively. The maximum airhole fill-fractions measured in RALFs
is reported to be 1.473\%, which is considerably lower than that of Ref.~\cite{SalmanOMEX}. Ref.~\cite{MingJun} reports 
the observation of transverse Anderson localization of light, despite the low airhole fill-fraction, and attributes
its observation to the substantially smaller diameter of the airholes compared with that reported in Ref.~\cite{SalmanOMEX}.

We also encourage the interested reader to consult a pioneering work by Pertsch \textit {et al}~\cite{Pertsch}
on light propagation in a disordered 2D array of mutually coupled optical fibers. They observed both
localized and delocalized modes and also analyzed the impact of nonlinearity on these modes.

\begin{highlight}
\textbf{Highlights:}
\begin{itemize}
\item Glass-air random optical fibers are highly desired because the large refractive index difference between glass and air can result
in narrowly localized beams, and also small variations around the mean beam radius.
\item All attempts at glass-air random fibers have so far resulted in undesirably small air fill-fractions.  
\end{itemize}
\end{highlight}
\section{Hyper-transport in longitudinally varying disordered waveguides}
As mentioned in section~\ref{Sec:numerics}, the paraxial propagation of an optical beam 
in a longitudinally invariant medium is described by the paraxial approximation to the Helmholtz 
equation, Eq.~\ref{eq:bpm}. This equation is formally equivalent to the Schr\"{o}dinger equation 
in quantum mechanics, where the time variable $t$ is replaced with the longitudinal coordinate 
$z$ and the potential is replaced by the term proportional to the refractive index profile
$n(x,y)^2-n_0^2$. The longitudinal invariance of the refractive index profile is taken into account
by explicitly showing that $n$ is only a function of the transverse coordinates $x$ and $y$.
The requirement for the longitudinal invariance of the refractive index profile is equivalent to
the potential being constant in time in the Schr\"{o}dinger equation. 

Presence of some form of time dependence in the random potential can {\em potentially} change the 
localization picture. This issue has been studied extensively over the years e.g., in 
Refs.~\cite{Zaslavskii,Jayannavar,Golubovic,Rosenbluth,Arvedson} for temporally random
or correlated potentials. An interesting finding has been the possibility of hyper-transport
in certain temporally varying random potentials, where the wavefunction expands even more 
rapidly than the ballistic expansion observed in free space. We recall our earlier discussions 
on ballistic, diffusive, and localized propagation of light in disordered optical waveguides. 
For the propagation of the optical beam, the beam width generally grows with the propagation 
distance according to $w(z)\propto z^p$, where $p=1$ for ballistic, $p=1/2$ for diffusive, 
and $p=0$ for localized propagation. An example of the ballistic expansion is the familiar 
formula for the width of a Gaussian beam as a function of the propagation distance~\cite{SalehTeich} 
\begin{equation}
w(z)=w_0\sqrt{1+\left(\dfrac{z}{z_0}\right)^2}, \qquad z_0=\dfrac{\pi w_0^2}{\lambda},
\end{equation}
where $w(z)\propto z$ for $z\gg z_0$. 

For a temporally random potential, Ref.~\cite{Jayannavar} showed analytically that the width
of the wavefunction, as measured by the second-moment method, expands with time according
to $t^{3/2}$. Hyper-expansion was later shown for correlated temporally fluctuating 
potentials as well in Refs.~\cite{Golubovic,Rosenbluth,Arvedson}. 

Hyper-transport for an optical beam was studied experimentally and theoretically in 
Ref.~\cite{Levi} and Ref.~\cite{Krivolapov}, respectively, where the temporal variations 
were replaced by rapid longitudinal 
fluctuations in the refractive index profile of the waveguide. The experiment reported in
Ref.~\cite{Levi} was performed in a similar setting to that of the first observation of
the AA mechanism earlier reported in section~\ref{sec:SegevAA} and Ref.~\cite{Schwartz}.
The disordered lattice was formed by means of the optical induction technique~\cite{Efremidis}
in a photo-refractive crystal. However, unlike Ref.~\cite{Schwartz}, where special care was taken 
to ensure the longitudinal invariance of the interference speckles and the resulting transversely 
random index profile, the refractive index profile created in the nonlinear crystal in
Ref.~\cite{Levi} was made to vary with $z$. The experiment proved that in the presence of 
a sufficiently large longitudinal variation of the index fluctuation, not only did the 
transverse localization cease to exist, but also the beam expanded at a rate faster than 
ballistic as it propagated through the random-index waveguide.
\begin{highlight}
\textbf{Highlights:}
\begin{itemize}
\item In the presence of rapid longitudinal fluctuations in the refractive index profile of 
a disordered waveguide, it is possible for the optical beam to expand even more 
rapidly than the ballistic expansion observed in free space.
\end{itemize}
\end{highlight}
\section{Anderson localization and nonlinearity}
The interplay between nonlinearity and disorder is of great interest to the studies of
transverse Anderson localization. The main question that is often asked is whether the 
presence of nonlinearity perseveres, enhances, undermines, or destroys Anderson localization.  
This issue has been explored over the years, where some of the studies have benefited from
the existing literature on systems with similar dynamical equations, such as the 
Bose-Einstein condensate in the presence of disorder (see e.g. Ref.~\cite{Pikovsky}
and the references therein). Here, we briefly highlight some of the results that are more 
relevant to the concept of transverse Anderson localization. For more details, we refer
the interested reader to an excellent review on this subject by Fishman, Krivolapov, and 
Soffer, in Ref.~\cite{Fishman1}.

Earlier, in section~\ref{sec:SegevAA}, we reviewed the numerical and experimental work
of Schwartz \textit {et al}.~\cite{Schwartz} that resulted in the observation of transverse 
Anderson localization of light for the AA mechanism. The authors also investigated the
transverse Anderson localization of light in the presence of Kerr nonlinearity, both 
numerically and experimentally. The defining equation for the nonlinear propagation of 
light is the nonlinear Schr\"{o}dinger equation (NLSE) and is identical to Eq.~\ref{eq:bpmNL}, 
with the addition of a Kerr nonlineaity term~\cite{AgrawalBook}: 
\begin{equation}
\label{eq:bpmNL}
i\dfrac{\partial A}{\partial z}+
\dfrac{1}{2n_0k_0}\left[\nabla^2_T A+k_0^2\left(n^2-n^2_0\right)A\right]+k_0 n_2 |A|^2 A=0,
\end{equation}
where $n_2$ is the nonlinear index, which is positive for self-focusing and negative for self-defocusing
nonlinearity.

As a case study, the authors considered a disordered lattice where the maximum contribution of the nonlinear 
term to the index change ``${\rm max}(|n_2|\times|A|^2)$'' was assumed to be a maximum of 15\% of the index contrast of the underlying 
periodic waveguide. They also varied the disorder level from 0\% to 30\%, where the disorder level was
defined as the magnitude of random index fluctuations relative to the index contrast of the underlying 
periodic waveguide. They observed that over this range, the self-defocusing nonlinearity ($n_2$) results in a 
moderate (nearly negligible) widening of the average beam profile. However, the self-focusing nonlinearity ($n_2>0$)
resulted in a substantial reduction of the average localized beam diameter. The enhancement of localization due
to the self-focusing nonlinearity was particularly noticeable when the disorder level was less than 15\%.

The experiments were carried out at 15\% disorder level. However, the maximum nonlinear contribution was taken to be
equal or higher (up to a factor of 3), compared with the index contrast of the underlying periodic waveguide.
This was achieved by making the probe intensity equal to or higher than the interference maxima of the 
lattice-writing beams (for a perfect lattice). The statistical analysis of the localized beam radius
clearly confirmed the expected reduction in the average beam radius due to the self-focusing nonlinearity.

Similar results were reported by Lahini \textit {et al}.~\cite{Lahini1} using disordered one-dimensional 
waveguide lattices. Their experiment consisted of a one-dimensional lattice of coupled optical waveguides patterned 
on an AlGaAs substrate. Light was injected into one or a few waveguides at the input, and light intensity distribution 
was measured at the output. We recall our earlier discussions in section~\ref{Sec:modeshapes}: we identified 
the highly localized eigenmodes near the top edge of the propagation constant band. The amplitude of these flat-phased 
modes, as they are referred to in Ref.~\cite{Lahini1}, are in-phase at all sites; perhaps only with a few flips 
as seen in Figures~\ref{fig:1D-mode-profiles}{\claop (a)} and {\claop (b)}. Another set of highly localized modes
were shown to exist near the top edge of the propagation constant band. The amplitude of these staggered 
modes, as they are referred to in Ref.~\cite{Lahini1}, have phase flips between adjacent sites, as 
shown in Figures~\ref{fig:1D-mode-profiles}{\claop (e)} and {\claop (f)}. 

In the weak nonlinear regime, Lahini \textit {et al} observed that nonlinearity enhances localization in
flat-phased modes and induces delocalization in the staggered modes. This behavior is explained as
follows: the presence of the weak nonlinearity perturbatively shifts (increases) the value of the 
propagation constant of each localized mode. For the flat-phased modes, the nonlinearity shifts the modes
outside the original linear spectrum. However, for the staggered, which belong to the bottom edge of the 
propagation constant band, a perturbative increase in the value of the propagation constant shifts 
it further inside the original linear spectrum. Therefore, the propagation constant of a staggered mode can 
cross and resonantly couple with other modes of the lattice, resulting in 
delocalization~\cite{Kopidakis1,Kopidakis2,Kopidakis3,Albanese}. 

In 2008, Pikovsky and Shepelyansky presented a somewhat different account of the interaction between
disorder and nonlinearity~\cite{Pikovsky}. Here we rephrase their main findings in a language more
consistent with our notation so far. In a disordered coupled waveguide lattice, they demonstrated that 
above a certain critical strength of nonlinearity the Anderson localization is destroyed and turns into
a subdiffusive spreading. They focused on the discrete Anderson nonlinear
Schr\"{o}dinger equation, which is essentially the same as Eq.~\ref{eq:coupledmode1} with the
addition of a third-order diagonal Kerr nonlinear term, expressed as 
\begin{align}
\Big(i\dfrac{\partial}{\partial z}+\beta_j\Big) A_j(z) + c_0 \Big[ A_{j+1}(z) + A_{j-1}(z) \Big] 
+ \gamma |A_j(z)|^2 A_j(z)= 0,\quad j=1,\cdots,N.
\label{eq:coupledmodeNL}
\end{align} 
They also assumed that the waveguide coupling coefficients $c_0$ is deterministic and is
identical for all waveguides. The disorder is introduced through the diagonal propagation constant terms, where 
they are assumed to be randomly distributed according to
\begin{align}
\beta_j\in{\rm unif}[\beta_0-\dfrac{\mathcal{B}}{2},\beta_0+\dfrac{\mathcal{B}}{2}].
\end{align} 
For the linear case of $\gamma=0$, the modes are exponentially localized due to the disorder.
In the nonlinear case where $\gamma\neq 0$, Pikovsky and Shepelyansky demonstrated that above 
a certain critical strength of nonlinearity, the Anderson localization is destroyed and 
the field spreads in a subdiffusive form indefinitely across the optical lattice. They also 
showed that the mode width calculated using the second moment method (see Eq.~\ref{eq:xi})
grows with the propagation distance as $z^\alpha$, where $0.3<\alpha<0.4$.

Pikovsky and Shepelyansky presented a theoretical argument in support of the
subdiffusive spreading of the beam with $\alpha=0.4$~\cite{Shepelyansky}; however,
their work is mainly based on numerical integration
of Eq.~\ref{eq:coupledmodeNL} and monitoring the results, up to $z=10^8/c_0$.
For the numerical simulation they used the boundary condition $A_j(z=0)=\delta_{j,{j_m}}$,
where $j_m$ represents the middle waveguide, and the integration is performed by the
operator splitting method. They also assumed that $\sum^{N}_{j=1} |A_j(z)|^2=1$,
without any loss of generality. In a sample set of simulations they chose the nonlinearity strength to be 
$\gamma=c_0$ for two cases: case 1 with $\mathcal{B}=2c_0$; and case 2 with $\mathcal{B}=4c_0$. The second moment 
was evaluated according to 
\begin{align}
\sigma(z)=\sum^{N}_{j=1} \Big(j-\langle j \rangle\Big)^2~|A_j(z)|^2, \qquad \langle j \rangle=\sum^{N}_{j=1} j\ |A_j(z)|^2.
\end{align} 
As expected, the initial expansion was ballistic for either case, but after some distance $z_0$, the expansion 
became subdiffusive. They fit the subdiffusive expansion to $\sigma(z)=\sigma_0 z^\alpha$ over the 
range $z_0<z<10^8/c_0$. In case 1, for different instances of randomness, they obtained
$0.32\le \alpha \le 0.39$; and for case 2 they reported $0.28\le \alpha \le 0.41$. Upon averaging 
over 8 independent realizations, they reported a fit of the form $57.5\times z^{0.344}$ for case 1 and 
$8.7\times z^{0.306}$ for case 2 over the subdiffusive range. They also reported a critical value of nonlinearity
$\gamma_c\approx 0.1 c_0$ above which this subdiffusive behavior is observed. 

Intuitively speaking, one may think that in a nonlinear disordered coupled waveguide system, the dynamics of the
beam is initially influenced by nonlinearity; and as the beam spreads, the effect of nonlinearity becomes 
weaker and the disorder dynamics takes over. Therefore, one should always expect Anderson localization after 
sufficiently long propagation. This is clearly in contrast with the findings of Pikovsky and Shepelyansky reported
above. Fishman \textit {et al}.~\cite{Fishman1} present a thorough survey of the many subtleties involved rearding
the interaction of nonlinearity and disorder. The conclusion is that the situation can best be described as inconclusive
at this point. For example, when using the numerical simulations, they caution that Eq.~\ref{eq:coupledmodeNL}
is chaotic with an exponential sensitivity to numerical errors. For long-distance propagation, it is not clear that 
reducing the $z$ step size can control the cumulative numerical error, given that the limit of zero $z$ step
may be singular. Details are beyond our intended scope and can best be found in Ref.~\cite{Fishman1}.

Other forms on nonlinearity besides Kerr can also interact with the disorder-induced localization. For example,
it was recently shown that a beam of light propagating in a pALOF (introduced in section~\ref{sec:MafiRLV})
exhibits self-focusing properties due to a  thermal nonlinearity~\cite{MarcoPRL}. The larger light absorption 
strength in PMMA than PS results in a inhomogeneous temperature distribution. The higher temperature in PMMA
translates into a decrease of its refractive index. The result is an increased refractive index mismatch
and stronger localization. The results are quite counter-intuitive, because the polymer materials used in the 
experiment have defocusing intrinsic nonlinear coefficients ($n_2<0$). In Ref,~\cite{MarcoPRL},
Leonetti \textit {et al}. demonstrated that transversally localized modes shrink when the pump intensity
is increased despite the fact that $n_2<0$ for the polymers. In a subsequent publication~\cite{MarcoAPL}, 
the authors provided further evidence of this behavior by analyzing the direct relation between the optical 
intensity and the localization length, and also demonstrated the disorder-induced focusing by a monochromatic
continuous wave (CW) laser.

The interested reader is also urged to study other aspects of the interaction between nonlinearity and disorder
not covered in this tutorial review e.g., on soliton propagation in random media~\cite{Tsoy,Garnier}. 
\begin{highlight}
\textbf{Highlights:}
\begin{itemize}
\item The main question of interest is whether the presence of nonlinearity perseveres, enhances, undermines, or destroys Anderson
localization. 
\item The answer to the above question can best be described as inconclusive at this point. 
\end{itemize}
\end{highlight}
\section{Coherence, classical and quantum light, and Anderson co-localization}
In the previous sections Anderson localization was explored for a temporally and spatially coherent light,
coupled to a disordered optical lattice. In this section some of the main issues related to the propagation
of partially incoherent light are explored. A discussion on the propagation of single photons and correlated
photons is also presented. 
  
Anderson localization of waves with imperfect coherence was reported in 2011 by 
\v{C}apeta, \textit {et al}~\cite{Capeta}. The main issue that was answered in
their paper was the extent to which transverse Anderson localization is affected
by the partial coherence of the in-coupling beam in a disordered linear lattice.
They observed that if all the eigenmodes of the disordered waveguide are exponentially 
localized, any partially incoherent beam exhibits localization with exponentially decaying 
tails, after sufficiently long propagation distances. The reported observation conforms with 
intuition, because an incoherent wave can be thought of as a superposition of coherent 
modes with stochastically varying coefficients. Because each coherent mode is expected to 
undergo localization, the entire beam should localize as well. However, localization is delayed 
by incoherence compared with the case of a coherent in-coupling beam: the more incoherent 
the wave is, the longer it diffusively spreads while propagating in the medium.

For an optical beam propagating in a 1D disordered optical lattice described by the propagation 
Eqs.~\ref{eq:coupledmode1} or~\ref{eq:bpm}, the state of the coherence of the beam is determined by the mutual 
coherence function $\Gamma^{(1)}(x_1,x_2,z) = \langle A^\ast(x_2,z)A(x_1,z)\rangle$~\cite{SalehTeich}, where $\langle \cdots \rangle$ is
the ensemble average, and $A$ is the stochastic field. For the disordered waveguide array of
Figure~\ref{fig:1D-array-fiber}, $x_1$ and $x_2$ are the waveguide array indexes. For example, the 
mutual coherence function of the partially coherent extension of the input Gaussian beam of Eq.~\ref{eq:exponentialInput} 
can be written as
\begin{equation}
\Gamma^{(1)}_{j,k}(z=0)=\exp[-\dfrac{(j - j_0)^2 + (k - j_0)^2}{4{\cal W}^2_0}]
\exp[-\dfrac{(j-k)^2}{{\cal S}^2_0}],\quad j,k=1,\cdots,N,
\label{eq:IncoherentInput}
\end{equation}
where $j_0$ is the index of the waveguide at the center of the input Gaussian beam;
and ${\cal W}_0$ and ${\cal S}_0$ are the spatial and the coherence widths of the beam, respectively.
We note that the optical intensity over the coupled waveguide array is given by the diagonal
element of the mutual coherence function $\Gamma^{(1)}_{j,j}$ ($j$ is the waveguide index)~\cite{SalehTeich}.    

Using Eq.~\ref{eq:coupledmode1} and following a similar procedure that resulted in Eq.~\ref{eq:coupledsol2},
we conclude that the propagation of the mutual coherence function is described by 
\begin{align}
\Gamma^{(1)}_{j,k}(z)=\sum^N_{{k^\prime}=1}\sum^N_{{j^\prime}=1}\ b_{{j^\prime},{k^\prime}}{\mathbb V}^{(j^\prime)}_j{\mathbb V}^{(k^\prime)}_k
\exp[-i({\bar \beta}_{j^\prime}-{\bar \beta}_{k^\prime}) z],
\end{align} 
where the propagation coefficients are determined at $z=0$ according to 
\begin{align}
b_{{j^\prime},{k^\prime}}=\sum^N_{{k^\prime}=1}\sum^N_{{j^\prime}=1}\ \Gamma^{(1)}_{j,k}(z=0) {\mathbb V}^{(j^\prime)}_j{\mathbb V}^{(k^\prime)}_k.
\end{align} 

\begin{figure}[htp]
\centering\includegraphics[width=2.5in]{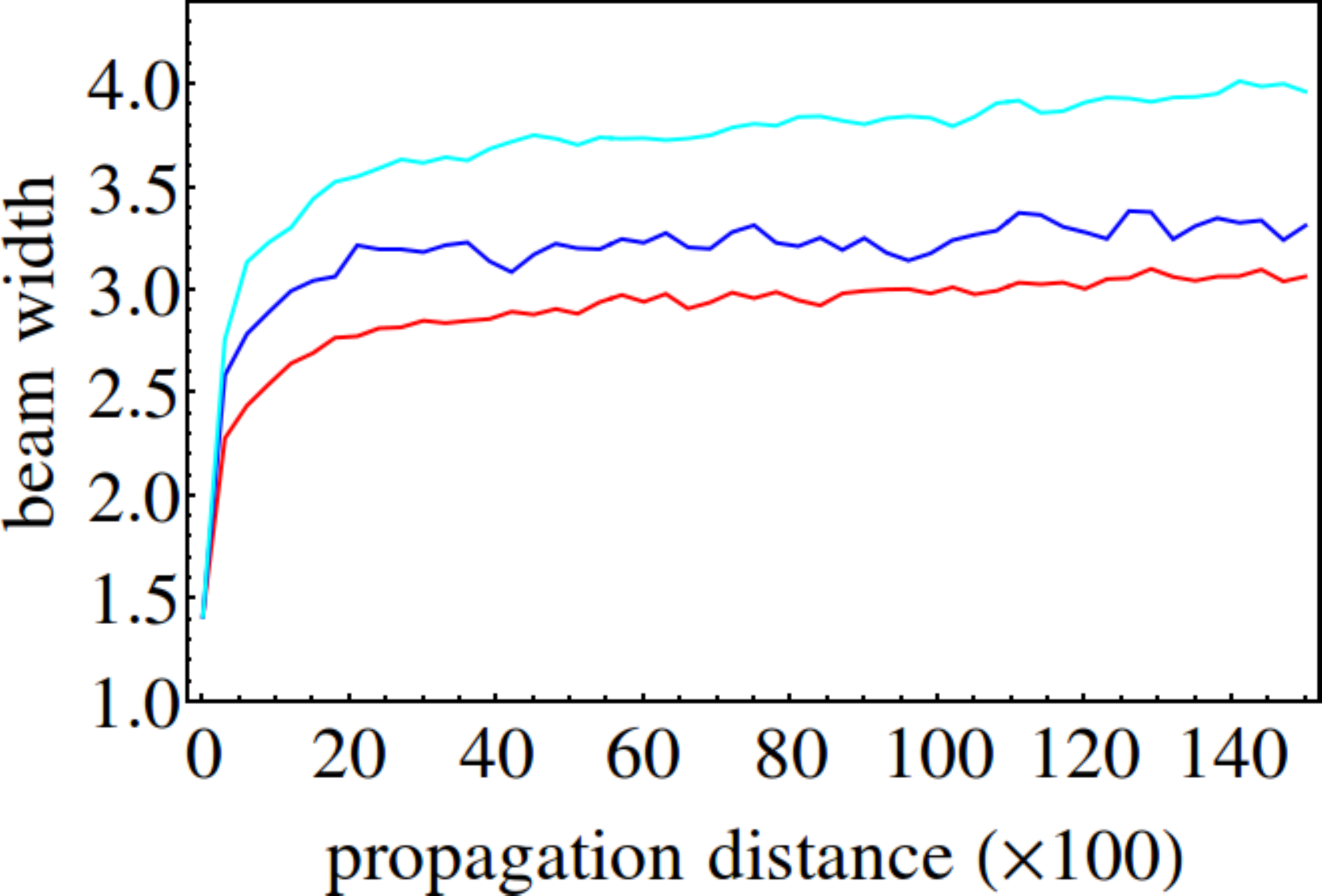}
  \caption{The average beam width as a function of the propagation distance in a disordered coupled waveguide 
           array is plotted in for the case of highly coherent 
           ${\cal S}_0=100$ in red, semi-coherent ${\cal S}_0=5$ in blue, and near-incoherent ${\cal S}_0=2$ in cyan.
           ${\cal W}_0=\sqrt{2}$ and $j_0=101$ have been used and the beam widths are averaged over 100 independent simulations. 
           The disordered coupled waveguide array is defined by $\beta_0 = 6$, $c_0=0.01$, $N=201$, $r_{j}\in{\rm unif}[-0.1,0.1]$,}
    \label{fig:beam-width-coherence}
\end{figure}

Let's consider the disordered coupled waveguide array in Figure~\ref{fig:randomCoupledWaveguides1D}, 
for $\beta_0 = 6$, $c_0=0.01$, $N=201$, $r_{j}\in{\rm unif}[-0.1,0.1]$, and $0 \le z \le 15000$, 
according to the notation used in the discussion following Eq.~\ref{eq:coupledmode1}.
The input light is assumed to be a partially coherent beam defined by Eq.~\ref{eq:IncoherentInput},
with ${\cal W}_0=\sqrt{2}$ and $j_0=101$. The average beam width as a function of the propagation distance is plotted in
Figure~\ref{fig:beam-width-coherence} for the case of highly coherent ${\cal S}_0=100$ in red,
semi-coherent ${\cal S}_0=5$ in blue, and near-incoherent ${\cal S}_0=2$ in cyan. The beam widths are averaged
over 100 independent simulations. The results agree
with the observations of \v{C}apeta, \textit {et al}~\cite{Capeta}, where localization happens for all three cases,
but is strongest for the more coherent input beam.

The propagation of non-classical light in disordered waveguides has also been a subject of interest over the past few years. 
For example, the propagation of a Fock number state, coherent input state, as well as a squeezed state in a disordered waveguide array
was studied by Thompson \textit {et al}~\cite{Thompson}. The quantum mechanical analogue of the disordered tight-binding model used 
for the propagation of light in a disordered waveguide array in Eq.~\ref{eq:coupledmode1} can be heuristically constructed by 
simply elevating the amplitudes of the optical field $A_{j}$ to photon annihilation operators  ${\hat a}_{j}$ in waveguide $j$, 
subject to the following commutation relations
\begin{align}
\left[{\hat a}_{j},{\hat a}_{k}\right]=0,\quad
\left[{\hat a}_{j},{\hat a}^\dagger_{k}\right]=\delta_{kj},\quad
j,k=1,\cdots,N.
\label{eq:commutation}
\end{align} 

The quantum mechanical analogue of Eq.~\ref{eq:coupledmode1} is the ``linear'' Heisenberg equation for the evolution of ${\hat a}_{j}(z)$ and
${\hat a}^\dagger_{j}(z)$ and the creation and annihilation operators at the output ports can be obtained from those at the input ports using 
the Green's function of the disordered medium as
\begin{align}
{\hat a}_{j}(z)=\sum^N_{k=1} G_{jk}(z) {\hat a}_{k}(0).
\label{eq:Greens1}
\end{align} 
The Green's function, in the language of the disordered couple waveguide array and Eq.~\ref{eq:coupledsol2}, can be expressed as
\begin{align}
G_{jk}(z)=\sum^N_{k^\prime=1} {\mathbb V}^{(k^\prime)}_k {\mathbb V}^{(k^\prime)}_j \exp[-i{\bar \beta}_{k^\prime} z].
\end{align} 

Using Eq.~\ref{eq:Greens1}, the input first-order quantum coherence function can be mapped to that at the output according to
\begin{align}
\Gamma^{(1)}_{j,j^\prime}(z)=
\sum^N_{k=1}\sum^N_{k^\prime=1}
\langle G^\ast_{jk}(z)G_{j^\prime k^\prime}(z) \rangle_{\rm dw} 
\Gamma^{(1)}_{k,k^\prime}(0),
\label{eq:quantumcoherence1}
\end{align} 
where $\Gamma^{(1)}_{k,k^\prime}(z)=\langle{\hat a}^\dagger_{k}(z){\hat a}_{k^\prime}(z)\rangle$. The averaging
on the Green's functions are carried over multiple realizations of
{\em disordered waveguides} hence the subscript ``dw'' in $\langle \cdots \rangle_{\rm dw}$.

Equation~\ref{eq:quantumcoherence1} is exactly what one would write for the propagation of the 
first-order classical coherence function as well. The quantum effects enter through the first 
order coherence function, which is defined by the density matrix $\rho$ as 
\begin{align}
\Gamma^{(1)}_{k,k^\prime}(0)={\rm Tr}\Big\{ \rho {\hat a}^\dagger_{k}(0){\hat a}_{k^\prime}(0) \Big\}.
\end{align} 
For example, if the input state to each waveguide is a Glauber coherent state, and the quantum state of the input light is 
defined by $|\alpha_1,\alpha_2,\cdots,\alpha_N\rangle$, we will have
\begin{align}
\Gamma^{(1)}_{k,k^\prime}(0)=\alpha^\ast_k\alpha_{k^\prime}.
\end{align} 
Alternatively, if the input state consists of Fock number states defined by $|n_1,n_2,\cdots,n_N\rangle$, we will have
\begin{align}
\Gamma^{(1)}_{k,k^\prime}(0)=n_k\delta_{k k^\prime}.
\end{align} 

The average output intensity in port $j$, ${\bar I}_j(z)$ is the diagonal element of the output first-order coherence function. For the
case of the coherent state above, we will have  
\begin{align}
{\bar I}_j(z)=\sum^N_{k=1}\sum^N_{k^\prime=1}
\langle G^\ast_{jk}(z)G_{jk^\prime}(z) \rangle_{\rm dw} 
~\alpha^\ast_k\alpha_{k^\prime}.
\label{eq:intalpha1}
\end{align} 
This formula is identical to the intensity we would expect to obtain at port $j$ when classical optical field of amplitude $\alpha_{k}$ 
are coupled at the input port $k$. For the Fock number states, the intensity is given by 
\begin{align}
{\bar I}_j(z)=\sum^N_{k=1}
\langle G^\ast_{jk}(z)G_{jk}(z) \rangle_{\rm dw}~n_k.
\label{eq:intfock1}
\end{align} 
The special case where the photons are coupled into a single input port, say middle port $M$, results 
from Eqs.~\ref{eq:intalpha1} and~\ref{eq:intfock1} can be simplified as
\begin{align}
{\bar I}_j(z)=\langle |G_{jM}(z)|^2 \rangle_{\rm dw}~|\alpha_M|^2,
\quad \ {\rm or}\quad \ =\langle |G_{jM}(z)|^2 \rangle_{\rm dw}~n_M.
\label{eq:intalphafock2}
\end{align} 

Equation~\ref{eq:intalphafock2} is notable--it shows that regardless of the statistics of the
input photons, if it is injected to a single port we will observe the same localization behavior.
In fact, localization is solely dictated by the coherent averaging of the Green's function of the 
disordered waveguide $\langle |G_{jM}(z)|^2 \rangle_{\rm dw}$. This is not the case when light is coupled 
into multiple ports, as can be observed in the difference between Eq.~\ref{eq:intalpha1} and Eq.~\ref{eq:intfock1}:
the coherent interference of the input coherent states in Eq.~\ref{eq:intalpha1} certainly affects the localization,
while Eq.~\ref{eq:intfock1} merely consists of multiple copies of Eq.~\ref{eq:intalphafock2} assembled incoherently 
together because of the total ambiguity in the phase information in the Fock number states. The bottom line
is that the localization is dictated by both the input photon statistics as well as the disordered waveguides, 
and their interaction can be complicated or simple depending on the situation.

\begin{figure}[htp]
\centering\includegraphics[width=2.5in]{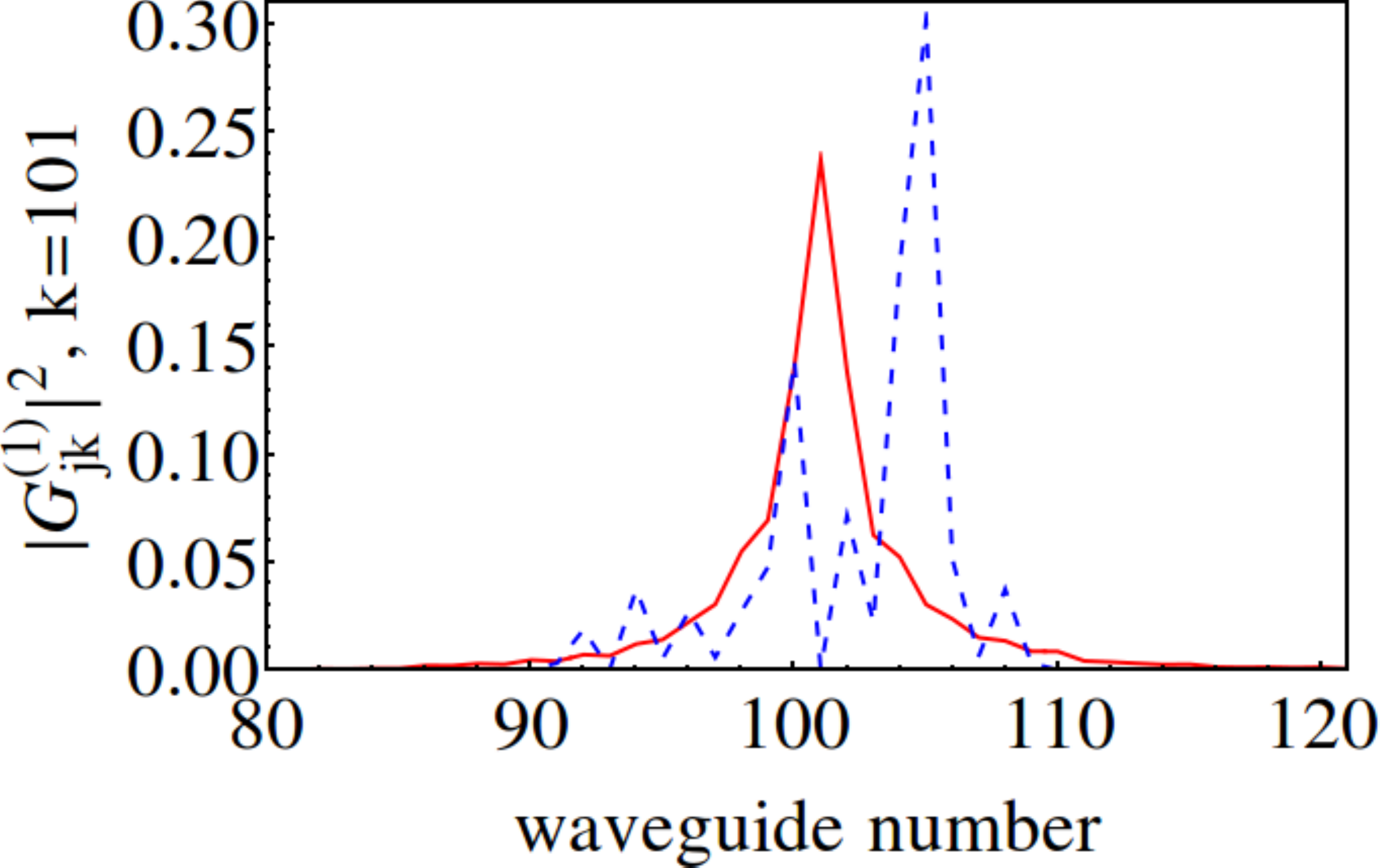}
  \caption{$|G_{jk}(z)|^2$ for $k=101$ is plotted as a function of the output waveguide number for a disordered coupled waveguide array.
           The dashed blue line represents a single simulations, while the
solid red line shows the result of averaging $\langle |G_{jM}(z)|^2 \rangle_{\rm dw}$ over 1000 simulations.}
    \label{fig:G1-coherence}
\end{figure}

In Figure~\ref{fig:G1-coherence} we plot the $|G_{jk}(z)|^2$ at $z=10,000$ for the disordered coupled waveguide array 
in Figure~\ref{fig:randomCoupledWaveguides1D}, for $\beta_0 = 6$, $c_0=0.01$, $N=201$, $r_{j}\in{\rm unif}[-0.1,0.1]$.
We also assume that $k=101$; therefore, the input light is assumed to be coupled only to the middle waveguide 
as discussed in Eq.~\ref{eq:intalphafock2}. $|G_{jk}(z)|^2$ in Figure~\ref{fig:G1-coherence} is plotted as a function
of $j$, which labels the output waveguide number. The dashed blue line represents a single simulation, while the
solid red line shows the result of averaging $\langle |G_{jM}(z)|^2 \rangle_{\rm dw}$ over 1000 simulations.
The exponential localization is clearly observed in Figure~\ref{fig:G1-coherence}.

Higher order quantum coherence functions can also be studied in disordered waveguides. In particular, the second-order quantum 
coherence function gives us information about two photon correlations and Hanbury Brown--Twiss (HBT) effect~\cite{HBT}.
The second-order quantum coherence function at the output of the disordered waveguide can be written as
\begin{align}
\label{eq:quantumcoherence2}
&\Gamma^{(2)}_{j,j^\prime}(z)=
\langle 
{\hat a}^\dagger_{j}(z)
{\hat a}^\dagger_{j^\prime}(z)
{\hat a}_{j^\prime}(z)
{\hat a}_{j}(z)
\rangle \\
\nonumber
&=\sum^N_{k=1}\sum^N_{k^\prime=1}
\langle 
G^\ast_{jk}(z)
G^\ast_{j^\prime k^\prime}(z)
G_{j^\prime l^\prime}(z) 
G_{j l}(z) 
\rangle_{\rm dw} 
\langle 
{\hat a}^\dagger_{k}(0)
{\hat a}^\dagger_{k^\prime}(0)
{\hat a}_{l^\prime}(0)
{\hat a}_{l}(0)
\rangle_{\rm qs}.
\end{align} 

We are interested in two particular scenarios. The first scenario is when the two photons are coupled 
to the same waveguide, say middle port $M$, and the input state is given 
by $({\hat a}^\dagger_{M})^2|{\rm vac}\rangle$. In this case, the second-order quantum 
coherence function is given by
\begin{align}
\Gamma^{(2)}_{j,j^\prime}(z)
=
\langle 
|G_{jM}(z)|^2
|G_{j^\prime M}(z)|^2
\rangle_{\rm dw}. 
\label{fig:G2same}
\end{align} 
The second scenario is when the two photons are coupled 
to two different waveguides, say ports $M$ and $M^\prime$, and the input state is given 
by ${\hat a}^\dagger_{M^\prime}{\hat a}^\dagger_{M}|{\rm vac}\rangle$. In this case, the second-order quantum 
coherence function is given by
\begin{align}
\Gamma^{(2)}_{j,j^\prime}(z)
=
\langle 
|G_{jM}(z)G_{j^\prime M^\prime}(z)+G_{jM^\prime}(z)G_{j^\prime M}(z)|^2
\rangle_{\rm dw}. 
\label{fig:G2different}
\end{align} 
In either scenario, we have indistinguishable photons that co-propagate in the lattice,
and are subject to the ``averaged'' correlations of the propagators of the disordered 
lattice, resulting in rather exotic quantum statistics behavior, as shown for example in 
Ref.~\cite{Lahini2}. 
In the following, we will adopt the same disordered waveguide array as that of Figure~\ref{fig:G1-coherence}
and illustrate the behavior of the second-order quantum coherence function.

\begin{figure}[t]
\centering\includegraphics[width=4.5in]{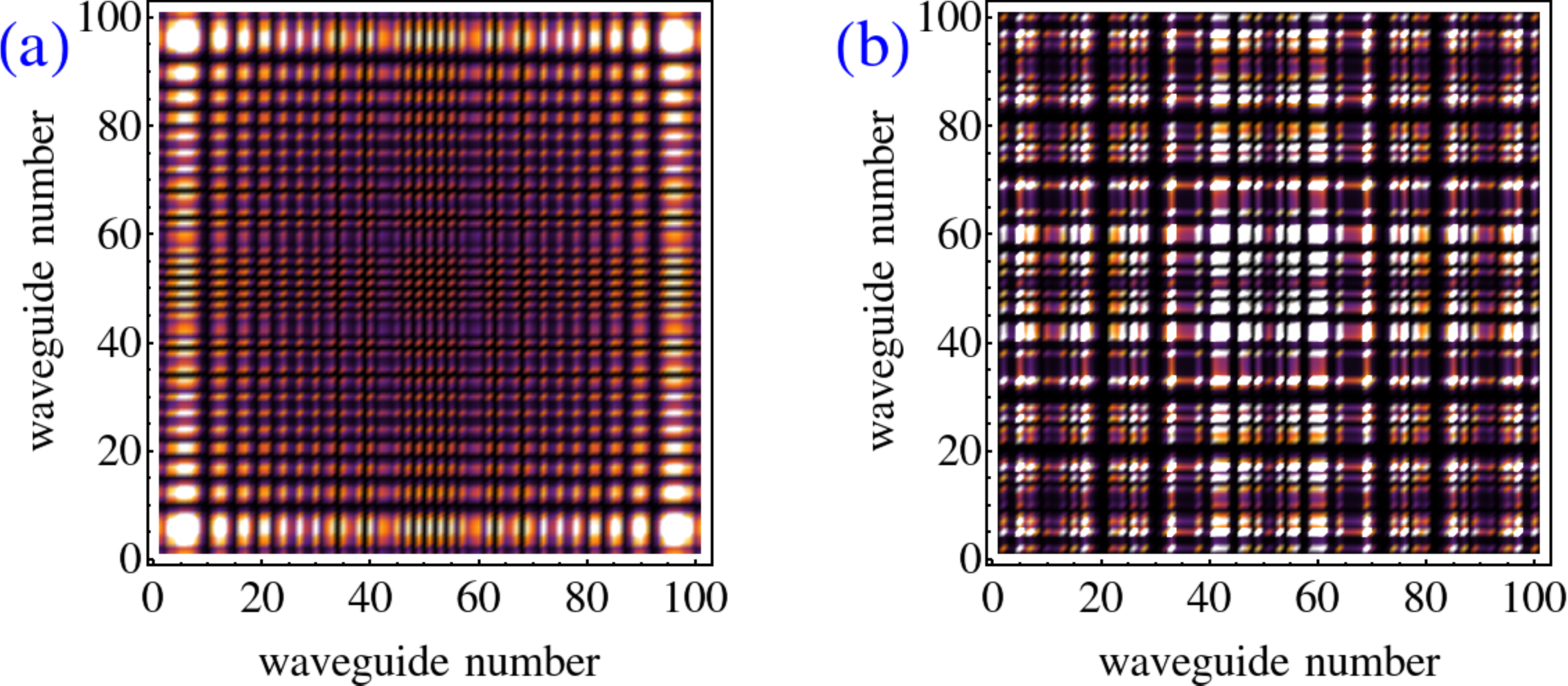}
  \caption{$\Gamma^{(2)}_{j,j^\prime}(z)$ of Eq.~\ref{fig:G2same} is plotted for $M=51$ ($N=101$ waveguides in the lattice) as a function of the output 
           waveguide numbers $j$ and $j^\prime$ for a disorder-free coupled waveguide array for {\claop (a)} $z=2400$, and {\claop (b)} $z=10,000$.}
    \label{fig:G2-same-waveguide-no-disorder}
\end{figure} 
In Figure~\ref{fig:G2-same-waveguide-no-disorder} we plot $\Gamma^{(2)}_{j,j^\prime}(z)$ of Eq.~\ref{fig:G2same}
for the case when the two photons are coupled to the middle waveguide $M=51$ (we use $N=101$ waveguides for easier
simulation), in the absence of any disorder ($r_{j}=0$),
where $z=2400$ and $z=10,000$ is assumed in Figures~\ref{fig:G2-same-waveguide-no-disorder}{\claop (a)} and {\claop (b)}, respectively.
The plots indicate the ballistic expansion of the photons, with larger probability to find the photons in the 
edge waveguides (Figure~\ref{fig:G2-same-waveguide-no-disorder}{\claop (a)}), and also reflections from the boundary for longer propagation
(Figures~\ref{fig:G2-same-waveguide-no-disorder}{\claop (b)}), resulting in strong interference effects.  

\begin{figure}[t]
\centering
\includegraphics[width=4.5in]{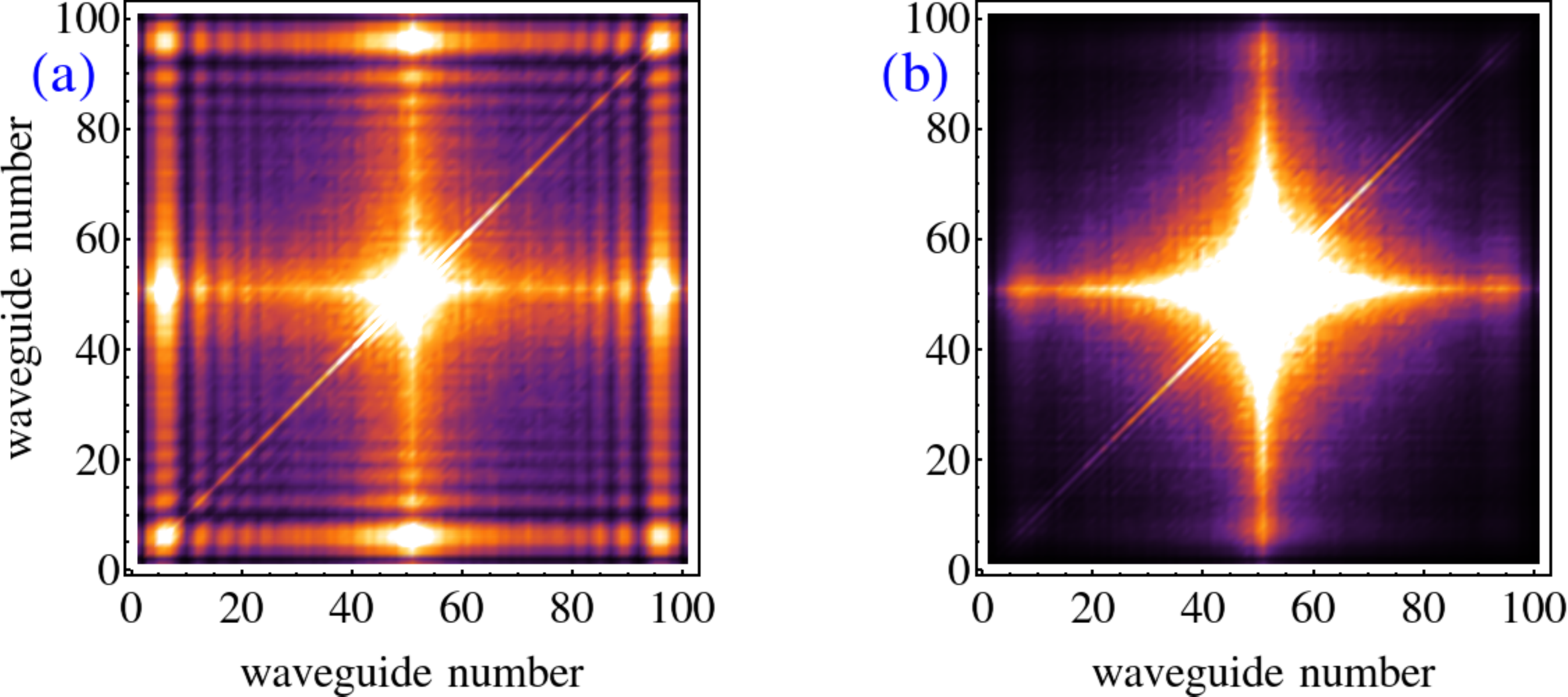}
\includegraphics[width=4.5in]{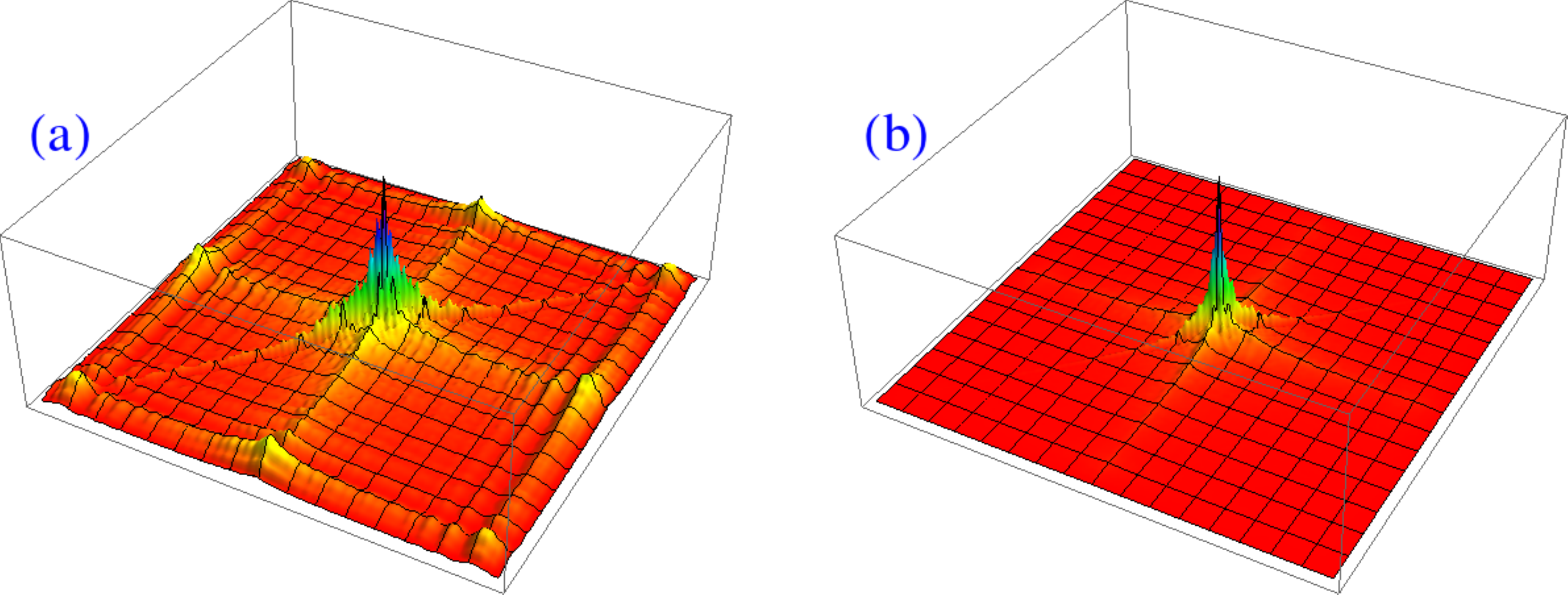}
  \caption{$\Gamma^{(2)}_{j,j^\prime}(z=2400)$ of Eq.~\ref{fig:G2same} is plotted for $M=51$ ($N=101$ waveguides in the lattice) as a function of the output
           waveguide numbers $j$ and $j^\prime$ for a disordered coupled waveguide array for 
           {\claop (a)} $r_{j}\in{\rm unif}[-0.002,0.002]$ and {\claop (b)} $r_{j}\in{\rm unif}[-0.004,0.004]$.
           Both density and 3D plots are presented in each case for easier comparison.}
    \label{fig:G2-same-waveguide-with-disorder}
\end{figure}

In Figure~\ref{fig:G2-same-waveguide-with-disorder} we present plots of the same scenario of 
Figure~\ref{fig:G2-same-waveguide-with-disorder}{\claop (a)}, except in the presence of disorder, averaged over 1000 independent simulations.
We have used $r_{j}\in{\rm unif}[-0.002,0.002]$ for Figure~\ref{fig:G2-same-waveguide-with-disorder}{\claop (a)} 
and $r_{j}\in{\rm unif}[-0.004,0.004]$ for Figure~\ref{fig:G2-same-waveguide-with-disorder}{\claop (b)}. 
We have also presented both density and 3D plots in each case for easier comparison. 
Note that we have taken $z=2400$ in order to prevent reflections from the outer waveguides; such reflections
complicate the analysis and are not observed as long as there is a sufficiently large number of coupled waveguides
in the array for a given propagation distance.

In the presence of low disorder as in Figure~\ref{fig:G2-same-waveguide-with-disorder}{\claop (a)}, it is possible for 
both photons to remain localized (center of the plot), one remains localized and one propagates freely (edge-centers
of the plot), or both freely propagate (corners of the plot), in decreasing order of probability. Once the disorder is
increased as in Figure~\ref{fig:G2-same-waveguide-with-disorder}{\claop (b)}, it is almost only possible to have localized
photons with small non-zero chance of having one localized and one freely propagating photon. Once the disorder
is further increased e.g., $r_{j}\in{\rm unif}[-0.01,0.01]$ (not plotted here), the only reasonable non-zero 
probability is to have both photons highly localized. We recall that the disorder-free case in 
Figure~\ref{fig:G2-same-waveguide-no-disorder}{\claop (a)} favors {\em both freely propagating} to 
{\em one localized and one freely propagating} to {\em both localized}, respectively, and the presence of disorder reverses this behavior.

\begin{figure}[t]
\centering\includegraphics[width=4.5in]{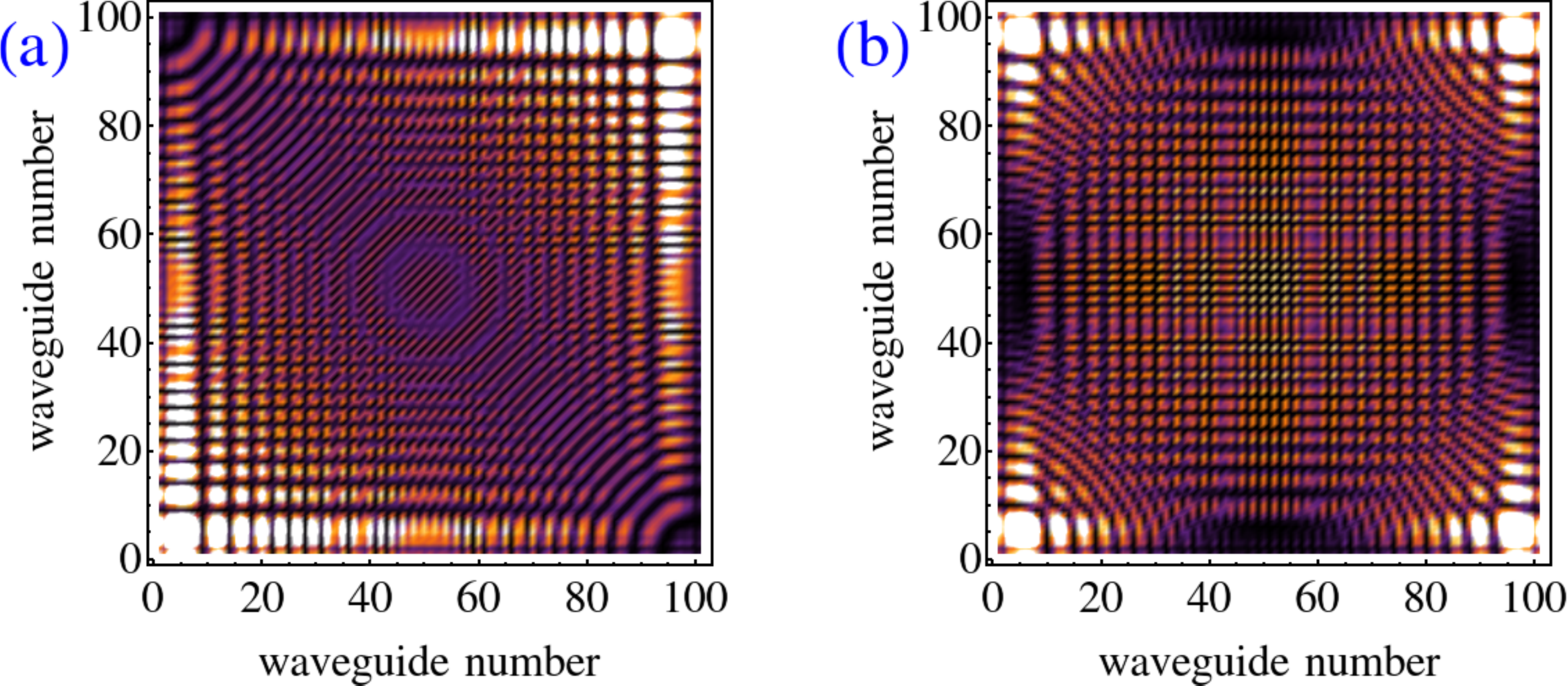}
  \caption{$\Gamma^{(2)}_{j,j^\prime}(z)$ of Eq.~\ref{fig:G2different} is plotted for {\claop (a)} $M=50$, $M^\prime=51$ ($N=101$ waveguides in the lattice), and 
          {\claop (b)} $M=50$, $M^\prime=52$, as a function of the output waveguide numbers $j$ and $j^\prime$ for a disorder-free 
          $N=101$ coupled waveguide array and $z=2400$.}
    \label{fig:G2-more-separation-no-disorder}
\end{figure} 
In Figure~\ref{fig:G2-more-separation-no-disorder} we plot $\Gamma^{(2)}_{j,j^\prime}(z)$ of Eq.~\ref{fig:G2different}
for the case where the photons are coupled to different waveguides:
$M=50$ and $M^\prime=51$ in Figure~\ref{fig:G2-more-separation-no-disorder}{\claop (a)} and 
$M=50$ and $M^\prime=52$ in Figure~\ref{fig:G2-more-separation-no-disorder}{\claop (b)}. 
Either case is in the absence of any disorder ($r_{j}=0$) for $z=2400$. The plots indicate the ballistic expansion of 
the photons, with larger probability to find the photons in the 
edge waveguides (Figure~\ref{fig:G2-same-waveguide-no-disorder}{\claop (a)}), and also reflections from the boundary for longer propagation
(Figures~\ref{fig:G2-same-waveguide-no-disorder}{\claop (b)}), resulting in strong interference effects.

When the two photons are coupled to neighboring waveguides as in Figure~\ref{fig:G2-more-separation-no-disorder}{\claop (a)},
the most likely scenario is that the two photons ballistically propagate to the same edge of the lattice,
and it is impossible to find the two photons in the opposite edges of the lattice; however, there is a small
yet non-vanishing probability to find one photon remaining in the center waveguide and the other in the edge lattice. When the two in-coupled photons 
are separated by one waveguide as in Figure~\ref{fig:G2-more-separation-no-disorder}{\claop (b)}, the most likely scenario is that 
the two photons ballistically propagate to either edge of the lattice, and it is impossible to find one photon remaining
in the center waveguide and the other in the edge.

\begin{figure}[t]
\centering
\includegraphics[width=4.5in]{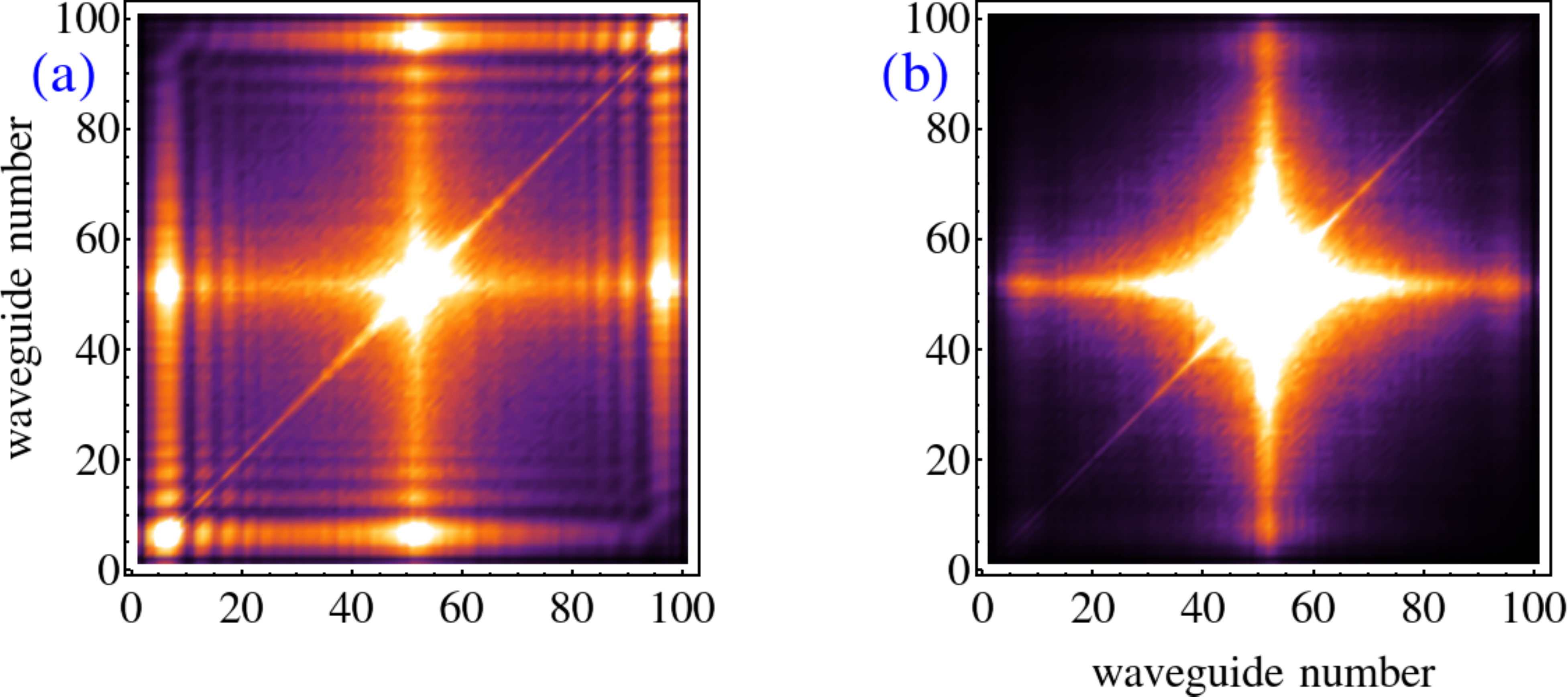}
\includegraphics[width=4.5in]{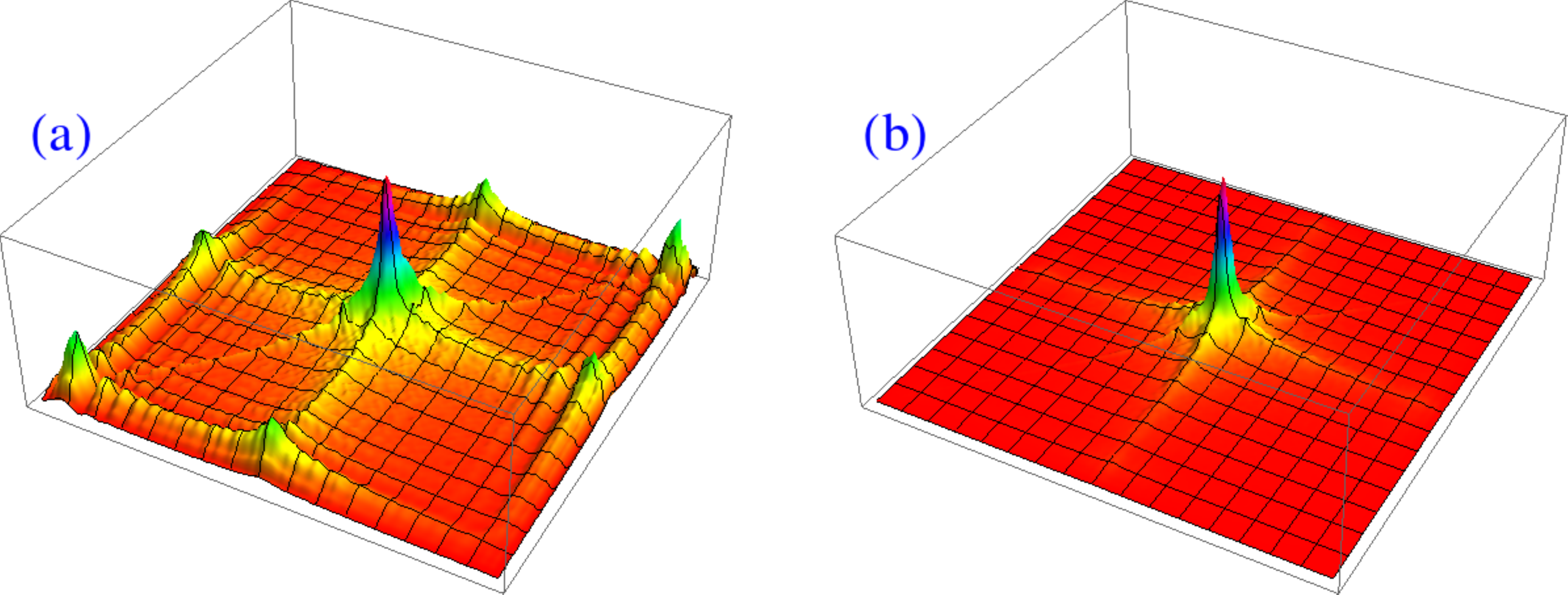}
  \caption{$\Gamma^{(2)}_{j,j^\prime}(z=2400)$ of Eq.~\ref{fig:G2different} is plotted for 
           $M=51$ and $M^\prime=52$ ($N=101$ waveguides in the lattice) as a function of the output
           waveguide numbers $j$ and $j^\prime$ for a disordered coupled waveguide array for 
           {\claop (a)} $r_{j}\in{\rm unif}[-0.002,0.002]$ and {\claop (b)} $r_{j}\in{\rm unif}[-0.004,0.004]$.
           Both density and 3D plots are presented in each case for easier comparison.}
    \label{fig:G2-neighbor-waveguide-with-disorder}
\end{figure}
In Figure~\ref{fig:G2-neighbor-waveguide-with-disorder} we present plots of the same scenario of 
Figure~\ref{fig:G2-more-separation-no-disorder}{\claop (a)}, except in the presence of disorder, averaged over 1000 independent simulations.
We have used $r_{j}\in{\rm unif}[-0.002,0.002]$ for Figure~\ref{fig:G2-neighbor-waveguide-with-disorder}{\claop (a)} 
and $r_{j}\in{\rm unif}[-0.004,0.004]$ for Figure~\ref{fig:G2-neighbor-waveguide-with-disorder}{\claop (b)}. 
We have also presented both density and 3D plots in each case for easier comparison.
We note that increasing the level of disorder compared with the disorder-free case in Figure~\ref{fig:G2-more-separation-no-disorder}{\claop (a)}
decreases the probability of ballistic co-propagation of photons to the same edge, while increasing
the probability of the localization of both photons in the center waveguide. 
Therefore, the disorder-free case of Figure~\ref{fig:G2-more-separation-no-disorder}{\claop (a)} favors {\em both freely propagating to the same edge} 
to {\em one localized and one freely propagating} to {\em both localized}, respectively, and the presence of disorder reverses this behavior.  

We encourage the interested reader to consult Ref.~\cite{Lahini2} for a more detailed account of the behavior of $\Gamma^{(2)}_{j,j^\prime}(z)$,
upon which much of the preceding discussion is based. 

The above discussion only covers the case of off-diagonal disorder in a disordered coupled array waveguide.
The comparison with the case of the diagonal disorder is performed in Ref.~\cite{Lahini3}. It is shown that
the evolution and localization of the photon density (optical intensity) is similar in the two cases of diagonal 
and off-diagonal disorder, as expected. However, the intensity correlation (photon density-density correlation 
or $\Gamma^{(2)}_{j,j^\prime}(z)$) carries a distinct signature of the type of disorder. The propagation of 
an entangled-photon pairs in a disordered waveguide array has also been explored in Ref.~\cite{Abouraddy}.
It has been shown that while neither photon is localized, the two-photon separation in coincidence space
is: this behavior is called Anderson colocalization. The increase in entanglement is accompanied by a
gradual evolution from Anderson-localization to Anderson colocalization. 
\begin{highlight}
\textbf{Highlights:}
\begin{itemize}
\item If all the eigenmodes of a disordered waveguide are exponentially localized, any partially incoherent beam
exhibits localization with exponentially decaying tails, after a sufficiently long propagation distances. 
\item The more incoherent the in-coupling beam is, the longer it diffusively spreads while propagating in the medium.
Therefore, localization is delayed by incoherence compared with the case of a coherent in-coupling beam.
\item The observed localization behavior in a disordered waveguide is affected by the quantum statistics of the in-coupling 
photons. The effect can be mainly observed in the second-order quantum coherence function, which gives us information about 
two photon correlations.
\end{itemize}
\end{highlight}
\section{Conclusions}
\label{sec:conclusions}
A tutorial review of the transverse Anderson localization of light in disordered waveguides is presented. In addition
to the intriguing fundamental science behind Anderson localization, there are potential applications for light interaction
with disordered dielectric systems. Examples such as spatially-multiplexed beam delivery and high-quality
image transport were explored. There are still plenty of open questions in both linear and nonlinear dynamics of disordered 
systems that have yet to be answered. There has been a recent surge of interest among the optics community in Anderson localization,
and it would have been impossible to cover them all even briefly in this survey. This brief tutorial review is intended to demystify 
Anderson localization for the broader audience and provide sufficient background to newcomers in this field to follow up
with studying other more in-depth reviews and research articles.
\section{Acknowledgments}
The author acknowledges support by Grant Number 1029547 from the National Science Foundation. 
The author is also grateful for the opportunity to collaborate on disordered optical fibers with 
Salman Karbasi, Ryan Frazier, Craig Mirr, Karl Koch, John Ballato, Thomas Hawkins,
Claudio Conti, and Marco Leonetti, over the past four years.

\newpage
\begin{wrapfigure}{l}{2.8cm}
\vspace*{-4mm}
  \begin{center}
    \includegraphics[width=2.6cm]{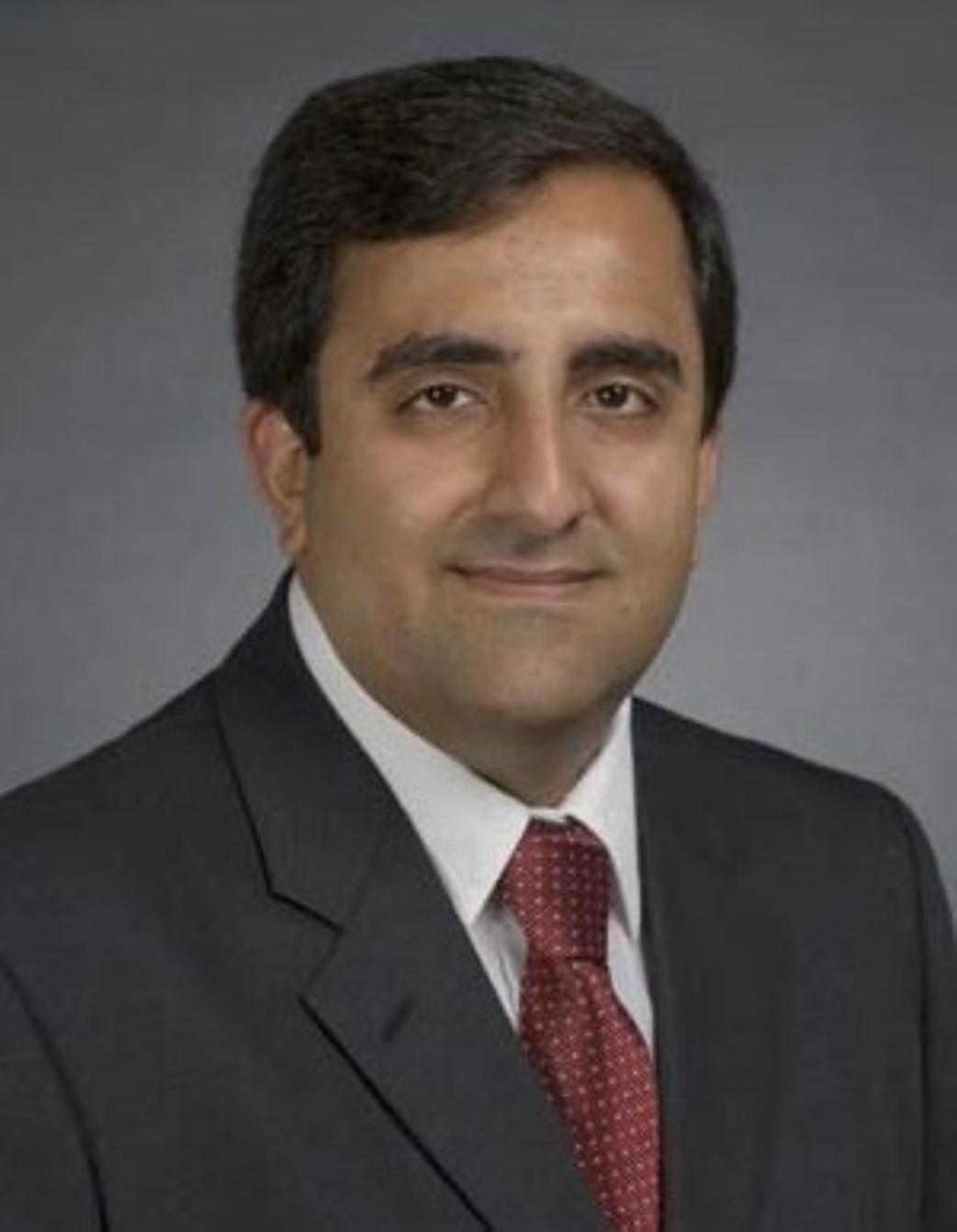}
  \end{center}
\vspace*{-5mm}
\end{wrapfigure}
\noindent
Arash Mafi is an Associate Professor of Physics and Astronomy and a 
member of the Center for High Technology Materials (CHTM) at the University of New Mexico.
He received his undergraduate degree in Physics from Sharif University of Technology in 1995,
and his Ph.D. degrees in Physics from The Ohio State University in 2001.
Following his postdoctoral appointments at the University of Arizona 
in Physics and The Optical Sciences Center, he joined Corning Inc. in 2005 as a Senior 
Research Scientist working on optical fibers and liquid crystal
displays. He moved to the University of Wisconsin-Milwaukee in 2008, where he
was an Associate Professor of Electrical Engineering and Computer 
Science, before joining the University of New Mexico in 2014. He was a recipient of 
the Early Career Development (CAREER) Award from the National Science Foundation in 
2013. His research interests include quantum and nonlinear behavior of optical 
waveguides, and light propagation in disordered media with applications in optical 
image transport.
\end{document}